\documentclass[a4paper,12pt]{article}

\usepackage[english]{babel}
\usepackage[utf8x]{inputenc}
\usepackage[T1]{fontenc}

\usepackage[a4paper,top=2cm,bottom=2.4cm,left=2cm,right=2cm,marginparwidth=1.75cm]{geometry}

\usepackage{amsmath}
\usepackage{amssymb}
\usepackage{graphicx}
\usepackage[colorinlistoftodos]{todonotes}
\usepackage[colorlinks=true, allcolors=blue]{hyperref}
\usepackage{latexsym}
\usepackage{mathtools}
\usepackage{comment}



\newpage
\begin{document}
\begin{flushright}
UT-Komaba/19-1
\end{flushright}
\begin{center}
\vspace{50mm}
\textbf{\large Open superstring field theory including the Ramond sector}\\
\vspace{2mm}
\textbf{ \large based on the supermoduli space}\\
\vspace{10mm}
{\large Tomoyuki Takezaki} \\
\vspace{10mm}
{\it \large Institute of Physics, The University of Tokyo,}\\
\vspace{1mm}
{\it \large Komaba, Meguro-ku, Tokyo 153-8902, Japan}\\
\vspace{10mm}
{takezaki@hep1.c.u-tokyo.ac.jp}
\end{center}
\bigskip
\begin{abstract}
We construct a gauge-invariant action for open superstring field theory up to quartic order including the Ramond sector based on the covering of the supermoduli space of super-Riemann surfaces, following the approach presented by Ohmori and Okawa.
Since our approach is based on the covering of the supermoduli space, the resulting action naturally has an $A_\infty$ structure. 
In our construction, adding stubs to the star product can be easily incorporated, and we explicitly construct an action for open superstring field theory including the Ramond sector with stubs up to quartic interactions.
\end{abstract}

\newpage
\tableofcontents

\section{Introduction}

String field theory is one approach to non-perturbative formulations of string theory.
A famous example of string field theory is open bosonic string field theory by Witten~\cite{Witten:1985cc}. Its action has a Chern-Simons-like gauge invariance.
One of the achievements in open bosonic string field theory is the construction of the analytic solution for the tachyon vacuum by Schnabl~\cite{Schnabl:2005gv}.
If we are interested in quantum aspects of string field theory, however, bosonic string field theory suffers from vacuum instability caused by tachyon fields, and it is desirable to consider superstring field theory.

One of the successful constructions of an action for the Neveu-Schwarz sector of open superstring field theory was presented by Berkovits~\cite{Berkovits:1995ab}. 
The key ingredient of this formulation is use of the large Hilbert space of the superconformal ghost~\cite{Friedan:1985ge}.
The fundamental string field is in the large Hilbert space, and the action is written in a closed form, realizing a Wess-Zumino-Witten-like gauge invariance.
However, its gauge fixing turned out to be very difficult. 
On the other hand, it is known that the gauge-fixing by the Batalin-Vilkovisky formalism is straightforward if an action has a structure called cyclic $A_\infty$, and an action for the Neveu-Schwarz sector of open superstring field theory with a cyclic $A_\infty$ structure was constructed by Erler, Konopka, and Sachs~\cite{Erler:2013xta}. 
A set of multi-string products with the $A_\infty$ relations is constructed in a recursive manner, and the action is not given in a closed form. 
Although these two actions~\cite{Berkovits:1995ab,Erler:2013xta} are different in appearance, it was proved that they are equivalent under partial gauge fixing and field redefinition~\cite{Erler:2015rra,Erler:2015uba,Erler:2015uoa}.

Recently, a complete action for open superstring field theory including the Ramond sector was constructed by Kunitomo and Okawa in~\cite{Kunitomo:2015usa}, extending the Wess-Zumino-Witten-like action for the Neveu-Schwarz sector~\cite{Berkovits:1995ab}. 
Furthermore, Kunitomo, Okawa, Sukeno, and the author showed that the Feynman rules derived from this action correctly reproduce on-shell four-point and five-point amplitudes involving fermions at the tree level~\cite{Kunitomo:2016bhc}. 
On the other hand, a complete action for open superstring field theory with the cyclic $A_\infty$ structure including the Ramond sector was constructed~\cite{Erler:2016ybs,Konopka:2016grr}, extending the action for the Neveu-Schwarz sector of open superstring field theory~\cite{Erler:2013xta}. 
Although these two complete actions are different in appearance, it was proved that they are equivalent under partial gauge fixing and field redifinition~\cite{Erler:2016ybs}. 

In the formulations~\cite{Kunitomo:2015usa,Erler:2016ybs,Konopka:2016grr}, the Ramond string field is in the restricted subspace, which preserves the BRST cohomology of the original Hilbert space of the Ramond sector.
As we will explain later, the restriction on the Ramond string field is closely related to the supermoduli space of super-Riemann surfaces.
Another approach to incorporate the Ramond sector in a covariant manner is to use an unrestricted Ramond string field and spurious free fields, and an action for closed superstring field theory is constructed by Sen~\cite{Sen:2015uaa}.
The approach by Sen can be applied to open superstring field theory.\footnote{The calculations in~\cite{Kunitomo:2016bhc} can be also interpreted as those for the formulation~\cite{Sen:2015uaa}.}

In the formulations of open superstring field theory which we mentioned so far, the use of the large Hilbert space played an important role.
In the Wess-Zumino-Witten-like formulation, the fundamental Neveu-Schwarz string field is in the large Hilbert space.
In the recursive construction of an action with the cyclic $A_\infty$ structure in~\cite{Erler:2013xta}, although its fundamental string field is in the small Hilbert space, the large Hilbert space is used in intermediate steps of the construction of multi-string products. 
However, the use of the large Hilbert space obscures the relation between gauge invariance and the covering of the supermoduli space. 
To improve the understanding of the relation between them, it would be useful to construct open superstring field theory without using the large Hilbert space. 
Recently, a new approach to formulating open superstring field theory based on the covering of the supermoduli space of super-Riemann surfaces was proposed by Ohmori and Okawa~\cite{Ohmori:2017wtx}. 
They explicitly constructed a gauge-invariant action in the Neveu-Schwarz sector up to quartic interactions.
Since their approach is based on the covering of the supermoduli space, the resulting action naturally has an $A_\infty$ structure.

In this paper, we extend the approach presented in~\cite{Ohmori:2017wtx} and construct a gauge-invariant action for open superstring field theory based on the supermoduli space including the Ramond sector up to quartic interactions.
Since our approach is based on the covering of the supermoduli space, our action also has an $A_\infty$ structure.
In our construction, adding stubs to the star product is easily incorporated, and we explicitly construct an action for open superstring field theory including the Ramond sector with stubs up to quartic interactions.

\section{Open bosonic string field theory \label{sec:stub}}
\setcounter{equation}{0}
In this section, we review open bosonic string field theory. 
After describing the cyclic $A_\infty$ structure and the extended BRST formalism, we illustrate the construction of an action of open bosonic string field theory with stubs based on the method presented in~\cite{Ohmori:2017wtx}.

\subsection{Cyclic \texorpdfstring{$A_\infty$}{Lg} structure in open bosonic string field theory}
The action of open bosonic string field theory by Witten~\cite{Witten:1985cc} is given by
\begin{equation}
S = {}-\frac{1}{2} \langle \, \Psi, Q\Psi \, \rangle 
{}-\frac{g}{3} \langle \, \Psi, \Psi \ast \Psi \, \rangle \,, \label{Witten_Open_bosonic_action}
\end{equation}
where the open bosonic string field $\Psi$ is a state in the Hilbert space of the boundary CFT, $Q$ is the BRST operator, and $A \ast B$ is the star product of string fields $A$ and $B$.
The string field $\Psi$ is a Grassmann-odd state carrying the ghost number $1$.
The BRST operator $Q$ and the star product satisfy the following equations:\footnote{Here and in what follows, a state in the exponent of $-1$ represents its Grassmann parity: it is 0 mod 2 for a Grassmann-even state and 1 mod 2 for a Grassmann-odd state.}
\begin{align}
Q^2 A_1 = & \ 0, \label{BRST charge is nilpotent} \\
Q (A_1 \ast A_2) = & \ QA_1 \ast A_2 +(-1)^{A_1} A_1 \ast QA_2 \,, \label{Leibnitz_Q_ast}\\
(A_1 \ast A_2) \ast A_3 = & \ A_1 \ast (A_2 \ast A_3) \label{associativity_of_star_product}\,.
\end{align}
The action~\eqref{Witten_Open_bosonic_action} is invariant under the following gauge transformation:
\begin{equation}
\delta \Psi = Q \Lambda -g \Lambda \ast \Psi + g \Psi \ast \Lambda \,,
\end{equation}
where $\Lambda$ is the gauge parameter. 
The string field $\Lambda$ is a Grassmann-even state of ghost number~0. In the proof of gauge invariance, we use the equations
\begin{align}
\langle \, A_1, A_2 \, \rangle = & \ (-1)^{A_1A_2} \langle \, A_2, A_1 \, \rangle \,, \label{BPZ_antisym} \\ 
\langle \, QA_1, A_2 \, \rangle = & \ {}-(-1)^{A_1} \langle \, A_1, QA_2 \, \rangle \,, \label{BPZ_Q_cyclic} \\
 \langle \, A_1 \ast A_2, A_3 \, \rangle = & \ \langle \, A_1, A_2 \ast A_3 \, \rangle \label{BPZ_ast_cyclic} 
\end{align}
together with the equations~\eqref{BRST charge is nilpotent}, \eqref{Leibnitz_Q_ast}, and \eqref{associativity_of_star_product}.

Gauge-invariant actions can be constructed in terms of a set of multi-string products with properties we will explain in the following. Let us consider an action of string field theory in the following form:
\begin{equation}
S = {}-\frac{1}{2} \langle \, \Psi, Q \Psi \, \rangle 
{}-\frac{g}{3} \langle \, \Psi, V_2( \Psi, \Psi) \, \rangle 
{}-\frac{g^2}{4} \langle \, \Psi, V_3 (\Psi, \Psi, \Psi) \, \rangle 
+ \mathcal{O}(g^3) \,. \label{A infty action}
\end{equation}
The two-string product $V_2(A_1, A_2)$ is defined for two string fields $A_1$ and $A_2$, and its Grassmann parity is
\begin{equation}
(-1)^{V_2(A_1,A_2)} = (-1)^{A_1+A_2} \,.
\end{equation}
The three-string product $V_3(A_1, A_2, A_3)$ is defined for three string fields $A_1$, $A_2$, and $A_3$, and its Grassmann parity is 
\begin{equation}
(-1)^{V_3(A_1,A_2,A_3)} = (-1)^{A_1+A_2+A_3+1} \,.
\end{equation}
We assume that the two-string product and the three-string product satisfy the following cyclicity equations:
\begin{align}
\langle \, A_1, V_2( A_2, A_3) \, \rangle 
= & \ \langle \, V_2(A_1, A_2), A_3 \, \rangle \label{V2 cyclicity} \,, \\
\langle \, A_1, V_3( A_2, A_3, A_4) \, \rangle 
= & \ {}-(-1)^{A_1} \langle \, V_3(A_1, A_2, A_3), A_4 \, \rangle \,, \label{V3 cyclicity}
\end{align}
or equivalently,
\begin{align}
\langle \, A_1, V_2 (A_2, A_3) \, \rangle 
= & \ (-1)^{A_1(A_2+A_3)} \langle \, A_2, V_2 (A_3, A_1) \, \rangle \,,\\ 
\langle \, A_1, V_3 (A_2, A_3, A_4) \, \rangle 
= & \, {}-(-1)^{A_1(A_2+A_3+A_4)} \langle \, A_2, V_3 (A_3, A_4, A_1) \, \rangle \,.
\end{align}
Using the cyclicity equations, the variation of the action~\eqref{A infty action} is given by
\begin{equation}
\delta S = {}- \langle \, \delta \Psi, Q \Psi \, \rangle 
{}-g^2 \langle \, \delta \Psi, V_3 (\Psi, \Psi, \Psi) \, \rangle 
+ \mathcal{O}(g^3) \,.
\end{equation}
The equation of motion is
\begin{equation}
0 = Q \Psi + g V_2 (\Psi, \Psi) + g^2 V_3(\Psi, \Psi, \Psi) + \mathcal{O}(g^3) \,. \label{A infty EOM}
\end{equation}
We can show that the action~\eqref{A infty action} is invariant under the following gauge transformation up to $\mathcal{O}(g^3)$:
\begin{equation}
\begin{split}
\delta \Psi = & \ Q \Lambda -g V_2 (\Lambda, \Psi) +g V_2 (\Psi, \Lambda) \\
& \quad -g^2 V_3 (\Lambda, \Psi, \Psi) +g^2 V_3 (\Psi, \Lambda, \Psi) -g^2V_3 (\Psi, \Psi, \Lambda) +\mathcal{O}(g^3) \,,
\end{split}
\end{equation}
if the BRST operator $Q$ and the multi-string products $V_2$ and $V_3$ satisfy the following equations:
\begin{align}
0 = & \ Q^2 A_1, \label{1st A infty} \\
0 = & \ QV_2(A_1, A_2) -V_2(QA_1, A_2) -(-1)^{A_1}V_2(A_1,QA_2) \,,  \label{2nd A infty}\\
0 = & \ QV_3(A_1,A_2,A_3) +V_3(QA_1,A_2,A_3) \notag  \\
& \quad + (-1)^{A_1}V_3(A_1,QA_2,A_3) +(-1)^{A_1+A_2}V_3(A_1,A_2,QA_3) \label{3rd A infty} \\
& \quad {}- V_2(V_2(A_1, A_2), A_3) + V_2(A_1, V_2(A_2, A_3)) \,. \notag
\end{align}
These equations can be extended to higher orders, and such the equations for multi-string products are called $A_\infty$ relations. 
If an action is constructed from a set of multi-string products which satisfy the $A_\infty$ relations and the cyclicity equations, the action has a cyclic $A_\infty$ structure.
Furthermore, quantization of string field theory based on the Batalin-Vilkovisky formalism is straightforward if the theory has the cyclic $A_\infty$ structure. 

Since the star product is associative~\eqref{associativity_of_star_product} and the BRST operator and the star product satisfy \eqref{Leibnitz_Q_ast}, the BRST operator and the star product satisfy the $A_\infty$ relations without introducing higher multi-string products $V_3, V_4, \cdots$. 
Since the star product satisfies the cyclicity equation~\eqref{BPZ_ast_cyclic}, the action~\eqref{Witten_Open_bosonic_action} has the cyclic $A_\infty$ structure.

In open superstring field theory or open bosonic string field theory with stubs, the two-string product $V_2$ is not associative:
\begin{equation}
V_2(V_2(A_1, A_2), A_3) \neq \ V_2(A_1, V_2 (A_2, A_3)) \,.
\label{non-associativity of V2}
\end{equation}
We can recover the gauge-invariance at quartic order by constructing a three-string product satisfying the $A_\infty$ relation~\eqref{3rd A infty}.

\subsection{Extended BRST formalism \label{subsec:extBRSTstub}}
The extended BRST transformation introduced by Witten in~\cite{Witten:2012bh} is useful when we construct an action for open superstring field theory. 
In this subsection, we explain the extended BRST formalism for the bosonic string.\footnote{The content of this subsection is based on the paper by Ohmori and Okawa~\cite{Ohmori:2017wtx}.
}
$\,$ Let us consider the tree-level four string amplitude. Using the upper-half plane (UHP), the tree-level four-point amplitude is given by
\begin{equation}
    \mathcal{A} 
    = \ g^2 \int_0^1 dt \,
    \langle \, \Psi_1 (0) \, b_{-1} \cdot \Psi_2 (t) \, \Psi_3 (1) \, \Psi_4 (\infty) \, \rangle_\text{UHP} \,,
    \label{tree-level four-string amplitude: b_(-1)}
\end{equation}
where $\Psi(t) = cV^\text{matter}(t)$ is an unintegrated vertex operator, and we used the relation 
\begin{equation}
    b_{-1} \cdot \Psi(t) = V^\text{matter}(t) \,,
\end{equation}
where $b_{-1}$ is the line integral of $b$ ghost:
\begin{equation}
    b_{-1} = \, \oint \, \frac{dz}{2 \pi i} \, b(z) \,.
\end{equation}
Using the line integral $L_{-1}$ of the energy-momentum tensor $T(z)$:
\begin{equation}
    L_{-1} = \, \oint \, \frac{dz}{2 \pi i} \, T(z) \,,
\end{equation}
we find
\begin{equation}
    L_{-1} \cdot V^\text{matter} (t) = \, \partial_t V^\text{matter} (t) \,,
    \label{action of L_(-1) on the vertex operator equals derivation}
\end{equation}
we can express this amplitude in the following form:
\begin{equation}
    \mathcal{A} 
    =  \ g^2 \int_0^1 dt \,
    \langle \, \Psi_1 (0) \, b_{-1}e^{tL_{-1}} \cdot \Psi_2 (0) \, \Psi_3 (1) \, \Psi_4 (\infty) \, \rangle_\text{UHP} \,.
\end{equation}
Following~\cite{Witten:2012bh}, We introduce a Grassmann-odd parameter $\tilde{t}$. Our convention for the integral over the Grassmann-odd variable is
\begin{equation}
    \int d\tilde{t} \, \tilde{t} = 1 \,.
\end{equation}
Then we can write the four-string amplitude as
\begin{equation}
    \mathcal{A} = {}-\, g^2 \int_0^1 dt \int d\tilde{t} \,
    \langle \, \Psi_1 (0) \, 
    e^{\tilde{t} b_{-1}+t L_{-1}}\cdot \Psi_2 (t) \, \Psi_3 (1) \, \Psi_4 (\infty) \, \rangle_\text{UHP} \,.
\end{equation}
In~\cite{Witten:2012bh}, the BRST operator is extended to act not only on the operators in boundary CFT, but also on the modulus $t$ and the Grassmann-odd variable $\tilde{t}$. We denote the extended BRST operator by $Q'$. It acts on $t$ and $\tilde{t}$ in the following manner:
\begin{equation}
    [ \, Q', t \, ] = \, \tilde{t} \,, \quad
    \{ \, Q', \tilde{t} \, \} = \, 0 \,,
\end{equation}
and its action on the boundary CFT is the same way as the BRST operator $Q$. For example,
\begin{equation}
    \{ \, Q', b_{-1} \, \} = \, L_{-1} \,, \quad
    [ \, Q', L_{-1} \, ] = \, 0 \,.
\end{equation}
We can show that the extended BRST operator is nilpotent:
\begin{equation}
    {Q'}^2 = \, 0 \,,
\end{equation}
and we can express $Q'$ as
\begin{equation}
    Q' = \, Q + \tilde{t} \, \partial_t \,.
\end{equation}
Note that the combination $\tilde{t}b_{-1} + tL_{-1}$ is annihilated by the action of $Q'$:
\begin{equation}
    [ \, Q', \tilde{t}b_{-1} + tL_{-1} \, ] 
    = \, {}-\tilde{t} L_{-1} + \tilde{t} L_{-1} 
    = \, 0 \,.
\end{equation}
In fact, this combination can be generated by the action of $Q'$ on $tb_{-1}$:
\begin{equation}
    \tilde{t}b_{-1} + tL_{-1} = \, \{ \, Q', t b_{-1} \} \,,
\end{equation}
and the amplitude can be written as the integration over $t$ and $\tilde{t}$ as
\begin{equation}
    \mathcal{A} = {}-\, g^2 \int_0^1 dt \int d\tilde{t} \,
    \langle \, \Psi_1 (0) \, 
    e^{ \{ \, Q', t b_{-1} \} }\cdot \Psi_2 (t) \, \Psi_3 (1) \, \Psi_4 (\infty) \, \rangle_\text{UHP} \,.
    \label{tree-level four-string amplitude: extended BRST}
\end{equation}
In the extended BRST formalism, we can show that the BRST invariance of the amplitude~\eqref{tree-level four-string amplitude: extended BRST} holds in the following manner. Unintegrated operators are invariant under the extended BRST transformation. Let us assume that $\Psi_4$ is BRST-exact. It can be written as
\begin{equation}
    \Psi_4(\infty) = \, Q \cdot \Lambda(\infty) = \, Q' \cdot \Lambda (\infty) \,.
\end{equation}
Then we find
\begin{equation}
\begin{split}
    \delta \mathcal{A} = & \, {}-\, g^2 \int_0^1 dt \int d\tilde{t} \,
    \langle \, \Psi_1 (0) \, 
    e^{ \{ \, Q', t b_{-1} \} }\cdot \Psi_2 (t) \, \Psi_3 (1) \, Q' \cdot \Lambda (\infty) \, \rangle_\text{UHP} \, \\
    = & \ g^2 \int_0^1 dt \int d\tilde{t} \,
    \langle \, Q' \cdot \Bigl( \, \Psi_1 (0) \, 
    e^{ \{ \, Q', t b_{-1} \} }\cdot \Psi_2 (t) \, \Psi_3 (1) \, \Lambda (\infty) \, \Bigr) \, \rangle_\text{UHP} \,.
\end{split}
\end{equation}
We can replace $Q'$ with $-\tilde{t}\partial_t$ by using the BRST invariance of the correlation function:
\begin{equation}
    \langle \, Q \cdot ( \, \dots \, ) \, \rangle = \, 0 \,.
\end{equation}
By integration over $\tilde{t}$, we find
\begin{equation}
\begin{split}
    \delta \mathcal{A} = & \, {}-\, g^2 \int_0^1 dt \int d\tilde{t} \,
    \langle \, \tilde{t} \, \partial_t \Bigl( \, \Psi_1 (0) \,  
    e^{ \{ \, Q', t b_{-1} \} }\cdot \Psi_2 (t) \, \Psi_3 (1) \,  \Lambda (\infty) \, \Bigr) \, \rangle_\text{UHP} \, \\
    = & \, {}\, g^2 \int_0^1 dt \,
    \langle \, \partial_t \Bigl( \Psi_1 (0) \, 
    \Psi_2 (t) \, \Psi_3 (1) \, \Lambda (\infty) \Bigr) \, \rangle_\text{UHP} \,.
\end{split}
\end{equation}
The integral over $t$ yields $\Psi_2(1) - \Psi_2(0)$.
The collision of two vertex operators corresponds to the boundary of the moduli space of Riemann surfaces, and this can be dropped by the canceled propagator argument.
Therefore we conclude that the amplitude~\eqref{tree-level four-string amplitude: extended BRST} is BRST-invariant. 

In the world-sheet theory of the bosonic string, a scattering amplitudes of the strings is given by correlation function of the vertex operators integrated over the moduli space of Riemann surfaces with associated ghost insertions. For example, the tree-level four-point amplitude is given in the equation~\eqref{tree-level four-string amplitude: b_(-1)}. In open string field theory~\cite{Witten:1985cc}, on the other hand, the integral over the moduli space is reproduced from the Feynman diagrams constructed from one propagator and two cubic vertices. The contribution from the $s$-channel covers the region $0 \leq t \leq 1/2$:
\begin{equation}
    \mathcal{A}_s = \, g^2 \int_0^{1/2} dt \, \langle \, \Psi_1 (0) \, 
    b_{-1}\cdot \Psi_2 (t) \, \Psi_3 (1) \, \Psi_4 (\infty) \, \rangle_\text{UHP} \,,
\end{equation}
and the contribution from the $t$-channel covers the region $1/2 \leq t \leq 1$:
\begin{equation}
    \mathcal{A}_t = \, g^2 \int_{1/2}^1 dt \, \langle \, \Psi_1 (0) \, 
    b_{-1}\cdot \Psi_2 (t) \, \Psi_3 (1) \, \Psi_4 (\infty) \, \rangle_\text{UHP} \,.
\end{equation}
The sum $\mathcal{A}_s + \mathcal{A}_t$ reproduces an integral over the moduli space of disks with four punctures. This is the situation in the open bosonic string field theory with the associative star product. If we use a non-associative two-string product, the sum $\mathcal{A}_s + \mathcal{A}_t$ does not reproduce the whole moduli space. In this subsection, we consider a toy model in which the amplitude $\mathcal{A}_s$ and $\mathcal{A}_t$ are given by
\begin{equation}
    \mathcal{A}_s = \, g^2 \int_0^{t_-} dt \, \langle \, \Psi_1 (0) \, 
    b_{-1}\cdot \Psi_2 (t) \, \Psi_3 (1) \, \Psi_4 (\infty) \, \rangle_\text{UHP} 
\end{equation}
and  
\begin{equation}
    \mathcal{A}_t = \, g^2 \int_{t_+}^1 dt \, \langle \, \Psi_1 (0) \, 
    b_{-1}\cdot \Psi_2 (t) \, \Psi_3 (1) \, \Psi_4 (\infty) \, \rangle_\text{UHP} \,,
\end{equation}
where $0<t_-<t_+<1$. The sum $\mathcal{A}_s + \mathcal{A}_t$ does not cover the region $t_-<t<t_+$ of the moduli space. We can show that the BRST invariance of the amplitude is broken. If the vertex operator $\Psi_4$ is BRST exact and written as $Q \cdot \Lambda$, the integrals in $\mathcal{A}_s$ and $\mathcal{A}_t$ localize at the boundary $t_-$ and $t_+$, respectively:
\begin{align}
    \mathcal{A}_s = & \ g^2 \,\langle \, \Psi_1 (0) \, 
    \Psi_2 (t_-) \, \Psi_3 (1) \, \Lambda (\infty) \, \rangle_\text{UHP} \,, \\
    \mathcal{A}_t = & \ {}-g^2 \,\langle \, \Psi_1 (0) \, 
    \Psi_2 (t_+) \, \Psi_3 (1) \, \Lambda (\infty) \, \rangle_\text{UHP} \,.
\end{align}
Since $t_- \neq t_+$, these surface terms do not cancel. We can cover the whole moduli space if we introduce a fundamental quartic vertex, which yields the amplitude of the form
\begin{equation}
    \mathcal{A}_4 = \, g^2 \int_{t_-}^{t_+} dt \, \langle \, \Psi_1 (0) \, 
    b_{-1}\cdot \Psi_2 (t) \, \Psi_3 (1) \, \Psi_4 (\infty) \, \rangle_\text{UHP} \,.
\end{equation}
How to construct such a fundamental quartic vertex? The key relation is
\begin{equation}
\begin{split}
    & \int_{t_-}^{t_+} dt \, 
    \langle \, \Psi_1 (0) \, \{ \, Q, b_{-1} e^{tL_{-1}} \, \} \cdot \Psi_2 (0) \, 
    \Psi_3 (1) \, \Psi_4 (\infty) \, \rangle_\text{UHP} \\
    = & \ \langle \, \Psi_1 (0) \, 
    \Psi_2 (t_+) \, \Psi_3 (1) \, \Psi_4 (\infty) \, \rangle_\text{UHP}
    -\langle \, \Psi_1 (0) \, 
    \Psi_2 (t_-) \, \Psi_3 (1) \, \Psi_4 (\infty) \, \rangle_\text{UHP} \,,
    \label{3rd A infty in terms of the covering of the moduli}
\end{split}
\end{equation}
where $\Psi_1$, $\Psi_2$, $\Psi_3$, and $\Psi_4$ are off-shell vertex operators.
The left-hand side of this equation corresponds to the four terms in the $A_\infty$ relation~\eqref{3rd A infty} which contain the BRST operator and the three-string product $V_3$, and the right-hand side of this equation corresponds to the two terms in~\eqref{3rd A infty} which contain the two-string products $V_2$. 
Using the equation~\eqref{action of L_(-1) on the vertex operator equals derivation}, the right-hand side of this equation can be written as 
\begin{equation}
\begin{split}
    & \ \langle \, \Psi_1 (0) \, 
    \Psi_2 (t_+) \, \Psi_3 (1) \, \Psi_4 (\infty) \, \rangle_\text{UHP}
    -\langle \, \Psi_1 (0) \, 
    \Psi_2 (t_-) \, \Psi_3 (1) \, \Psi_4 (\infty) \, \rangle_\text{UHP} \\
    = & \ \langle \, \Psi_1 (0) \, 
    e^{(t_+ -t_0)L_{-1}} \cdot \Psi_2 (t_0) \, \Psi_3 (1) \, \Psi_4 (\infty) \, \rangle_\text{UHP} \\
    & \quad -\langle \, \Psi_1 (0) \, 
    e^{(t_- -t_0)L_{-1}} \cdot \Psi_2 (t_0) \, \Psi_3 (1) \, \Psi_4 (\infty) \, \rangle_\text{UHP} \,,
\end{split}
\end{equation}
where $t_0$ is an arbitrary reference point in the region $0< t_0 < t_-$. We express this equation using the line integral $L_{-1}[f]$ defined for a conformal transformation $f(\xi)$ by
\begin{equation}
    L_{-1} [f] = \, \oint \frac{dz}{2 \pi i} 
    \left( \, \frac{df^{-1}(z)}{dz} \, \right)^{-1} T(z) 
    \quad \text{with} \quad 
    z = f(\xi) \,.
\end{equation}
Then we have
\begin{equation}
\begin{split}
    & \ \langle \, \Psi_1 (0) \, 
    \Psi_2 (t_+) \, \Psi_3 (1) \, \Psi_4 (\infty) \, \rangle_\text{UHP}
    -\langle \, \Psi_1 (0) \, 
    \Psi_2 (t_-) \, \Psi_3 (1) \, \Psi_4 (\infty) \, \rangle_\text{UHP} \\
    = & \ \langle \, \Psi_1 (0) \, 
    (\mathcal{L}_t - \mathcal{L}_s) \Psi_2 (t_0) \, \Psi_3 (1) \, \Psi_4 (\infty) \, \rangle_\text{UHP} \,,
\end{split}
\end{equation}
where $\mathcal{L}_t$ and $\mathcal{L}_s$ are
\begin{equation}
    \mathcal{L}_t = \, e^{(t_+ - t_0)L_{-1} [f_0]} \,, \quad
    \mathcal{L}_s = \, e^{(t_- - t_0)L_{-1} [f_0]} \,,
\end{equation}
with $f_0 (\xi) = \xi - t_0$. Then the relation~\eqref{3rd A infty in terms of the covering of the moduli} can be written as
\begin{equation}
    Q \cdot \mathcal{B} = \, \mathcal{L}_t - \mathcal{L}_s \,,
\end{equation}
where the operator $\mathcal{B}$ is given by
\begin{equation}
    \mathcal{B} = b_{-1} [f_0] \,\int_{t_-}^{t_+} dt \, e^{(t-t_0)L_{-1}[f_0]} \,
\end{equation}
with the line integral $b_{-1}[f]$ defined for a conformal transformation $f(\xi)$ by 
\begin{equation}
    b_{-1} [f] = \, \oint \frac{dz}{2 \pi i} 
    \left( \, \frac{df^{-1}(z)}{dz} \, \right)^{-1} T(z) 
    \quad \text{with} \quad 
    z = f(\xi) \,.
\end{equation}
Let us consider the following equation
\begin{equation}
    Q \cdot \mathcal{B} = e^{\{ Q, \, \mathcal{F}(t_+) \}} 
    - e^{\{Q, \, \mathcal{F}(t_-)\}} \,,
    \label{A infty relation for general B}
\end{equation}
where $\mathcal{F}(t)$ is a Grassmann-odd operator. We can show that the Grassmann-odd operator $\mathcal{B}$ satisfying this equation can be constructed as
\begin{equation}
    \mathcal{B} = \, \int_{t_-}^{t_+} dt \int d\tilde{t} \,
    e^{\{ Q', \, \mathcal{F}(t) \}} \,.
    \label{general interpolation B}
\end{equation}
We find
\begin{equation}
    Q \cdot \mathcal{B} 
    = \, {}-\int_{t_-}^{t_+} dt \int d\tilde{t} \, Q \cdot e^{ \{ \, Q', \mathcal{F}(t) \, \} } 
    = \, \int_{t_-}^{t_+} dt \int d\tilde{t} \, \tilde{t} \, \partial_t \, e^{\{ \, Q', \mathcal{F}(t)\, \}} \,,
\end{equation}
where we replaced the action of $Q$ on $e^{ \{ \, Q', \mathcal{F}(t) \, \} }$ by that of $-\tilde{t}\partial_t$. Then we find
\begin{equation}
    Q \cdot \mathcal{B} 
    = \, \int_{t_-}^{t_+} dt \, \partial_t 
    e^{\{ \, Q', \, \mathcal{F}(t) \, \}} \Bigr|_{\tilde{t} = 0}
    = \, \int_{t_-}^{t_+} dt \, \partial_t 
    e^{\{ \, Q, \, \mathcal{F}(t) \, \}}
    = \, e^{\{ \, Q, \, \mathcal{F}(t_+) \, \}} 
    -e^{\{ \, Q, \, \mathcal{F}(t_-) \, \}} \,.
\end{equation}
We will use this in the construction of open bosonic string field theory with stubs.

\subsection{Cubic vertex for open bosonic string field theory with stubs}
In this subsection and the following subsection, we give an explicit construction of an action for open bosonic string field theory including stubs up to quartic order. We define a cubic vertex with stubs by
\begin{equation}
S_3 = \, \frac{g}{3} \, \langle \, e^{-wL_0}\Psi \,,  \, ( e^{-wL_0} \Psi) \ast ( e^{-wL_0}\Psi)  \, \rangle \,,
\label{cubic vertex with stubs}
\end{equation}
where the Grassmann-even operator $L_0$ is the zero-mode of the energy-momentum tensor $T(z)$:
\begin{equation}
    L_0 = \, \oint \frac{dz}{2\pi i} z \, T(z) \,.
\end{equation}
By varying the cubic vertex $S_3$ with respect to the string field $\Psi$, we find
\begin{equation}
    \delta S_3 = \, \langle \, \delta \Psi \,,  e^{-wL_0} ( \, ( e^{-wL_0} \Psi) \ast ( e^{-wL_0}\Psi) \, ) \, \rangle \,,
\end{equation}where we used the fact that $L_0$ is BPZ-even:
\begin{equation}
    L_0 = \, L_0^\star \,.
    \label{L_0 is BPZ even}
\end{equation}
Motivated by this equation, we define the two-string product $V_2^\text{stub}(A_1, A_2)$ by
\begin{equation}
V_2^\text{stub} (A_1, A_2)  
= \ e^{-wL_0} ( \, ( e^{-wL_0} A_1) \ast ( e^{-wL_0} A_2) \, ) \,. \label{two-string product with stubs}
\end{equation}
We can show that the two-string product $V_2^\text{stub}$ satisfies the cyclicity equation. We find
\begin{equation}
\begin{split}
\langle \, A_1, V_2^\text{stub} (A_2, A_3) \, \rangle 
= & \ \left \langle \, A_1, e^{-wL_0} ( \, ( e^{-wL_0} A_2) \ast ( e^{-wL_0} A_3) \, ) \, \right \rangle \\
= & \ \langle \, e^{-wL_0} A_1, ( e^{-wL_0} A_2) \ast ( e^{-wL_0} A_3) \, \rangle \\
= & \ \langle \, (e^{-wL_0} A_1) \ast ( e^{-wL_0} A_2), e^{-wL_0} A_3 \, \rangle \\
= & \ \left \langle \, e^{-wL_0} ( \, (e^{-wL_0} A_1) \ast ( e^{-wL_0} A_2) \, ), A_3 \, \right \rangle \\
= & \ \langle \, V_2^\text{stub} (A_1, A_2), A_3 \, \rangle \,,
\label{cyclicity_of_stub_product}
\end{split}
\end{equation}
where we used the fact that $L_0$ is BPZ-even~\eqref{L_0 is BPZ even}. We can show that the two-string product $V_2^\text{stub}$ satisfies the $A_\infty$ relation~\eqref{2nd A infty}. We find
\begin{equation}
\begin{split}
QV_2^\text{stub}(A_1,A_2)
= & \ e^{-wL_0} ((e^{-wL_0} QA_1)\ast(e^{-wL_0} A_2)) \\
& \quad +(-1)^{A_1} e^{-wL_0} ((e^{-wL_0} A_1)\ast(e^{-wL_0} QA_2)) \\
= & \ V_2^\text{stub}(QA_1,A_2) + (-1)^{A_1} V_2^\text{stub}(A_1,QA_2) \,
\label{Leibnitz_Q_stub},
\end{split}
\end{equation}
where we used the fact that BRST operator $Q$ commutes with $L_0$:
\begin{equation}
    [\, Q \,, L_0 \, ] = \, 0 \,.
\end{equation}
We note that the two-string product $V_2^\text{stub}$~\eqref{two-string product with stubs} is not associative. We find
\begin{equation}
V_2^\text{stub} (V_2^\text{stub} (A_1, A_2), A_3) 
= \, e^{-wL_0} \Bigl[ \, (e^{-2wL_0} ((e^{-wL_0} A_1)\ast(e^{-wL_0} A_2)) \, ) \ast(e^{-wL_0} A_3)\Bigr] \,    
\end{equation}
and
\begin{equation}
V_2^\text{stub} (A_1, V_2^\text{stub} (A_2, A_3))
= \, e^{-wL_0} \Bigl[ (e^{-wL_0} A_1) \ast \, ( e^{-2wL_0} ((e^{-wL_0} A_2)\ast(e^{-wL_0} A_3)) \, ) \Bigr] \,.
\end{equation}
The operator $e^{-2wL_0}$ is inserted differently, therefore $V_2^\text{stub}$ is not associative:
\begin{equation}
V_2^\text{stub} (V_2^\text{stub} (A_1, A_2), A_3) \neq \ V_2^\text{stub} (A_1, V_2^\text{stub} (A_2, A_3)) \,.
\label{non-associativity of V2 stub}
\end{equation}
We can recover the gauge-invariance at quartic order by constructing a three-string product satisfying the $A_\infty$ relation~\eqref{3rd A infty}.

\subsection{Quartic vertex for open bosonic string field theory with stubs}
\begin{figure}[htb]
    \centering
    \includegraphics[width=4cm]{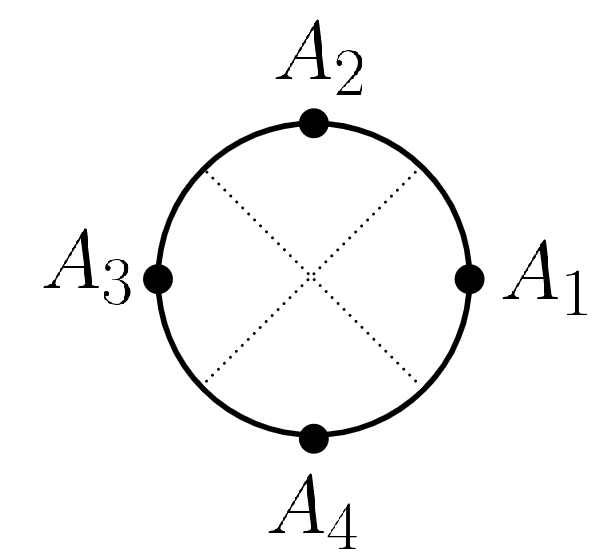}
    \caption{A pictorial representation of the correlation function on the disk~\eqref{4pt_correlation_function}. The insertions of the vertex operators $g_1 \circ A_1(0)$, $g_2 \circ A_2(0)$, $g_3 \circ A_3(0)$, and $g_4 \circ A_4(0)$ are represented by dots on the boundary of the disk.}
    \label{fig:4pt_function}
\end{figure}

We construct a three-string product $V_3^\text{stub} (A_1, A_2, A_3)$ satisfying the $A_\infty$ relation~\eqref{3rd A infty}. In the following, we express the BPZ inner product in terms of the CFT correlation function on a unit disk $D$, instead of the UHP. For example, the BPZ inner product $\langle \, A_1, A_2 \ast A_3 \ast A_4 \, \rangle$ for arbitrary string fields $A_1$, $A_2$, $A_3$, and $A_4$ can be expressed as
\begin{equation}
\langle \, A_1, A_2 \ast A_3 \ast A_4 \, \rangle
= \langle \, g_1 \circ A_1(0) \, g_2 \circ A_2(0) \, g_3 \circ A_3(0) \, g_4 \circ A_4(0) \, \rangle_D \,,
\label{4pt_correlation_function}
\end{equation}
where $A_1 (0)$, $A_2 (0)$, $A_3 (0)$, and $A_4 (0)$ are vertex operators corresponding to the string fields $A_1$, $A_2$, $A_3$, and $A_4$, respectively. The subscript $D$ stands for correlation functions on the unit disk, and functions $g_1(\xi)$, $g_2(\xi)$, $g_3(\xi)$, and $g_4(\xi)$ are defined by
\begin{equation}
\begin{split}
& g_1 (\xi) = \ \left( \frac{1+i\xi}{1-i\xi} \right)^{1/2}, \quad 
g_2 (\xi) = \ e^{\pi i/2} \left( \frac{1+i\xi}{1-i\xi} \right)^{1/2}, \\
& g_3 (\xi) = \ e^{\pi i} \left( \frac{1+i\xi}{1-i\xi} \right)^{1/2}, \quad 
g_4 (\xi) = \ e^{3\pi i/2} \left( \frac{1+i\xi}{1-i\xi} \right)^{1/2} \,.
\end{split}
\end{equation}
In the following, we use a pictorial representation of the correlation function on the unit disk in Figure~\ref{fig:4pt_function}.

\begin{figure}[htb]
    \centering
    \includegraphics[width=10cm]{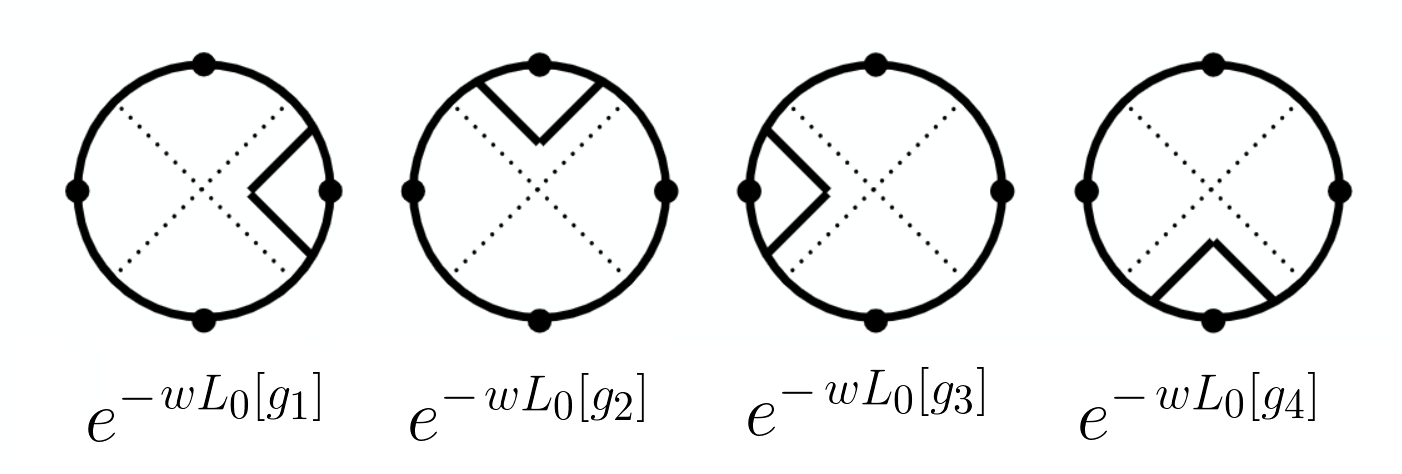}
    \caption{Pictorial representations of correlation functions on the disk $D$ with $e^{-wL_0[g_1]}$, $e^{-wL_0[g_2]}$, $e^{-wL_0[g_3]}$, and $e^{-wL_0[g_4]}$, respectively.}
    \label{fig:integration contour g}
\end{figure}

\begin{figure}[htb]
    \centering
    \includegraphics[width=10cm]{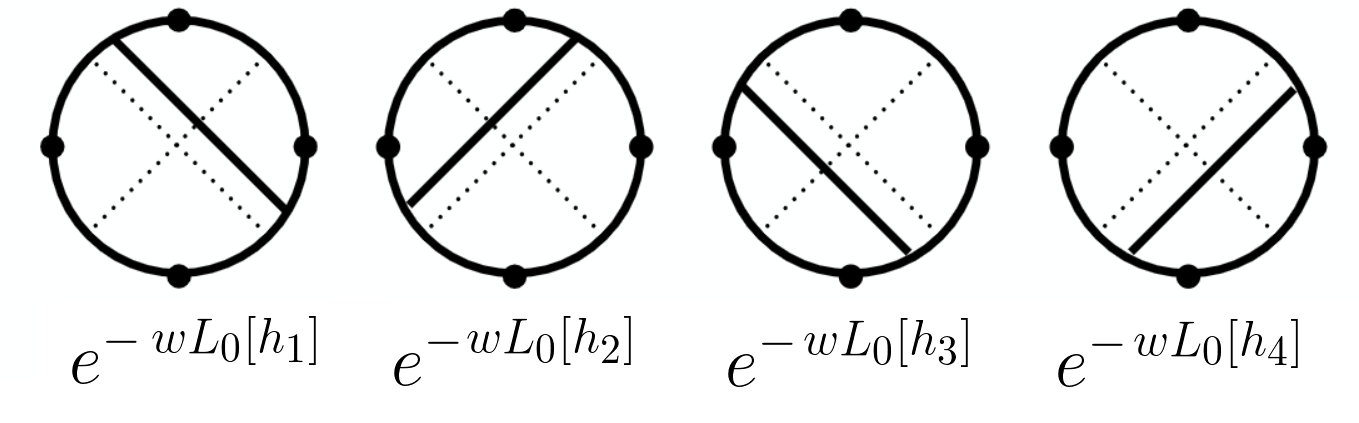}
    \caption{Pictorial representations of correlation functions on the disk $D$ with $e^{-wL_0[h_1]}$, $e^{-wL_0[h_2]}$, $e^{-wL_0[h_3]}$, and $e^{-wL_0[h_4]}$, respectively.}
    \label{fig:integration contours h}
\end{figure}

\begin{figure}[htb]
    \centering
    \includegraphics[width=9cm]{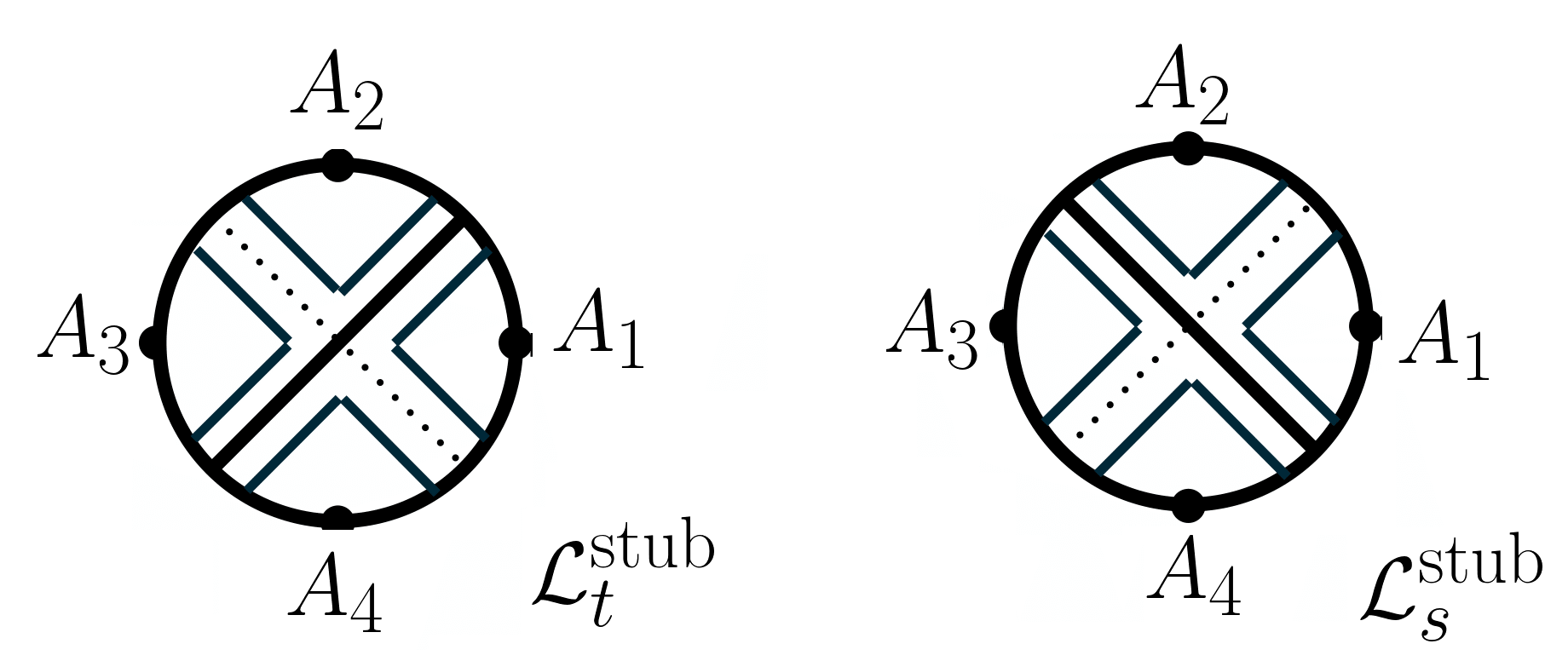}
    \caption{Pictorial representations of correlation functions on the disk with $\mathcal{L}_t^\text{stub}$ and $\mathcal{L}_s^\text{stub}$, respectively.}
    \label{fig:t_channel and s_channel of stub}
\end{figure}

\begin{figure}[tb]
    \centering
    \includegraphics[width=11cm]{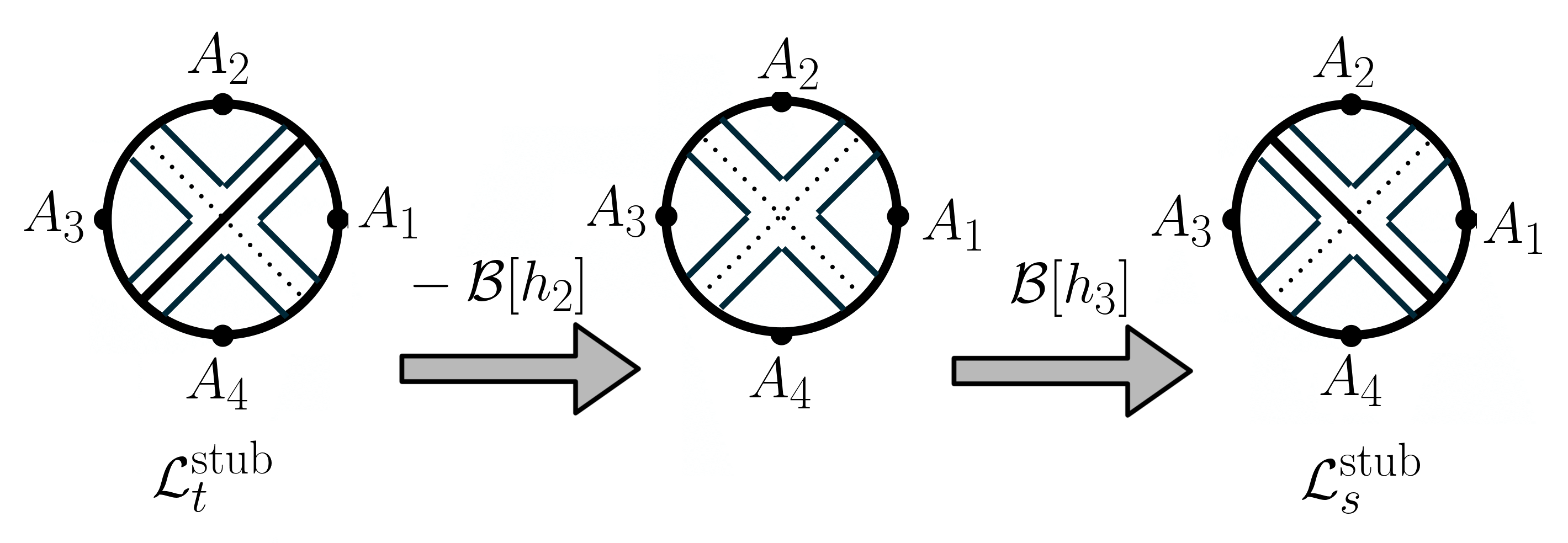}
    \caption{A pictorial representation of the interpolation between the $t$-channel and the $s$-channel. The stub operator $e^{-2wL_0[h_2]}$ in the $t$-channel is deleted by the action of the stub annihilation operator $-\mathcal{B}[h_2]$, then the stub operator $e^{-2wL_0[h_3]}$ in the $s$-channel is created by the action of the stub creation operator $\mathcal{B}[h_3]$.}
    \label{fig:interpolation_stub}
\end{figure}

 We define a Grassmann-even operator $\mathcal{L}_t^\text{stub}$ by
\begin{equation}
\langle \, A_1, V_2^\text{stub} (V_2^\text{stub} (A_2, A_3), A_4) \, \rangle
= \, \langle \, \mathcal{L}_t^\text{stub} g_1 \circ A_1(0) \, g_2 \circ A_2(0) \, g_3 \circ A_3(0) \, g_4 \circ A_4(0) ) \, \rangle_D 
\end{equation}
and $\mathcal{L}_s^\text{stub}$ by
\begin{equation}
\langle \, A_1, V_2^\text{stub} (A_2, V_2^\text{stub} (A_3, A_4)) \, \rangle 
= \, \langle \, \mathcal{L}_s^\text{stub} g_1 \circ A_1(0) \, g_2 \circ A_2(0) \, g_3 \circ A_3(0) \, g_4 \circ A_4(0) ) \, \rangle_D \,.
\end{equation}
Let us calculate $\mathcal{L}_t^\text{stub}$. We find
\begin{equation}
    \langle \, A_1, V_2^\text{stub}(V_2^\text{stub}(A_2, A_3), A_4) \, \rangle
    = \, \langle \, e^{-wL_0}A_1, e^{-2wL_0} \, (e^{-wL_0}A_2 \ast  e^{-wL_0}A_3) \, \ast  e^{-wL_0}A_4 \, \rangle \,. 
\end{equation}
We would like to express the right-hand side of this equation in terms of the correlation function on the disk $D$. Therefore, we introduce the line integral $L_0[f]$ for a conformal transformation $f(\xi)$ by
\begin{equation}
    L_0[f] = \oint \frac{dz}{2\pi i} \, f^{-1}(z)
    \left( \frac{df^{-1}(z)}{dz} \right)^{-1} \, T(z) \quad \text{with} \quad z=f(\xi) \,.
\end{equation}
then we have
\begin{align*}
        \langle \, e^{-wL_0} A_1, A_2 \ast A_3 \ast A_4 \, \rangle
    = & \ \langle \, e^{-wL_0[g_1]} \, g_1 \circ A_1(0) \, g_2 \circ A_2(0) \, g_3 \circ A_3(0) \, g_4 \circ A_4(0) \, \rangle_D \,, \\
        \langle \, A_1, (e^{-wL_0} A_2) \ast A_3 \ast A_4 \, \rangle
    = & \ \langle \, e^{-wL_0[g_2]} \, g_1 \circ A_1(0) \, g_2 \circ A_2(0) \, g_3 \circ A_3(0) \, g_4 \circ A_4(0) \, \rangle_D \,,\\
        \langle \, A_1, A_2 \ast (e^{-wL_0} A_3) \ast A_4 \, \rangle
    = & \ \langle \, e^{-wL_0[g_3]} \, g_1 \circ A_1(0) \, g_2 \circ A_2(0) \, g_3 \circ A_3(0) \, g_4 \circ A_4(0) \, \rangle_D \,,\\
        \langle \, A_1, A_2 \ast A_3 \ast (e^{-wL_0} A_4) \, \rangle
    = & \ \langle \, e^{-wL_0[g_4]} \, g_1 \circ A_1(0) \, g_2 \circ A_2(0) \, g_3 \circ A_3(0) \, g_4 \circ A_4(0) \, \rangle_D \,.
\end{align*}
Here and in what follows, we express a correlation function with insertions of these operators by a disk with solid lines. See Figure~\ref{fig:integration contour g}.
Furthermore, we use other functions $h_1(\xi)$, $h_2(\xi)$, $h_3(\xi)$, and $h_4(\xi)$ on the disk $D$ defined by
\begin{equation}
\begin{split}
& h_1 (\xi) = e^{\pi i/4} \left( \frac{1+i\xi}{1-i\xi} \right) \,, \quad
h_2 (\xi) = e^{3\pi i/4} \left( \frac{1+i\xi}{1-i\xi} \right) \,, \\
& h_3 (\xi) = e^{5\pi i/4} \left( \frac{1+i\xi}{1-i\xi} \right) \,, \quad
h_4 (\xi) = e^{7\pi i/4} \left( \frac{1+i\xi}{1-i\xi} \right) \,.
\label{integration contour h}
\end{split}
\end{equation}
Then we have
\begin{align*}
    \langle \, e^{-wL_0}(A_1 \ast A_2), A_3 \ast A_4 \, \rangle
    = & \ \langle \, e^{-wL_0[h_1]} \, g_1 \circ A_1(0) \, g_2 \circ A_2(0) \, g_3 \circ A_3(0) \, g_4 \circ A_4(0) \, \rangle_D \,, \\
    \langle \, A_1, e^{-wL_0}(A_2 \ast A_3 ) \ast A_4 \, \rangle
    = & \ \langle \, e^{-wL_0[h_2]} \, g_1 \circ A_1(0) \, g_2 \circ A_2(0) \, g_3 \circ A_3(0) \, g_4 \circ A_4(0) \, \rangle_D \,, \\
    \langle \, A_1, A_2 \ast e^{-wL_0}(A_3 \ast A_4) \, \rangle
    = & \ \langle \, e^{-wL_0[h_3]} \, g_1 \circ A_1(0) \, g_2 \circ A_2(0) \, g_3 \circ A_3(0) \, g_4 \circ A_4(0) \, \rangle_D \,, \\
    \langle \, A_1, (e^{-wL_0})^\star (A_2 \ast A_3) \ast A_4 \, \rangle
    = & \ \langle \, e^{-wL_0[h_4]} \, g_1 \circ A_1(0) \, g_2 \circ A_2(0) \, g_3 \circ A_3(0) \, g_4 \circ A_4(0) \, \rangle_D \,.
\end{align*}
These correlation functions are pictorially represented in Figure~\ref{fig:integration contours h}.
\\
Using the definition of the two-string product $V_2^\text{stub}$ \eqref{two-string product with stubs}, we find
\begin{equation}
\mathcal{L}_t^\text{stub} = \ e^{-wL_0[g_1]} \, e^{-2wL_0[h_2]} \, e^{-wL_0[g_2]} \, e^{-wL_0[g_3]} \, e^{-wL_0[g_4]} \,. \label{definition of L_t}
\end{equation}
By a similar calculation, we find
\begin{equation}
\mathcal{L}_s^\text{stub} = \ e^{-wL_0[g_1]} \, e^{-wL_0[g_2]} \, e^{-2wL_0[h_3]} \, e^{-wL_0[g_3]} \, e^{-wL_0[g_4]} \,. \label{definition of L_s}
\end{equation}
Pictorial representations of $\mathcal{L}_t^\text{stub}$ and $\mathcal{L}_s^\text{stub}$ are given in Figure~\ref{fig:t_channel and s_channel of stub}.
We define the three-string product $V_3^\text{stub}$
in terms of a Grassmann-odd operator $\mathcal{B}_3^\text{stub}$:
\begin{equation}
\langle \, A_1, V_3^\text{stub} (A_2, A_3, A_4) \, \rangle 
= \, \langle \, g_1 \circ A_1(0) \, \mathcal{B}_3^\text{stub} \, g_2 \circ A_2(0) \, g_3 \circ A_3(0) \, g_4 \circ A_4(0) \, \rangle_D \,. \label{definition of V3 with stub}
\end{equation}
Then the $A_\infty$ relation for the three-string product $V_3^\text{stub}$
\begin{equation}
\begin{split}
0 = & \ QV_3^\text{stub}(A_1,A_2,A_3) +V_3^\text{stub}(QA_1,A_2,A_3)  \\
& \quad + (-1)^{A_1}V_3^\text{stub}(A_1,QA_2,A_3) +(-1)^{A_1+A_2}V_3^\text{stub}(A_1,A_2,QA_3) \\
& \quad {}- V_2^\text{stub}(V_2^\text{stub}(A_1, A_2), A_3) 
+ V_2^\text{stub}(A_1, V_2^\text{stub}(A_2, A_3)) 
\label{3rd A_infty open bosonic with stubs}
\end{split}
\end{equation}
is translated into the following equation:
\begin{equation}
Q \cdot \mathcal{B}_3^\text{stub} = \ \mathcal{L}_t^\text{stub} - \mathcal{L}_s^\text{stub} \label{A infty relation for B3} \,.
\end{equation}
To construct $\mathcal{B}_3^\text{stub}$, it is convenient to use the extended BRST formalism~\cite{Witten:2012bh}. 
The different factors between $\mathcal{L}_t$ and $\mathcal{L}_s$ are $e^{-2wL_0[h_2]}$ and $e^{-2wL_0[h_3]}$, so we focus on these factors.
Following the construction~\eqref{general interpolation B} in subsection~\ref{subsec:extBRSTstub}, a naive solution is
\begin{equation}
    \mathcal{B}^\text{naive}
    = \, \int_0^1 dt \int d\tilde{t} \,
    e^{\{ \, Q' \,, \, \mathcal{F}^\text{naive}(t) \, \} } \,
\end{equation}
with
\begin{equation}
    \mathcal{F}^\text{naive}(t) = \,  {}-2wt b_0 [h_2]-2w(1-t) b_0 [h_3] \,,
\end{equation}
where we defined the line integral $b_0[f]$ for a conformal transformation $f(\xi)$ by
\begin{equation}
b_0 [f] = \, \oint \frac{dz}{2\pi i} \, f^{-1}(z) \,
\Bigl( \frac{df^{-1}(z)}{dz} \Bigr)^{-1} b(z)  \quad \text{with} \quad z=f(\xi) \,.
\end{equation}
Using
\begin{align}
    \{ \, Q' \,, \, \mathcal{F}^\text{naive}(t=1) \, \} 
    = & \ \{ \, Q \,, -2w b_0 [h_2] \, \}
    = {}-2wL_0[h_2] \,, \\
    \{ \, Q' \,, \, \mathcal{F}^\text{naive}(t=0) \, \} 
    = & \ \{ \, Q \,, -2w b_0 [h_3] \, \}
    = {}-2wL_0[h_3] \,,
\end{align}
we can show that this operator interpolates between $e^{-2wL_0[h_2]}$ and $e^{-2wL_0[h_3]}$:
\begin{equation}
    Q \cdot \mathcal{B}^\text{naive} = \, e^{-2wL_0[h_2]} - e^{-2wL_0[h_3]} \,.
\end{equation}
Instead of the operator $\mathcal{B}^\text{naive}$, we introduce the operator $\mathcal{B}[f]$ in the following form:
\begin{equation}
\mathcal{B}[f] = \ \int_0^1 dt \int d\tilde{t} \, e^{\{Q', t(-2w)b_0[f] \} } \,. \label{stub_creation_operator}
\end{equation}
The operator $\mathcal{B}[f]$ interpolates between the stub operator $e^{-2wL_0[f]}$ and the unit operator $1$. Let us carry out the integration over $\tilde{t}$. Since
\begin{align}
\{ \, Q' \,, t(-2w)b_0[f] \, \} = & \ \tilde{t}(-2w)b_0[f] + t(-2w)L_0[f] \,,
\end{align}
we find
\begin{align}
\mathcal{B}[f] 
= & \, {}-\int_0^1 dt \, (-2w)b_0[f] e^{t(-2w)L_0[f]} \,.
\label{stub_creation_operator_explicit}
\end{align}
The BRST transformation of $\mathcal{B}[f]$ is then
\begin{equation}
\begin{split}
 \, Q \cdot \mathcal{B}[f] 
= & \, -\int_0^1 dt \, (-2w)L_0[f] e^{t(-2w)L_0[f]} \\
= & \, -\int_0^1 dt \, \partial_t e^{t(-2w)L_0[f]} \\
= & \ 1 - e^{-2wL_0[f]} \,.
\label{stub_creation_relation}
\end{split}
\end{equation}
Since the operator $\mathcal{B}[f]$ interpolates the unit operator $1$ and the stub operator $e^{-2wL_0[f]}$, we regard $\mathcal{B}[f]$ as an operator which creates the stub operator $e^{-2wL_0}$, and we call this the stub creation operator. On the other hand, the operator $-\mathcal{B}[f]$ interpolates the unit operator and the stub operator in the reverse direction:
\begin{equation}
 \, Q \cdot (-\mathcal{B}[f]) 
= \, e^{-2wL_0[f]} -1 \,.
\label{stub_annihilation_relation}
\end{equation}
Therefore, we call $-\mathcal{B}[f]$ the stub annihilation operator. Using the stub creation and annihilation operators, we find that $\mathcal{B}_3^\text{stub}$ is given by
\begin{equation}
\mathcal{B}_3^\text{stub} = (-\mathcal{B}[h_2] + \mathcal{B}[h_3]) \, e^{-wL_0[g_1]} e^{-wL_0[g_2]} e^{-wL_0[g_3]} e^{-wL_0[g_4]} \,.
\label{mathcal B_3 answer}
\end{equation}
The interpolation by $\mathcal{B}_3^\text{stub}$ is illustrated in Figure~\ref{fig:interpolation_stub}.
Then the action of the BRST operator on $\mathcal{B}_3$ is given by
\begin{equation}
\begin{split}
Q \cdot \mathcal{B}_3^\text{stub}
= & \ \Bigl( (e^{-2wL_0[h_2]} -1)  + (1-e^{-2wL_0[f]}) \Bigr) e^{-wL_0[g_1]} e^{-wL_0[g_2]} e^{-wL_0[g_3]} e^{-wL_0[g_4]} \\
= & \ \mathcal{L}_t^\text{stub} - \mathcal{L}_s^\text{stub}
\,.    
\end{split}
\end{equation}
Therefore, the three-string product $V_3^\text{stub}$ constructed from $\mathcal{B}_3^\text{stub}$ \eqref{mathcal B_3 answer} satisfies the $A_\infty$ relation~\eqref{3rd A_infty open bosonic with stubs}. In appendix~\ref{app:B}, we show that the three-string product $V_3^\text{stub}$ also satisfies the cyclicity equation
\begin{equation}
    \langle \, A_1, V_3^\text{stub}( A_2, A_3, A_4) \, \rangle 
= {}-(-1)^{A_1} \langle \, V_3^\text{stub}(A_1, A_2, A_3), A_4 \, \rangle \,.
\label{V3 cyclicity for stubs}
\end{equation}
Therefore, the action constructed from the two-string product $V_2^\text{stub}$~\eqref{two-string product with stubs} and the three-string product $V_3^\text{stub}$~\eqref{definition of V3 with stub} has the cyclic $A_\infty$ structure up to quartic order.

In the construction of open bosonic string field theory, the form of the $b$ ghost insertion in~\eqref{stub_creation_operator_explicit} is not complicated, and we may be able to construct it without using the extended BRST formalism.
The extended BRST formalism plays a crucial role in the extension to the superstring.

\section{The Neveu-Schwarz sector of open superstring field theory \label{sec:NS}}
\setcounter{equation}{0}
In this section, we review the Neveu-Schwarz sector of open superstring field theory. 
Starting with an explanation of the extended BRST formalism for the superstring, we review the construction of an action for the Neveu-Schwarz sector of open superstring field theory based on the supermoduli space presented by Ohmori and Okawa~\cite{Ohmori:2017wtx}.

\subsection{Extended BRST formalism \label{subsec:extBRSTsuper}}
Let us consider tree-level scattering amplitudes in the Neveu-Schwarz sector of the open superstring. If we use the superspace, the location of a vertex operator on the boundary is described by one Grassmann-even coordinate and one Grassmann-odd coordinate. Using global super-conformal transformations on the disk, we can fix three even coordinates and two odd coordinates. Therefore, disks with $n$ Neveu-Schwarz punctures on the boundary have $n-3$ even moduli and $n-2$ odd moduli. The integration over a Grassmann-odd modulus yields the local picture-changing operator. 
In the construction of superstring field theory, it is useful to consider other parametrizations of the supermoduli space of super-Riemann surfaces. 
Disks with three Neveu-Schwarz punctures have one odd modulus, and we denote it by $\zeta$.
We consider a parametrization of the supermoduli space of super-Riemann surfaces in the following form:
\begin{equation}
    \int d\zeta e^{\zeta G_{-1/2}} 
    \quad \text{with} \quad
    G_{-1/2} = \, \oint \frac{dz}{2 \pi i} \, T_\text{F} (z) \,,
\end{equation}
where $T_F(z)$ is the supercurrent.
The integration over the odd modulus requires an associated ghost insertion. Following~\cite{Witten:2012bh}, we introduce a Grassmann-even variable $\tilde{\zeta}$, and define an action of an extended BRST operator $Q'$ by
\begin{equation}
    \{ \, Q' , \zeta \, \} = \, \tilde{\zeta} \,, \quad
    [ \, Q' , \tilde{\zeta} \, ] = \, 0 \,.
\end{equation}
We can show that the extended BRST operator is nilpotent:
\begin{equation}
    Q'^2 = 0\,,
\end{equation}
and we can express $Q'$ as
\begin{equation}
    Q' = \, Q + \tilde{\zeta} \partial_\zeta \,.
\end{equation}
The actions of the extended BRST operator on the states of the CFT are the same as those of the BRST operator $Q$. In particular, we have
\begin{equation}
[\, Q', \beta_{-1/2} \, ] = G_{-1/2} \,, \quad \{ \, Q', G_{-1/2} \, \} = 0 \,,
\end{equation}
where $\beta_{-1/2}$ is given by
\begin{equation}
    \beta_{-1/2} = \, \oint \frac{dz}{2 \pi i} \beta (z) \,.
\end{equation}
Note that the combination $-\tilde{\zeta}\beta_{-1/2}+\zeta G_{-1/2}$ anti-commutes with $Q'$:
\begin{equation}
    \{\, Q' \,, -\tilde{\zeta}\beta_{-1/2}+\zeta G_{-1/2} \,\}
    = \, {}-\tilde{\zeta}G_{-1/2} +\tilde{\zeta} G_{-1/2}
    = \, 0 \,.
\end{equation}
In fact, this combination can be generated by the action of $Q'$ on $-\zeta \beta_{-1/2}$:
\begin{equation}
    -\tilde{\zeta}\beta_{-1/2} + \zeta G_{-1/2}
    = \, \{\, Q' \,, -\zeta \beta_{-1/2} \,\} \,.
\end{equation}
The integration over the odd modulus $\zeta$ can be combined with the associated ghost insertion to yield the picture-changing operator\footnote{The subscript N denotes that this is the picture-changing operator which acts on Neveu-Schwarz string fields. In the next section, we will introduce the picture-changing operator $X_\text{R}$ which acts on Ramond string fields.} 
\begin{equation}
X_\text{N} = \, \int d\zeta d\tilde{\zeta} \mathcal{X}_\text{N} (\zeta ,\tilde{\zeta}) \,, 
\label{PCO for NS sector}
\end{equation}
where 
\begin{equation}
\mathcal{X}_\text{N} (\zeta ,\tilde{\zeta}) 
= \ e^{- \{ Q', \, \zeta\beta_{-1/2} \, \} }
= \ e^{-\tilde{\zeta}\beta_{-1/2}+\zeta G_{-1/2}} \,.
\end{equation}
By using the Baker-Hausdorff-Campbell formula and carrying out an integration over $\zeta$, the explicit expression of the operator $X_\text{N}$ is given by
\begin{eqnarray}
X_\text{N} = G_{-1/2} \, \delta(\beta_{-1/2}) +b_{-1} \, \delta'(\beta_{-1/2}) \,,
\end{eqnarray}
where the delta function operators are defined by 
\begin{eqnarray}
\delta (\beta_{-1/2}) = \, \int d\tilde{\zeta} e^{-\tilde{\zeta} \beta_{-1/2}} \,, \quad 
\delta' (\beta_{-1/2}) = {}- \int d\tilde{\zeta} \, \tilde{\zeta} \, e^{-\tilde{\zeta} \beta_{-1/2}} \,.
\end{eqnarray}
It is emphasized in~\cite{Witten:2012bh} that the integration over a Grassmann-even variable $\tilde{\zeta}$ should be understood as an algebraic operation, and we give its definition in appendix~\ref{app:A}.
We treat the measure $d\tilde{\zeta}$ as a Grassmann-odd object for the reason we will explain in appendix~\ref{app:A}, therefore the delta functions $\delta (\beta_{-1/2})$ and $\delta' (\beta_{-1/2})$ are Grassmann-odd.
The picture-changing operator $X_\text{N}$ carries the picture number $1$, and $X_\text{N}$ commutes with the BRST operator:
\begin{equation}
\begin{split}
[ \, Q, X_\text{N} \, ]
= & \ \int \int d\zeta d\tilde{\zeta} \, \{ \, Q, \mathcal{X}_\text{N} (\zeta ,\tilde{\zeta}) \, \} \\
= & \ - \int \int d\zeta d\tilde{\zeta} \tilde{\zeta} \partial_\zeta  \mathcal{X}_\text{N} (\zeta ,\tilde{\zeta}) \\
= & \ 0 \,, \label{Q_commutes_with_XN}
\end{split}
\end{equation}
where we replaced $Q$ acting on $\mathcal{X}_\text{N} (\zeta, \tilde{\zeta})$ by $-\tilde{\zeta}\partial_\zeta$ since $\mathcal{X}_\text{N} (\zeta, \tilde{\zeta})$ anti-commutes with $Q'$.

\subsection{Kinetic term for the Neveu-Schwarz sector}
Let us consider the Neveu-Schwarz sector of open superstring field theory. 
The Neveu-Schwarz string field $\Psi_\text{N}$ is a state in the Hilbert space of the boundary CFT describing the open superstring background we consider, which consists of the matter sector, the $bc$ ghost sector, and the $\beta \gamma$ ghost sector. 
We choose the Neveu-Schwarz string field $\Psi$ to be a Grassmann-odd state of ghost number $1$ and picture number $-1$. 
The BPZ inner product for the superstring has the same properties as the BPZ inner product of bosonic string field theory, except that the BPZ inner product for the superstring is well defined only when the sum of the picture numbers of inputs is equal to $-2$. 
The kinetic term of the Neveu-Schwarz string field is 
\begin{equation}
{}-\frac{1}{2} \langle \, \Psi_\text{N}, Q \Psi_\text{N} \, \rangle \,,
\end{equation}
where $Q$ is the BRST operator of the world-sheet theory of the superstring. This kinetic term is invariant under the gauge transformation:
\begin{equation}
\delta \Psi_\text{N} = Q \Lambda_\text{N} \,,
\end{equation}
where $\Lambda_\text{N}$ is a Grassmann-even state of ghost number $0$ in the $-1$ picture.
\subsection{Cubic vertex for the Neveu-Schwarz sector}
Let us consider a cubic vertex for the Neveu-Schwarz string fields. We construct a cubic vertex for the Neveu-Schwarz string fields in terms of the two-string product which satisfies the cyclicity equation
\begin{equation}
    \langle \, N_1, V_2( N_2, N_3) \, \rangle 
    = \, \langle \, V_2(N_1, N_2), N_3 \, \rangle
\end{equation}
and the $A_\infty$ relation
\begin{equation}
0 = \ QV_2(N_1, N_2) -V_2(QN_1, N_2) -(-1)^{N_1}V_2(N_1,QN_2) \,,
\end{equation}
where $N_1$ and $N_2$ are Neveu-Schwarz string fields in the $-1$ picture. 
As we have seen in subsection~\ref{subsec:extBRSTsuper}, disks with three Neveu-Schwarz punctures have one odd modulus, and we denote it by $\zeta$. In the construction by Ohmori and Okawa in~\cite{Ohmori:2017wtx}, an integral over the odd modulus is implemented by an insertion of the picture-changing operator $X_\text{N}$ \eqref{PCO for NS sector}. Then we consider the cubic vertex for the Neveu-Schwarz string fields of the form
\begin{equation}
    S_\text{N} = \, \langle \, X_\text{N} \Psi_\text{N}, \Psi_\text{N} \ast \Psi_\text{N} \, \rangle \,.
\end{equation}
By varying the cubic vertex $S_\text{N}$ with respect to the Neveu-Schwarz string field $\Psi_\text{N}$, we find
\begin{equation}
    \delta S_\text{N} = \,
    \langle \, \delta \Psi, {X_\text{N}}^\star (\Psi_\text{N} \ast \Psi_\text{N}) + X_\text{N} \Psi_\text{N} \ast \Psi_\text{N} + \Psi_\text{N} \ast X_\text{N} \Psi_\text{N} \, \rangle \,,
\end{equation}
where $\mathcal{O}^\star$ represents the BPZ conjugate of an operator $\mathcal{O}$.
In the construction of an action by Ohmori and Okawa in~\cite{Ohmori:2017wtx}, the two-string product $V_2$ in the following form is introduced:
\begin{equation}
V_2 (N_1, N_2) = \ \frac{1}{3} \Bigl( {X_\text{N}}^\star (N_1 \ast N_2) + X_\text{N} N_1 \ast N_2 + N_1 \ast X_\text{N} N_2 \Bigr) \,. \label{V2 N-N}
\end{equation}
We can show that this two-string product \eqref{V2 N-N} satisfies the cyclicity equation. We find
\begin{equation}
\begin{split}
 \ \langle \, N_1, V_2 (N_2, N_3) \, \rangle 
= & \ \frac{1}{3} ( \, \langle \, N_1, X_\text{N}^\star (N_2 \ast N_3) \, \rangle 
+ \langle \, N_1, X_\text{N} N_2 \ast N_3 \, \rangle 
+ \langle \, N_1, N_2 \ast X_\text{N} N_3 \, \rangle \, ) \\
= & \ \frac{1}{3} ( \, \langle \, X_\text{N} N_1, N_2 \ast N_3 \, \rangle 
+ \langle \, N_1 \ast X_\text{N} N_2, N_3 \, \rangle
+ \langle \, N_1 \ast N_2, X_\text{N} N_3 \, \rangle ) \\
= & \ \frac{1}{3} ( \, \langle \, X_\text{N} N_1 \ast N_2, N_3 \, \rangle 
+ \langle \, N_1, X_\text{N} N_2 \ast N_3 \, \rangle
+ \langle \, X_\text{N} ^\star (N_1 \ast N_2), N_3 \, \rangle \, ) \\
= & \ \langle \, N_1, V_2 (N_2, N_3) \, \rangle \,,
\end{split}
\end{equation}
where we used the equation~\eqref{BPZ_ast_cyclic}. We can show that this product satisfies the $A_\infty$ relation. We find
\begin{equation}
\begin{split}
 Q V_2 (N_1, N_2) 
= & \ \frac{1}{3} ( \, Q X_\text{N}^\star (N_1 \ast N_2) +Q(X_\text{N} N_1 \ast N_2) +Q (N_1 \ast X_\text{N} N_2) \, ) \\
= & \ \frac{1}{3} ( \, X_\text{N}^\star Q(N_1 \ast N_2) + Q X_\text{N} N_1 \ast N_2 + (-1)^{N_1} X_\text{N} N_1 \ast Q N_2 \\
 & \quad  + Q N_1 \ast X_\text{N} N_2 + (-1)^{N_1} N_1 \ast Q X_\text{N} N_2 \, ) \\
= & \ \frac{1}{3} ( \, X_\text{N}^\star (Q N_1 \ast N_2) + (-1)^{N_1} X^\star (N_1 \ast Q N_2) + X_\text{N} Q N_1 \ast N_2 \\
 & \quad  + (-1)^{N_1} X_\text{N} N_1 \ast Q N_2 + Q N_1 \ast X_\text{N} N_2 + (-1)^{N_1} N_1 \ast X_\text{N} Q N_2 \, ) \\
= & \ V_2 (Q N_1, N_2) + (-1)^{N_1} V_2 (N_1, Q N_2) \,,
\end{split}
\end{equation}
where we used the equations~\eqref{Leibnitz_Q_ast} and \eqref{Q_commutes_with_XN}. 

Let us see that the two-string product~\eqref{V2 N-N} is not associative. We find
\begin{equation}
\begin{split}
& V_2 (V_2(N_1, N_2), N_3) \\
= & \ \frac{1}{9} \Bigl( \, X_\text{N}^\star X_\text{N}^\star(N_1 \ast N_2) \ast N_3 
+ X_\text{N} X_\text{N}^\star (N_1 \ast N_2) \ast N_3 
+ X_\text{N}^\star (N_1 \ast N_2) \ast X_\text{N} N_3 \\
& \quad + X_\text{N}^\star (X_\text{N} N_1 \ast N_2 \ast N_3) 
+ X_\text{N} (X_\text{N} N_1 \ast N_2) \ast N_3 
+ X_\text{N} N_1 \ast N_2 \ast X_\text{N} N_3 \\
& \quad + X_\text{N}^\star (N_1 \ast X_\text{N} N_2 \ast N_3) 
+ X_\text{N} (N_1 \ast X_\text{N} N_2) \ast N_3 
+ N_1 \ast X_\text{N} N_2 \ast X_\text{N} N_3 \, \Bigr) 
\label{NNN associator t-channel} \\
\end{split}
\end{equation}
and
\begin{equation}
\begin{split}
& V_2 (N_1, V_2(N_2, N_3) ) \\
= & \ \frac{1}{9} \Bigl( \, X_\text{N}^\star N_1 \ast X_\text{N}^\star(N_2 \ast N_3) 
+ X_\text{N} N_1 \ast X_\text{N}^\star (N_2 \ast N_3) 
+N_1 \ast X_\text{N} X_\text{N}^\star(N_2 \ast N_3) \\
& \quad + X_\text{N}^\star (N_1 \ast X_\text{N} N_2 \ast N_3) 
+ X_\text{N} N_1 \ast X_\text{N} N_2 \ast N_3 
+ N_1 \ast X_\text{N} (X_\text{N} N_2 \ast N_3) \\
& \quad + X_\text{N}^\star (N_1 \ast N_2 \ast X_\text{N} N_3) 
+ X_\text{N} N_1 \ast N_2 \ast X_\text{N} N_3 
+ N_1 \ast X_\text{N} (N_2 \ast X_\text{N} N_3) \, \Bigr) \,. 
\label{NNN associator s-channel} \\
\end{split}
\end{equation}
Since the picture-changing operators are inserted in a different manner in~\eqref{NNN associator t-channel} and \eqref{NNN associator s-channel}, the two-string product~\eqref{V2 N-N} is not associative:
\begin{equation}
V_2 (V_2(N_1, N_2), N_3) \neq V_2 (N_1, V_2(N_2, N_3) ) \,. \label{nonassociativity of V_2 NNN}
\end{equation}
We can recover the gauge-invariance at quartic order by constructing a three-string product $V_3 (N_1, N_2, N_3)$ satisfying the $A_\infty$ relation~\eqref{3rd A infty}.

\subsection{Quartic vertex for the Neveu-Schwarz sector}
Let us consider a quartic vertex for the Neveu-Schwarz sector. We express a quartic vertex in terms of a three-string product $V_3 (N_1, N_2, N_3)$. As we demonstrated in the case of open bosonic string field theory with stubs, we construct three-string product by interpolating the $t$-channel contribution and the $s$-channel contribution. 

We introduce the line integral $X_\text{N}[f]$ by
\begin{align}
X_\text{N}[f] = \int d\zeta d\tilde{\zeta} e^{- \{Q', \zeta \beta_{-1/2} [f] \} }
 \quad \text{with} \quad z=f(\xi) \,,
\end{align}
where $\beta_{-1/2} [f]$ is the line integral mapped from $\beta_{-1/2}$ by the conformal transformation $f(\xi)$:
\begin{align}
& \beta_{-1/2} [f] = \oint \frac{dz}{2\pi i}  \Bigl( \frac{df^{-1}(z)}{dz} \Bigr)^{-1/2} \beta(z) \quad \text{with} \quad z=f(\xi) \,.
\end{align}
Using the line integral $X_\text{N}$, we can express the picture-changing operator in the BPZ inner product by an operator insertion on the correlation function. For example, we find
\begin{align}
    \ \langle \, N_1, X_\text{N} N_2 \ast N_3 \ast N_4) \, \rangle 
    = & \ \langle \, X_\text{N} [g_2] \, g_1 \circ N_1(0) \, g_2 \circ N_2(0) \, g_3 \circ N_3(0) \, g_4 \circ N_4(0) \, \rangle_D \,, \\
    \ \langle \, N_1, N_2 \ast X_\text{N} (N_3 \ast N_4) ) \, \rangle 
    = & \ \langle \, X_\text{N}[h_3] \, g_1 \circ N_1(0) \, g_2 \circ N_2(0) \, g_3 \circ N_3(0) \, g_4 \circ N_4(0) \, \rangle_D \,.
\end{align}
We define a Grassmann-even operator $X_t$ by
\begin{equation}
 \ \langle \, N_1, V_2( V_2 (N_2, N_3), N_4) \, \rangle 
= \, \langle \, X_t \, g_1 \circ N_1(0) \, g_2 \circ N_2(0) \, g_3 \circ N_3(0) \, g_4 \circ N_4(0) \, \rangle_D \,, \label{definition of X_t for NS} \\
\end{equation}
and we can calculate $X_t$ from the definition of $V_2 (N_1, N_2)$~\eqref{V2 N-N}. We find that
\begin{align}
X_t = & \ \frac{1}{9} \Bigl( \, X_\text{N}[h_4] +  X_\text{N}[g_2] +  X_\text{N}[g_3] \, \Bigr)\Bigl( \, X_\text{N}[g_1] +  X_\text{N}[h_2] +  X_\text{N}[g_4] \, \Bigr) \,. 
\end{align}
We define a Grassmann-even operator $X_s$ by
\begin{equation}
 \ \langle \, N_1, V_2( N_2, V_2(N_3, N_4)) \, \rangle 
= \, \langle \, X_s \, g_1 \circ N_1(0) \, g_2 \circ N_2(0) \, g_3 \circ N_3(0) \, g_4 \circ N_4(0) \, \rangle_D \label{definition of X_s for NS} \,,
\end{equation}
and we find
\begin{align}
X_s = & \ \frac{1}{9} \Bigl( \, X_\text{N}[h_1] +  X_\text{N}[g_3] +  X_\text{N}[g_4] \, \Bigr)\Bigl( \, X_\text{N}[g_2] +  X_\text{N}[h_3] +  X_\text{N}[g_1] \, \Bigr) \,.
\end{align}
We express the three-string product $V_3 (N_1, N_2, N_3)$ in terms of a Grassmann-odd operator $\Xi$:
\begin{equation}
 \ \langle \, N_1, V_3(N_2, N_3, N_4) \, \rangle 
= \, \langle \, g_1 \circ N_1(0) \, \Xi \, g_2 \circ N_2(0) \, g_3 \circ N_3(0) \, g_4 \circ N_4(0) \, \rangle_D \,. \label{definition of Xi NNNN}
\end{equation}
Then the $A_\infty$ relation 
\begin{equation}
\begin{split}
0 = & \ QV_3(N_1,N_2,N_3) +V_3(QN_1,N_2,N_3)  \\
& \quad + (-1)^{N_1}V_3(N_1,QN_2,N_3) +(-1)^{N_1+N_2}V_3(N_1,N_2,QN_3) \\
& \quad {}- V_2(V_2(N_1, N_2), N_3) + V_2(N_1, V_2(N_2, N_3)) 
\label{3rd A_infty relation for NS}
\end{split}
\end{equation}
is translated into the following equation:
\begin{align}
Q \cdot \Xi = & \ X_t - X_s \,. \label{3rd A_infty in term of Xi_NNNN}
\end{align}
In~\cite{Ohmori:2017wtx}, the following operator is introduced as building blocks for the construction of $\Xi$:
\begin{equation}
\Xi [a, b] = \int_0^1 dt \int d\tilde{t} \int d\zeta d\tilde{\zeta} \ e^{-\{ Q', t\zeta \beta_{-1/2} [a] \} } \ e^{-\{ Q', (1-t) \zeta \beta_{-1/2} [b]\} } \,, \label{Xi_NN definition}
\end{equation}
where
\begin{equation}
Q' = \, Q + \tilde{\zeta}\partial_\zeta + \tilde{t}\partial_{t} \,,
\end{equation}
and $a$, $b$ are some integration contours $g_i$, $h_i$ $(i = 1,2,3,4)$ on the disk $D$. This operator
interpolates between the picture-chainging operators $X_\text{N}[a]$ and $X_\text{N}[b]$:
\begin{equation}
Q \cdot \Xi [a, b] = X_\text{N}[a] - X_\text{N}[b] \,, \label{Xi_NN interpolate}
\end{equation}
and it is anti-symmetric about the integration contours $a$ and $b$: 
\begin{equation}
\Xi [a, b] = - \Xi [b, a] \,, \label{Xi_NN antisym}
\end{equation}
The equation~\eqref{Xi_NN interpolate} can be shown in the following way:
\begin{equation}
\begin{split}
 \, Q \cdot \Xi [a, b]  
= & \, \int_0^1 dt \int d\tilde{t} \int d\zeta d\tilde{\zeta} \ Q \cdot \left( \, e^{-\{ Q', t\zeta \beta_{-1/2} [a] \} } \ e^{-\{ Q', (1-t) \zeta \beta_{-1/2} [b]\} } \, \right) \\
= & \, {}-\int_0^1 dt \int d\tilde{t} \int d\zeta d\tilde{\zeta} \ \tilde{t}\partial_{t} \left( \, e^{-\{ Q', t\zeta \beta_{-1/2} [a] \} } \ e^{-\{ Q', (1-t) \zeta \beta_{-1/2} [b]\} } \, \right) \\
& \, {}-\int_0^1 dt \int d\tilde{t} \int d\zeta d\tilde{\zeta} \ \tilde{\zeta}\partial_{\zeta} \left( \, e^{-\{ Q', t\zeta \beta_{-1/2} [a] \} } \ e^{-\{ Q', (1-t) \zeta \beta_{-1/2} [b]\} } \, \right)  \,, 
\end{split}
\end{equation}
where we replaced the action of $Q$ on $e^{-\{ Q', t\zeta \beta_{-1/2} [a] \} } \ e^{-\{ Q', (1-t) \zeta \beta_{-1/2} [b]\} }$ by that of  $-\tilde{t}\partial_t - \tilde{\zeta}\partial_\zeta$. 
The second term on the right-hand side of this equation vanishes, since it is an integration over total derivative of the Grassmann-odd variable.
Then we find
\begin{equation}
\begin{split}
 \, Q \cdot \Xi [a, b]  
= & \, \int d\zeta d\tilde{\zeta} \, e^{-\{ Q', t\zeta \beta_{-1/2} [a] \} } \ e^{-\{ Q', (1-t) \zeta \beta_{-1/2} [b]\} } \,\Bigr|_{\tilde{t}=0}  \\
& \quad \, {}-\int d\zeta d\tilde{\zeta} \, e^{-\{ Q', t\zeta \beta_{-1/2} [a] \} } \ e^{-\{ Q', (1-t) \zeta \beta_{-1/2} [b]\} } \, \Bigr|_{\tilde{t}=0}  \\
= & \, \int d\zeta d\tilde{\zeta} \  e^{-\{ Q'', \zeta \beta_{-1/2} [a] \} } 
\, {}-\int d\zeta d\tilde{\zeta} \ e^{-\{ Q'',\zeta \beta_{-1/2} [b]\} } \,,
\end{split}
\end{equation}
where
\begin{equation}
Q'' = Q + \tilde{\zeta} \partial_\zeta \,.
\end{equation}
Therefore, the equation~\eqref{Xi_NN interpolate} is confirmed. 
To look at the operator $\Xi[a,b]$ in detail, let us expand $\Xi[a,b]$ and carry out the integral over $\tilde{t}$ and $\zeta$. Since we have
\begin{align}
    \{ \, Q', t\zeta \beta_{-1/2}[a] \, \} 
    = & \ {}-\tilde{t} \zeta \beta_{-1/2}
    -t \tilde{\zeta} \beta_{-1/2}
    -t \zeta G_{-1/2} \,, \\
    \{ \, Q', (1-t)\zeta \beta_{-1/2}[a] \, \} 
    = & \ \tilde{t} \zeta \beta_{-1/2}
    -(1-t) \tilde{\zeta} \beta_{-1/2}
    -(1-t) \zeta G_{-1/2}  \,,
\end{align}
we find
\begin{equation}
    \Xi [a,b] = \, \int_0^1 dt \, (\beta_{-1/2}[a] - \beta_{-1/2}[b]) \,
    \delta(t\beta_{-1/2}[a] + (1-t) \beta_{-1/2}[b]) \,.
    \label{Xi_NN explicit}
\end{equation}

Correlation functions which contain the operator $\delta(t\beta_{-1/2}[a] + (1-t) \beta_{-1/2}[b])$ look unfamiliar. 
An example of such correlation functions were explicitly calculated in~\cite{Ohmori:2017wtx}, and it was found that the correlation function develops a double pole at $t=1/2$. 
In general, when the gauge fixing of the world-sheet supergravity fails, the superconformal ghost sector develops a pole and such a pole is called a spurious pole.
In the case of~\cite{Ohmori:2017wtx}, the double pole can be avoided by deforming the contour, and the resulting integral does not depend on the defomation of the contour, since the correlation function does not develop a single pole.
Therefore, correlation functions which contain the operator $\delta(t\beta_{-1/2}[a] + (1-t) \beta_{-1/2}[b])$ are unambiguously defined.

In~\cite{Ohmori:2017wtx}, an operator $\Xi$ satisfying~\eqref{3rd A_infty in term of Xi_NNNN} is constructed. The final result is given by
\begin{equation}
\begin{split}
\Xi = \ \frac{1}{18} \Bigl( & \ \Xi[h_2, h_4; h_3, h_1] + \Xi[h_2, g_4; h_3, g_3] + \Xi[h_2, g_3; h_3, g_4] \\
 & + \Xi[g_4, h_4; g_1, h_1] + \Xi[g_4, g_2; g_1, g_3] + \Xi[g_4, g_3; g_1, g_4] \\
 & + \Xi[g_1, h_4; g_2, h_1] + \Xi[g_1, g_2; g_2, g_3] + \Xi[g_1, g_3; g_2, g_4] \\
 & - \Xi[h_3, h_1; h_4, h_2] - \Xi[h_3, g_3; h_4, g_4] - \Xi[h_3, g_4; h_4, g_1] \\
 & - \Xi[g_1, h_1; g_2, h_2] - \Xi[g_1, g_3; g_2, g_4] - \Xi[g_1, g_4; g_2, g_1] \\
 & - \Xi[g_2, h_1; g_3, h_2] - \Xi[g_2, g_3; g_3, g_4] - \Xi[g_2, g_4; g_3, g_1] \ \Bigr) \,, \label{Xi_NNNN answer}
\end{split}
\end{equation}
where 
\begin{equation}
\begin{split}
\Xi[a_1, a_2; b_1, b_2] =
\frac{1}{2} \Bigl( & \, \Xi[a_1, b_1] X_\text{N}[a_2] + X_\text{N}[b_1] \Xi[a_2, b_2] \\
& +X_\text{N}[a_1] \Xi[a_2, b_2]  +  \Xi[a_1, b_1] X_\text{N}[b_2]\, \Bigr) \,,
\end{split}
\end{equation}
and $a_1$, $b_1$, $a_2$, $b_2$ are integration contours $g_i, h_i \ (i = 1,2,3,4)$  on the disk. The two-string product $V_2 (N_1, N_2)$~\eqref{V2 N-N} and the three-string product $V_3 (N_1, N_2, N_3)$~\eqref{definition of Xi NNNN} constructed from the Grassmann-odd operator $\Xi$ satisfies the $A_\infty$ relation~\eqref{3rd A_infty relation for NS} and the cyclicity equation
\begin{equation}
    \langle \, N_1, V_3( N_2, N_3, N_4) \, \rangle 
= {}-(-1)^{N_1} \langle \, V_3(N_1, N_2, N_3), N_4 \, \rangle \,.
\label{V3 cyclicity for NS}
\end{equation}
Therefore, the action constructed from the two-string product $V_2 (N_1, N_2)$~\eqref{V2 N-N} and the three-string product $V_3 (N_1, N_2, N_3)$~\eqref{definition of Xi NNNN} has the cyclic $A_\infty$ structure up to quartic order.

\section{Open superstring field theory including the Ramond sector\label{sec:NSR}}
\setcounter{equation}{0}
In this section, we consider open superstring field theory including the Ramond sector.
Following the paper~\cite{Kunitomo:2015usa}, we review the kinetic term of the Ramond sector of open superstring field theory by comparison with closed bosonic string field theory. After describing the cubic vertex including the Ramond sector, we construct the quartic vertices including the Ramond sector in the same manner as the Neveu-Schwarz sector.

\subsection{The kinetic term for the Ramond sector}
Recentry, a complete action for open superstring field theory including the Ramond sector was constructed by Kunitomo and Okawa~\cite{Kunitomo:2015usa}. We explain an interpretation of the kinetic term for the Ramond sector in the context of the supermoduli space of super-Riemann surfaces by comparing it with closed bosonic string field theory.\footnote{The content of this subsection is based on the paper by Kunitomo and Okawa~\cite{Kunitomo:2015usa}.}

The propagator strip in open bosonic string field theory~\cite{Witten:1985cc} can be generated by the zero mode of Virasoro operator $L_0$ as $e^{-t L_0}$, and the parameter $t$ is the modulus corresponding to the length of the strip. The integration over this modulus is implemented by the propagator in Siegel gauge as
\begin{equation}
    \frac{b_0}{L_0} = \, \int_0^\infty dt b_0 e^{-tL_0} \,,
\end{equation}
where $b_0$ is the ghost insertion associated with the integration over the modulus $t$.

The propagatpor surfece in closed bosonic string field theory~\cite{Zwiebach:1992ie} can be generated by the Virasoro generators $L_0 + \tilde{L_0}$ and $i(L_0 - \tilde{L_0})$ as $e^{-t(L_0 + \tilde{L_0})+ \theta i(L_0 - \tilde{L_0})}$, where $\tilde{L}_0$ is the zero mode of the anti-holomorphic part of the energy momentum tensor, and $t$ and $\theta$ are moduli. 
In closed string field theory, the integration over $t$ is implemented by the propagator 
\begin{equation}
    \frac{b_0^+}{L_0^+} = \, \int_0^\infty b_0^+ e^{-tL_0^+} 
\end{equation}
in the Siegel gauge
\begin{equation}
    b_0^+ \Psi = \, 0 \,,
\end{equation}
where
\begin{equation}
    b_0^+ = \, b_0 + \tilde{b_0}\,, \quad  L_0^+ = L_0 + \tilde{L_0} \,,
\end{equation}
and $\tilde{b}_0$ is the zero mode of the anti-holomorphic part of the $b$ ghost. The ghost insertion associated with the integration over the modulus $t$ is $b_0^+$. 
The integration over $\theta$ is implemented as a restriction on the Hilbert space of closed string fields. 
The integration over $\theta$ yields the operator of the form
\begin{equation}
    B = \, b_0^- \, \int_0^{2 \pi} \, \frac{d\theta}{2 \pi} e^{i \theta L_0^-} \,,
\end{equation}
where
\begin{equation}
    b_0^- = \, b_0 - \tilde{b_0}\,, \quad  L_0^- = L_0 - \tilde{L_0} \,,
\end{equation}
and $b_0^-$ is the ghost insertion associated with integration over the modulus $\theta$. The operator $B$ can be schematically understood as $\delta(b_0^-) \, \delta(L_0^-)$. The closed bosonic string field $\Psi$ of ghost number $2$ is restricted to satisfy
\begin{equation}
    b_0^- \Psi = \, 0\,, \quad L_0^- \Psi = \, 0 \,.
\end{equation}
It is known that the BRST cohomology on this restricted space gives the correct spectrum of the closed bosonic string. 
The appropriate inner product for restricted string fields $\Psi_1$ and $\Psi_2$ can be written as
\begin{equation}
    \langle \, \Psi_1, c_0^- \Psi_2 \, \rangle \,,
\end{equation}
where
\begin{equation}
    c_0^- = \, \frac{1}{2} (c_0 - \tilde{c}_0) \,,
\end{equation}
and $\tilde{c}_0$ is the zero mode of the anti-holomorphic part of the $c$ ghost. The kinetic term of the restricted string field is given by
\begin{equation}
    S = {}-\frac{1}{2} \langle \, \Psi, c_0^- Q \Psi \, \rangle \,,
\end{equation}
where $Q$ is the BRST operator of the closed bosonic string. The operator $B$ can also be written as
\begin{equation}
    B = \, -i \, \int_0^{2 \pi} \, \frac{d\theta}{2 \pi} \int d\tilde{\theta}
        e^{i \theta L_0^- + i \tilde{\theta}b_0^-} \,,
\end{equation}
where $\tilde{\theta}$ is a Grassmann-odd variable. We can introduce the extended BRST operator $Q'$ for closed superstring field theory, which maps $\theta$ to $\tilde{\theta}$ and $\tilde{\theta}$ to $0$. Then the operator $B$ can be written in the following form:
\begin{equation}
    B = \, -i \, \int_0^{2 \pi} \, \frac{d\theta}{2 \pi} \int d\tilde{\theta}
        e^{\{ \, Q' \,, \, i\theta b_0^- \, \} } \,.
\end{equation}
The closed string field satisfying the restriction can be characterized as
\begin{equation}
    Bc_0^- \Psi = \, \Psi \,.
    \label{restriction on closed bosonic string field}
\end{equation}

Let us consider the kinetic term for the Ramond sector of open superstring field theory. The fermionic direction of the moduli space can be parameterized as $e^{\zeta G_0}$, where $\zeta$ is the odd modulus. As is the case with the Neveu-Schwarz sector of the open superstring~\eqref{PCO for NS sector}, the integration over $\zeta$ with an associated ghost insertion yields the operator $X_\text{R}$\footnote{The subscript R denotes that this is the picture-changing operator for the Ramond sector.} 
\begin{equation}
    X_\text{R} = \int d\zeta \int d\tilde{\zeta} e^{\zeta G_0 - \tilde{\zeta}\beta_0} \,,
    \label{PCO for R sector}
\end{equation}
where $\tilde{\zeta}$ is a Grassmann-even variable, $\beta_0$ is the zero mode of the $\beta$ ghost, and $G_0$ is the zero mode of the supercurrent:
\begin{equation}
    \beta_0 = \ \oint \frac{dz}{2 \pi i} z^{1/2} \, \beta(z) \,, \quad 
    G_0 = \oint \frac{dz}{2 \pi i} z^{1/2} \, T_\text{F}(z) \,.
\end{equation}
We note that $X_\text{R}$ is BPZ even:
\begin{equation}
X_\text{R}^\star = \ X_\text{R} \,. \label{X_R is BPZ even}
\end{equation}
The appropriate innter product of restrected string fields $R_1$ and $R_2$ in the $-1/2$ picture can be written as
\begin{equation}
    \langle \, R_1, Y_\text{R} R_2 \, \rangle 
\end{equation}
with
\begin{equation}
    Y_\text{R} = \, {}-c_0 \delta'(\gamma_0) \,,
\end{equation}
where $\delta'(\gamma_0)$ is defined by an integral over a Grassmann-even variable $\sigma$:
\begin{equation}
    \delta'(\gamma_0) = \int d\sigma \sigma e^{\sigma \gamma_0} \,.
\end{equation}
As in the case of $\delta(\beta_{-1/2})$ and $\delta'(\beta_{-1/2})$, $\delta'(\gamma_0)$ is also a Grassmann-odd operator. 
We choose the Ramond string field $\Psi_\text{R}$ to be a Grassmann-odd state of ghost number $1$ and picture number $-1/2$. 
It is known that the BRST cohomology on this restricted space gives the correct spectrum of the Ramond sector of open superstring~\cite{Kazama:1985hd,Kazama:1986cy,Terao:1986ex}. In~\cite{Yamron:1986nb,Kazama:1985hd,Terao:1986ex}, the kinetic term for the restricted Ramond string field is given by
\begin{equation}
    {}-\frac{1}{2} \langle \, \Psi_\text{R}, Y_\text{R} Q \Psi_\text{R} \, \rangle \,. \label{kinetic term for the Ramond sector}
\end{equation}
The Ramond string field in the restricted space is characterized by the following equation~\cite{Kugo:1988mf}:
\begin{equation}
    X_\text{R} Y_\text{R} \Psi_\text{R} = \, \Psi_\text{R} \,.
    \label{restriction on Ramond string field}
\end{equation}
This is analogous to the characterization for the closed bosonic string field~\eqref{restriction on closed bosonic string field}. 

As is the case with the picture-changing operator for the Neveu-Schwarz sector, we can write $X_\text{R}$ using the extended BRST operator $Q'$:
\begin{equation}
X_\text{R} = \ \int d\zeta d\tilde{\zeta} \mathcal{X}_\text{R} (\zeta, \tilde{\zeta}) \,, 
\quad \mathcal{X}_\text{R} (\zeta, \tilde{\zeta}) = \ e^{ \{ Q', -\zeta \beta_0 \} } \,. 
\end{equation}
Then we can show that
\begin{equation}
[ \, Q, X_\text{R} \, ] = 0 \,. \label{Q commutes with X_R}
\end{equation}
The explicit form of $X_\text{R}$ is given by
\begin{equation}
    X_\text{R} = \, G_0 \delta(\beta_0) + b_0 \delta'(\gamma_0) \,,
\end{equation}
where $\delta (\beta_0)$ and $\delta'(\beta_0)$ are defined by
\begin{align}
    \delta (\beta_0) = & \, \int d\tilde{\zeta} e^{-\tilde{\zeta} \beta_0} \,, \\
    \delta' (\beta_0) = & \, {}-\int d\tilde{\zeta}\tilde{\zeta} e^{-\tilde{\zeta} \beta_0} \,.
\end{align}
We note that the action of $\delta(\beta_0)$ is well-defined when it acts on states in the $-3/2$ picture. Otherwise, we encounter singular products such as $\delta(\beta_0) \, \delta(\beta_0)$. The same applies to $\delta(\gamma_0)$: the action of $\delta(\gamma_0)$ is well-defined when it acts on the states in the $-1/2$ picture. 

In appendix~\ref{app:A}, we show that the operators $X_\text{R}$ and $Y_\text{R}$ satisfy the following equations:
\begin{align}
    X_\text{R} Y_\text{R} X_\text{R} = & \ X_\text{R} \,, \label{XYX=X} \\
    Y_\text{R} X_\text{R} Y_\text{R} = & \ Y_\text{R} \,. \label{YXY=Y}
\end{align}
If the Ramond string field $\Psi_R$ is in the restricted space, $Q\Psi$ is also in it:
\begin{equation}
\begin{split}
    X_\text{R} Y_\text{R} Q \Psi_\text{R} 
    = & \ X_\text{R} Y_\text{R} Q X_\text{R} Y_\text{R} \Psi_\text{R} \\
    = & \ X_\text{R} Y_\text{R} X_\text{R} Q Y_\text{R} \Psi_\text{R} \\
    = & \ X_\text{R} Q Y_\text{R} \Psi_\text{R} \\
    = & \ Q X_\text{R} Y_\text{R} \Psi_\text{R} \\
    = & \ Q \Psi_\text{R} \,,
    \label{Q Psi_R is in the restriceted space}
\end{split}
\end{equation}
where we used the equations~\eqref{restriction on Ramond string field}, \eqref{Q commutes with X_R}, and \eqref{XYX=X}.
The kinetic term is invariant under a gauge transformation of the form:
\begin{equation}
    \delta \Psi_\text{R} = \, Q \Lambda_\text{R} \,,
\end{equation}
where $\Lambda_\text{R}$ is a gauge parameter of ghost number $0$ and picture number $-1/2$, and $\Lambda_\text{R}$ is also in the restricted subspace:
\begin{equation}
    X_\text{R} Y_\text{R} \Lambda_\text{R} = \, \Lambda_\text{R} \,.
    \label{restriction on Ramond gauge parameter}
\end{equation}
By varying the kinetic term~\eqref{kinetic term for the Ramond sector} with respect to the Ramond string field $\Psi_\text{R}$, we find
\begin{equation}
\begin{split}
    & \, {}-\frac{1}{2} \langle \, \delta \Psi_\text{R}, Y_\text{R} Q \Psi_\text{R} \, \rangle  
    {}-\frac{1}{2} \langle \, \Psi_\text{R}, Y_\text{R} Q \delta \Psi_\text{R} \, \rangle \\
    = & \, {}-\frac{1}{2} \langle \, Q\Lambda_\text{R}, Y_\text{R} Q \Psi_\text{R} \, \rangle 
    {}-\frac{1}{2} \langle \, \Psi_\text{R}, Y_\text{R} Q^2 \Lambda_\text{R} \, \rangle \,.
\end{split}
\end{equation}
The second term on the right-hand side vanishes since the BRST operator is nilpotent. We find that the first term also vanishes: 
\begin{equation}
\begin{split}
    {}-\frac{1}{2} \langle \, 
    Q \Lambda_\text{R}, Y_\text{R} Q \Psi_\text{R} \, \rangle 
    = & \, {}-\frac{1}{2} \langle \, 
    Q X_\text{R} Y_\text{R} \Lambda_\text{R}, 
    Y_\text{R} Q \Psi_\text{R} \, \rangle \\
    = & \, {}-\frac{1}{2} \langle \, 
    X_\text{R} Q Y_\text{R} \Lambda_\text{R}, 
    Y_\text{R} Q \Psi_\text{R} \, \rangle \\
    = & \, {}-\frac{1}{2} \langle \, Y_\text{R} \Lambda_\text{R}, Q X_\text{R} Y_\text{R} Q \Psi_\text{R} \, \rangle \\
    = & \, {}-\frac{1}{2} \langle \, Y_\text{R} \Lambda_\text{R}, Q^2  \Psi_\text{R} \, \rangle \\
    = & \ 0 \,,
\end{split}
\end{equation}
where we used the equations~\eqref{BPZ_Q_cyclic}, \eqref{X_R is BPZ even}, \eqref{Q commutes with X_R}, \eqref{Q Psi_R is in the restriceted space}, and \eqref{restriction on Ramond gauge parameter}.

\subsection{Cubic vertex including the Ramond sector}
Let us consider a cubic vertex for one Neveu-Schwarz and two Ramond string fields. Disks with two Ramond punctures and one Neveu-Schwarz puncture have no moduli, therefore we can simply contruct the cubic vertex using the star product:
\begin{equation}
S_\text{NRR} = \alpha \, \langle \, \Psi_\text{N}, \Psi_\text{R} \ast \Psi_\text{R} \, \rangle \,,
\label{the cubic vertex for the Ramond sector}
\end{equation}
where $\alpha$ is a non-zero constant to be determined.\footnote{Its value will be determined from the gauge invariance of the action at quartic order. The value of $\alpha$ will be given by \eqref{alpha = 1} in subsection~\ref{subsec:alpha}.}
$\,$ In the following, we consider the following two-string products:
\begin{equation}
V_2 (R_1, R_2) \,, \quad V_2 (N_1, R_1) \,, \quad V_2 (R_1, N_1) \,, 
\label{V2s for NS and R}
\end{equation}
where $N_1$ is the Neveu-Schwarz string fields in the $-1$ picture, and $R_1$ and $R_2$ are the restricted Ramond string fields in the $-1/2$ picture:
\begin{equation}
    X_\text{R} Y_\text{R} R_1 = \, R_1 \,, \quad 
    X_\text{R} Y_\text{R} R_2 = \, R_2 \,.
\end{equation}
We would like to construct the two-string products in~\eqref{V2s for NS and R} which satisfy the $A_\infty$ relations
\begin{align}
0 = & \ QV_2 (R_1, R_2) - V_2(QR_1, R_2) + (-1)^{R_1} V_2 (R_1, QR_2) \,, 
\label{2nd A_infty for RR} \\
0 = & \ QV_2 (N_1, R_1) - V_2(QN_1, R_1) + (-1)^{N_1} V_2 (N_1, QR_1) \,,
\label{2nd A_infty for NR} \\
0 = & \ QV_2 (R_1, N_1) - V_2(QR_1, N_1) + (-1)^{R_1} V_2 (R_1, QN_1) 
\label{2nd A_infty for RN}
\end{align}
and the cyclicity equations
\begin{align}
\langle \, N_1, V_2( R_1, R_2) \, \rangle = & \ \langle \, V_2(N_1, R_1), Y_\text{R} R_2 \, \rangle \,, 
\label{V2 cyclicity for NRR} \\
\langle \, R_1, Y_\text{R} V_2( N_1, R_2) \, \rangle = & \ \langle \, V_2(R_1, N_1), Y_\text{R} R_2 \, \rangle \,,
\label{V2 cyclicity for RNR} \\
\langle \, R_1, Y_\text{R} V_2( R_2, N_1) \, \rangle = & \ \langle \, V_2(R_1, R_2), N_2 \, \rangle \,.
\label{V2 cyclicity for RRN}
\end{align}
Furthermore, we also require that the two-string products $V_2 (N_1, R_1)$ and $V_2 (R_1, N_1)$ be consistent with the restriction on the Ramond string fields:
\begin{align}
    X_\text{R} Y_\text{R} V_2 (N_1, R_1) = & \, V_2 (N_1, R_1) \,,\\
    X_\text{R} Y_\text{R} V_2 (R_1, N_1) = & \, V_2 (R_1, N_1) \,.
\end{align}

Let us read the two-string products in~\eqref{V2s for NS and R} from the cubic vertex $S_\text{NRR}$. By varying the cubic vertex $S_\text{NRR}$ with respect to the Neveu-Schwarz string field $\Psi_\text{N}$, we find
\begin{equation}
\delta S_\text{NRR} = \alpha \, \langle \, \delta \Psi_\text{N}, \Psi_\text{R} \ast \Psi_\text{R} \, \rangle \,.
\end{equation}
Motivated by this equation, we define the two-string product $V_2 (R_1, R_2)$ by
\begin{equation}
V_2 (R_1, R_2) = \alpha \, R_1 \ast R_2 \,. \label{V2 R-R} 
\end{equation}
By varying the cubic vertex $S_\text{NRR}$ with respect to the Ramond string field  $\Psi_\text{R}$, we find
\begin{equation}
\begin{split}
\delta S_\text{NRR} = & \ \alpha \, \langle \, \Psi_\text{N}, \delta \Psi_\text{R} \ast \Psi_\text{R} \, \rangle
+ \alpha \, \langle \, \Psi_\text{N}, \Psi_\text{R} \ast \delta \Psi_\text{R} \, \rangle \\
= & \ \alpha \, \, \langle \, \delta \Psi_\text{R}, Y_\text{R} X_\text{R} (\Psi_\text{N} \ast \Psi_\text{R}) \, \rangle 
+ \alpha \, \langle \, \delta \Psi_\text{R}, Y_\text{R} X_\text{R} (\Psi_\text{R} \ast \Psi_\text{N}) \, \rangle \,, 
\end{split}
\end{equation}
where we used the equations~\eqref{BPZ_antisym}, \eqref{BPZ_ast_cyclic}, and  \eqref{restriction on Ramond string field}, and the fact that $X_\text{R}$ is BPZ even~\eqref{X_R is BPZ even}. Motivated by this equation, we define the two-string products $V_2 (N_1, R_1)$ and $V_2 (R_1, N_1)$ by
\begin{align}
V_2 (N_1, R_1) = & \ \alpha \, X_\text{R} (N_1 \ast R_1) \,, \label{V2 N-R} \\
V_2 (R_1, N_1) = & \ \alpha \, X_\text{R} (R_1 \ast N_1) \,. \label{V2 R-N} 
\end{align}
We can show that the two-string products in~\eqref{V2s for NS and R} satisfy the cyclicity equations. We find 
\begin{equation}
\begin{split}
\langle \, R_1, Y_\text{R} V_2 (N_1, R_2) \, \rangle 
= & \ \alpha \, \langle \, R_1, Y_\text{R} X_\text{R} (N_1 \ast R_2) \, \rangle \\
= & \ \alpha \, \langle \, X_\text{R} Y_\text{R} R_1, Y_\text{R} X_\text{R} (N_1 \ast R_2) \, \rangle \\
= & \ \alpha \, \langle \, Y_\text{R} R_1, X_\text{R} Y_\text{R} X_\text{R} (N_1 \ast R_2) \, \rangle \\
= & \ \alpha \, \langle \, Y_\text{R} R_1, X_\text{R} (N_1 \ast R_2) \, \rangle \\
= & \ \alpha \, \langle \, X_\text{R} Y_\text{R} R_1, N_1 \ast R_2 \, \rangle \\
= & \ \alpha \, \langle \, R_1, N_1 \ast R_2 \, \rangle \,,\\
\end{split}
\end{equation}
where we used the identity~\eqref{XYX=X}, the restriction on the Ramond string field, and the fact that $X_\text{R}$ is BPZ-even~\eqref{X_R is BPZ even}. Using the associativity of the star product~\eqref{associativity_of_star_product}, we find
\begin{equation}
\begin{split}
\langle \, R_1, Y_\text{R} V_2 (N_1, R_2) \, \rangle 
= & \ \alpha \, \langle \, R_1 \ast N_1, R_2 \, \rangle \,,\\
= & \ \alpha \, \langle \, X_\text{R}(R_1 \ast N_1), Y_\text{R} R_2 \, \rangle \\
= & \ \langle \, V_2(R_1,N_1), Y_\text{R} R_2 \, \rangle \,.
\end{split}
\end{equation}
We can show that the cyclicity equations for the remaining two-string products~\eqref{V2 cyclicity for NRR} and~\eqref{V2 cyclicity for RRN} hold in the same manner.

The $A_\infty$ relation~\eqref{2nd A_infty for RR} for the two-string product $V_2 (R_1, R_2)$ follows from the equation~\eqref{Leibnitz_Q_ast} for the star product. 
We can show that the remaining two-string products also satisfy the $A_\infty$ relations~\eqref{2nd A_infty for NR} and \eqref{2nd A_infty for RN}. We find
\begin{equation}
\begin{split}
Q V_2 (N_1, R_1) 
= & \ \alpha \, QX_\text{R} (N_1 \ast R_1) \\
= & \ \alpha \, X_\text{R} Q(N_1 \ast R_1) \\
= & \ \alpha \, X_\text{R} (QN_1 \ast R_1) +  \alpha \, (-1)^{N_1} (N_1\ast QR_1) \\
= & \ V_2 (QN_1, R_1) + (-1)^{N_1} V_2 (N_1, QR_1) \,,
\end{split}
\end{equation}
where we used the equations~\eqref{Leibnitz_Q_ast} and \eqref{Q commutes with X_R}. We can show that the $A_\infty$ relation~\eqref{2nd A_infty for NR} holds in the same manner. 
We can show that the two-string products $V_2 (N_1,R_1)$ and $V_2 (R_1,N_1)$ satisfy the restriction on the Ramond string field:
\begin{align}
    X_\text{R}Y_\text{R} V_2 (N_1, R_1) 
    = \, X_\text{R}Y_\text{R} X_\text{R} (N_1 \ast R_1) 
    = \, X_\text{R} (N_1 \ast R_1) 
    = \, V_2 (N_1, R_1) \,, \\
    X_\text{R}Y_\text{R} V_2 (R_1, N_1) 
    = \, X_\text{R}Y_\text{R} X_\text{R} (R_1 \ast N_1) 
    = \, X_\text{R} (R_1 \ast N_1) 
    = \, V_2 (R_1, N_1) \,,
\end{align}
where we used the equation~\eqref{XYX=X}\,. 
Let us see whether the two-string products in~\eqref{V2s for NS and R} are associative or not. We find
\begin{equation}
V_2 ( V_2(N_1,R_1), N_2) 
= \, \alpha^2 X_\text{R} ( X_\text{R} (N_1 \ast R_1) \ast N_2) \,
\end{equation}
and
\begin{equation}
V_2 (N_1,V_2 (R_1, N_2)) 
= \, \alpha^2 X_\text{R} (N_1 \ast X_\text{R} ( R_1 \ast N_2)) \,.
\end{equation}
The picture-changing operator is inserted in a different manner, therefore we find
\begin{equation}
V_2 (V_2(N_1, R_1), N_2) \neq \ V_2 (N_1, V_2(R_1, N_2) ) \,. \label{nonassociativity of V_2 NRN}
\end{equation}
By similar calculations, we can show that the two-string product $V_2$ is not associative for other inputs of Neveu-Schwarz and Ramond string fields, except for three Ramond string fields:
\begin{equation}
V_2( V_2(R_1,R_2), R_3) = \, V_2 (R_1, V_2 (R_2, R_3)) \,. \label{V2 R-R is associative}
\end{equation}
Non-associativity of two-string products mean incorrect integration over the moduli space of punctured disks. Therefore, we need appropriate quartic vertices with two Neveu-Schwarz and two Ramond string fields. Since we have \eqref{V2 R-R is associative}, we do not need any quartic vertex for four Ramond string fields.\footnote{If we use the two-string stub product, we may need a quartic vertex for four Ramond string fields. In section~\ref{sec:NSRstub}, we consider this case.}

\subsection{Quartic vertices including the Ramond sector}
Let us consider quartic vertices including the Ramond sector. We express quartic vertices for various Neveu-Schwarz and Ramond orderings in terms of the following three-string products:
\begin{equation}
\begin{split}
&V_3 (R_1, N_1, N_2) \,, \quad V_3 (N_1, R_1, N_2) \,, 
\quad V_3 (N_1, N_2, R_1) \,, \\
&V_3 (R_1, R_2, N_1) \,, \quad V_3 (R_1, N_1, R_2) \,,
\quad V_3 (N_1, R_1, R_2) \,. \label{V3s NS and R}
\end{split}
\end{equation}
The cyclicity equations for these three-string products can be divided into two groups. The equations in the first group are
\begin{align}
\langle \, R_1, Y_\text{R} V_3( N_2, R_2, N_2) \, \rangle 
= & \ {}-(-1)^{R_1} \langle \, V_3(R_1, N_1, R_2), N_2 \, \rangle \label{V3 cyclicity for RNRN} \,, \\
\langle \, N_1, V_3( R_1, N_2, R_2) \, \rangle 
= & \ {}-(-1)^{N_1} \langle \, V_3(N_1, R_1, N_2), Y_\text{R} R_2 \, \rangle \,, \label{V3 cyclicity for NRNR} 
\end{align}
and the equations in the second group are
\begin{align}
\langle \, R_1, Y_\text{R} V_3( R_2, N_1, N_2) \, \rangle 
= & \ {}-(-1)^{R_1} \langle \, V_3(R_1, R_2, N_1), N_2 \, \rangle \label{V3 cyclicity for RRNN} \,, \\
\langle \, N_1, V_3( R_1, R_2, N_2) \, \rangle 
= & \ {}-(-1)^{N_1} \langle \, V_3(N_1, R_1, R_2), N_2 \, \rangle \label{V3 cyclicity for NRRN} \,, \\
\langle \, N_1, V_3( N_2, R_1, R_2) \, \rangle 
= & \ {}-(-1)^{N_1} \langle \, V_3(N_1, N_2, R_1), Y_\text{R} R_2 \, \rangle \label{V3 cyclicity for NNRR} \,, \\
\langle \, R_1, Y_\text{R} V_3( N_1, N_2, R_2) \, \rangle 
= & \ {}-(-1)^{R_1} \langle \, V_3(R_1, N_1, N_2), Y_\text{R} R_2 \, \rangle \,. \label{V3 cyclicity for RNNR} 
\end{align}
We call a set of quartic vertices which appear in the first group the RNRN group, and we call a set of quartic vertices which appear in the second group the RNRN group.
The three-string products in~\eqref{V3s NS and R} are also divided into the following two groups 
\begin{equation}
V_3 (R_1, N_1, R_2) \,, \quad V_3 (N_1, R_1, N_2) \label{V3 RNRN and V3 NRNR}
\end{equation}
which appear in the quartic vertices in the RNRN group, and
\begin{equation}
V_3 (N_1, N_2, R_1) \,, \quad  V_3 (R_1, R_2, N_1) \,, \quad
V_3 (R_1, N_1, N_2) \,, \quad V_3 (N_1, R_1, R_2) 
\label{V3s for other NS R combination}
\end{equation}
which appear in the quartic vertices in the RRNN group. In this subsection, we construct three-string products in~\eqref{V3 RNRN and V3 NRNR}. We will construct three-string products in~\eqref{V3s for other NS R combination} in the next subsection and appendix~\ref{app:C1}.

Let us consider three-string products in~\eqref{V3 RNRN and V3 NRNR}. As before, we express the three-string product $V_3(N_1, R_1, N_2)$ in terms of a Grassmann-odd operator $\Xi_\text{RNRN}$:
\begin{equation}
 \ \langle \, R_1, Y_\text{R} V_3(N_1, R_2, N_2) \, \rangle 
= \, \langle \, g_1 \circ R_1(0) \ \Xi_\text{RNRN} \, g_2 \circ N_1(0) \, g_3 \circ R_2(0) \, g_4 \circ N_2(0) \, \rangle_D  \,. \label{RNRN definition}
\end{equation}
As before, we define a Grassmann-even operator $(X_t)_\text{RNRN}$ by
\begin{equation}
\langle \, R_1, Y_\text{R} V_2( V_2 (N_1, R_2), N_2) \, \rangle
= \, \langle \, (X_t)_\text{RNRN} \, g_1 \circ R_1(0) \, g_2 \circ N_1(0) \, g_3 \circ R_2(0) \, g_4 \circ N_2(0) \, \rangle_D \label{definition of X_t RNRN} \end{equation}
and $(X_s)_\text{RNRN}$ by
\begin{equation}
\langle \, R_1, Y_\text{R} V_2( N_1, V_2(R_2, N_2)) \, \rangle
= \, \langle \, (X_s)_\text{RNRN} \, g_1 \circ R_1(0) \, g_2 \circ N_1(0) \, g_3 \circ R_2(0) \, g_4 \circ N_2(0) \, \rangle_D \label{definition of X_s RNRN} \,.
\end{equation}
Then the $A_\infty$ relation for $V_3(N_1,R_1,N_2)$,
\begin{equation}
\begin{split}
0 = & \ QV_3(N_1,R_1,N_2) +V_3(QN_1,R_1,N_2)  \\
& \quad + (-1)^{N_1}V_3(N_1,QR_1,N_2) +(-1)^{N_1+R_1}V_3(N_1,R_1,QN_2) \\
& \quad - V_2(V_2(N_1, R_1), N_2) + V_2(N_1, V_2(R_1, N_2)) \,, \label{3rd A_infty for NRN}
\end{split}
\end{equation}
is translated into the following equation:
\begin{equation}
Q \cdot \Xi_\text{RNRN} = (X_t)_\text{RNRN} - (X_s)_\text{RNRN} \,. \label{3rd A_infty for Xi_RNRN}
\end{equation}
From the definition of the two-string products~\eqref{V2 N-R} and \eqref{V2 R-N}, we have
\begin{align}
\langle \, R_1,Y_\text{R} V_2(V_2(N_1,R_2), N_2) \, \rangle 
= \, \alpha^2 \, \langle \, R_1, X_\text{R} (N_1 \ast R_2 ) \ast N_2 \, \rangle \,, \\
\langle \, R_1,Y_\text{R} V_2(N_1,V_2(R_2, N_2)) \, \rangle
= \, \alpha^2 \, \langle \, R_1, N_1 \ast X_\text{R} ( R_2 \ast N_2) \, \rangle \,.
\end{align}
Then we obtain
\begin{equation}
(X_t)_\text{RNRN} = \alpha^2 \, X_\text{R} [h_2] \,, \quad (X_s)_\text{RNRN} = \alpha^2 \, X_\text{R} [h_3] \,,
\end{equation}
where the line integral $X_\text{R}[f]$ for a conformal transformation $f(\xi)$ is defined by
\begin{equation}
X_\text{R} [f] = \ \int d\zeta d\tilde{\zeta} e^{- \{Q', \zeta \beta_0 [f] \} }  \quad \text{with} \quad z=f(\xi) \,, 
\end{equation}
and the line integral $\beta_0 [f]$ for a conformal transformation $f(\xi)$ is defined by
\begin{equation}
\beta_0 [f] = \, \oint \frac{dz}{2\pi i} (f^{-1}(z))^{1/2} \Bigl( \frac{df^{-1}(z)}{dz} \Bigr)^{-1/2} \beta(z)  \quad \text{with} \quad z=f(\xi) \,.    
\end{equation}
Then the $A_\infty$ relation~\eqref{3rd A_infty for Xi_RNRN} can be written as
\begin{equation}
Q \cdot \Xi_\text{RNRN} = \alpha^2 \, X_\text{R} [h_2] - \alpha^2 \, X_\text{R} [h_3] \,. \label{3rd A_infty for Xi_RNRN explicit}
\end{equation}
Since the right-hand side of \eqref{3rd A_infty for Xi_RNRN explicit} is written in the difference of the picture-changing operator $X_\text{R}$, we can interpolate them by introducing the following operator:
\begin{equation}
\Xi_\text{RR} [a, b] = \int dt \int d\tilde{t} \int d\zeta d\tilde{\zeta} \ e^{-\{ Q', t \zeta \beta_0 [a] \} } \ e^{-\{ Q', (1-t) \zeta \beta_0 [b] \} } \,, \label{Xi_RR definition}
\end{equation}
where $a$ and $b$ are integration contours $g_i, h_i \ (i = 1,2,3,4)$. As in the case of the Neveu-Schwarz sector, we can show that $\Xi_\text{RR}$ interpolates the picture-changing operators:
\begin{align}
 Q \cdot \Xi_\text{RR} [a, b] = & \ X_\text{R} [a] - X_\text{R} [b] \,, \label{Xi_RR interpolate}
\end{align}
and $\Xi_\text{RR}$ is anti-symmetric:
\begin{align}
\Xi_\text{RR} [a, b] = & \ {}-\Xi_\text{RR} [b, a] \,. \label{Xi_RR antisym}
\end{align}
Using $\Xi_\text{RR} [a,b]$, a naive solution for the $A_\infty$ relation~\eqref{3rd A_infty for Xi_RNRN explicit} is given by
\begin{equation}
\Xi_\text{RNRN} \overset{?}{=} \alpha^2 \,  \Xi_\text{RR} [h_2, h_3] \,. \label{Xi RNRN naive}
\end{equation}
As mentioned in~\cite{Ohmori:2017wtx}, this interpolation generates potentially singular products of line integrals with intersecting contours such as $G_0 [h_2] \, \delta (\beta_0[h_3])$. We avoid operators of the form $\Xi_\text{RR}[h_i, h_j] \, (i \neq j)$ by introducing  $\check{\Xi}_\text{RR} [h_i, h_j]$:
\begin{equation}
\begin{split}
\check{\Xi}_\text{RR} [h_i, h_j]
= & \, \frac{1}{4} \, ( \Xi_\text{RR} [h_i, g_1] + \Xi_\text{RR} [g_1, h_j] ) 
+ \frac{1}{4} \, ( \Xi_\text{RR} [h_i, g_2] + \Xi_\text{RR} [g_2, h_j] ) \\
& + \frac{1}{4} \, ( \Xi_\text{RR} [h_i, g_3] + \Xi_\text{RR} [g_3, h_j] )
+ \frac{1}{4} \, ( \Xi_\text{RR} [h_i, g_4] + \Xi_\text{RR} [g_4, h_j] ) \,. 
\label{Xi_RR check}
\end{split}
\end{equation}
We can show that the operator $\check{\Xi}_\text{RR} [h_i, h_j]$ also interpolates the picture-changing operators:
\begin{align}
 Q \cdot \check{\Xi}_\text{RR} [h_i, h_j] = & \ X_\text{R} [h_i] - X_\text{R} [h_j] \,, \label{check_Xi_RR interpolate}
\end{align}
and $\check{\Xi}_\text{RR} [h_i, h_j]$ is anti-symmetric:
\begin{align}
\check{\Xi}_\text{RR} [h_i, h_j] = & \ {}-\check{\Xi}_\text{RR} [h_j, h_i] \,, \label{check_Xi_RR antisym}
\end{align}
which follows from \eqref{Xi_RR interpolate} and \eqref{Xi_RR antisym}. Then we choose $\Xi_\text{RNRN}$ in the following form:
\begin{equation}
\Xi_\text{RNRN} =  \alpha^2 \,  \check{\Xi}_\text{RR} [h_2, h_3] \,.
\label{Xi RNRN answer}
\end{equation}
Let us consider the three-string product $V_3(R_1, N_1, R_2)$. We express this product in terms of a Grassmann-odd operator $\Xi_\text{NRNR}$:
\begin{equation}
\langle \, N_1, V_3(R_1, N_2, R_2) \, \rangle 
= \, \langle \, g_1 \circ N_1(0) \ \Xi_\text{NRNR} \, g_2 \circ R_1(0) \, g_3 \circ N_2(0) \, g_4 \circ R_2(0) \, \rangle_D \,. \label{definition of Xi NRNR}
\end{equation}
We define Grassmann-even operators $(X_t)_\text{NRNR}$ and $(X_s)_\text{NRNR}$ by
\begin{align}
\langle \, N_1, V_2( V_2 (R_1, N_2), R_2) \, \rangle
= \, \langle \, (X_t)_\text{NRNR} \, g_1 \circ N_1(0) \, g_2 \circ R_1(0) \, g_3 \circ N_2(0) \, g_4 \circ R_2(0) \, \rangle_D \label{definition of X_t NRNR} \,, \\
\langle \, N_1, V_2( R_1, V_2(N_2, R_2)) \, \rangle 
= \, \langle \, (X_s)_\text{NRNR} \, g_1 \circ N_1(0) \, g_2 \circ R_1(0) \, g_3 \circ N_2(0) \, g_4 \circ R_2(0) \, \rangle_D \label{definition of X_s NRNR} \,.
\end{align}
Then the $A_\infty$ relation,
\begin{equation}
\begin{split}
0 = & \ QV_3(R_1,N_1,R_2) +V_3(QR_1,N_1,R_2)  \\
& \quad + (-1)^{R_1}V_3(R_1,QN_1,R_2) +(-1)^{R_1+N_1}V_3(R_1,N_1,QR_2) \\
& \quad - V_2(V_2(R_1, N_1), R_2) + V_2(R_1, V_2(N_1, R_2)) \,, \label{3rd A_infty for RNR}
\end{split}
\end{equation}
is translated into the following equation:
\begin{equation}
Q \cdot \Xi_\text{NRNR} = \, (X_t)_\text{NRNR} - (X_s)_\text{NRNR} \,.
\end{equation}
Using the definition of two-string products~\eqref{V2 R-R}, \eqref{V2 N-R}, and \eqref{V2 R-N}, we find
\begin{equation}
(X_t)_\text{NRNR} = \alpha^2 \, X_\text{R}[h_2] \,, \quad (X_s)_\text{NRNR} = \alpha^2 \, X_\text{R}[h_3] \,.
\end{equation}
As before, the solution is
\begin{equation}
\quad \Xi_\text{NRNR} = \alpha^2 \, \check{\Xi}_\text{RR} [h_2,h_3] \,. \label{Xi NRNR answer}
\end{equation}

\subsection{Cyclicity equations}
Let us consider the cyclicity equations~\eqref{V3 cyclicity for RNRN} and \eqref{V3 cyclicity for NRNR}. These equations are translated into
\begin{equation}
    \omega \circ \Xi_\text{RNRN} = {}- \Xi_\text{NRNR}\,, \quad
        \omega \circ \Xi_\text{NRNR} = {}- \Xi_\text{RNRN}\,,
        \label{Xi cyclicity for RNRN and NRNR}
\end{equation}
where $\omega (z)$ is the conformal transformation for the rotation of degree $\pi/2$ on the disk $D$,
\begin{equation}
\omega (z) = e^{2\pi i / 4} = iz,
\end{equation}
and we denote the operator $\mathcal{O}$ transformed under the rotation $\omega (z)$ by $\omega \circ \mathcal{O}$. For example, we have the following equations:
\begin{align}
\omega \circ X_\text{R} [g_i] = & X_\text{R} [g_{i+1}] \,, \quad \omega \circ X_\text{R} [h_i] = X_\text{R} [h_{i+1}] \,, \\
\omega \circ \beta_\text{R} [g_i] = & \beta_\text{R} [g_{i+1}] \,, \quad \omega \circ \beta_\text{R} [h_i] = \beta_\text{R} [h_{i+1}] \,,
\end{align}
for $i = 1,2,3,4$ with the understanding that $g_5 \equiv g_1$ and $h_5 \equiv h_1$.
Let us consider the action of $\omega$ on the operator $\Xi_\text{RR}[g_i, g_j]$. We find
\begin{equation}
\begin{split}
\omega \circ \Xi_\text{RR}[g_i,g_j] 
= & \, \int_0^1 dt \int d\tilde{t} \int d\zeta d\tilde{\zeta} \ e^{-\{ Q', t\zeta \omega \circ \beta_0 [g_i] \} } \ e^{-\{ Q', (1-t) \zeta \omega \circ \beta_0 [g_j]\} } \\
= & \, \int_0^1 dt \int d\tilde{t} \int d\zeta d\tilde{\zeta} \ e^{-\{ Q', t\zeta \beta_0 [g_{i+1}] \} } \ e^{-\{ Q', (1-t) \zeta \beta_0 [g_{j+1}]\} } \\
= & \ \Xi_\text{RR}[g_{i+1},g_{j+1}] \,. \label{rotation of Xi_RR}
\end{split}
\end{equation}
Similarly, we find
\begin{align}
    \omega \circ \Xi_\text{RR}[g_i,h_j] 
= \, \Xi_\text{RR}[g_{i+1},h_{j+1}] \,, \quad
    \omega \circ \Xi_\text{RR}[h_i,g_j] 
= \, \Xi_\text{RR}[h_{i+1},g_{j+1}] \,.
\end{align}
Furthermore, we can show that $\check{\Xi}_\text{RR}$ \eqref{Xi_RR check} satisfies
\begin{equation}
\omega \circ \check{\Xi}_\text{RR}[h_i,h_j] 
= \, \check{\Xi}_\text{RR}[h_{i+1},h_{j+1}] \,. \label{rotation of check_Xi_RR}
\end{equation}
Then the cyclicity equation for $V_3(N_1, R_1, N_2)$ in terms of the operator $\Xi_\text{RNRN}$ \eqref{Xi cyclicity for RNRN and NRNR} can be shown in the following way:
\begin{equation}
\begin{split}
\omega \circ \Xi_\text{RNRN} 
= & \ \omega \circ (\alpha^2 \, \check{\Xi}_\text{RR} [h_2, h_3]) \\
= & \, \alpha^2 \, \check{\Xi}_\text{RR} [h_3, h_4] \\
= & {}- \alpha^2 \, \check{\Xi}_\text{RR} [h_4, h_3] \,,
\end{split}
\end{equation}
where we used the anti-symmetry of $\check{\Xi}_\text{RR}$ \eqref{check_Xi_RR antisym}. 
In the proof of the cyclicity equation, we use the following relations:
\begin{align}
    &\beta_0 [h_1] = \beta_0 [h_3], \quad \beta_0 [h_2] = \beta_0 [h_4] \,.
    \label{beta_0 is BPZ even}
\end{align}
We can prove the equations~\eqref{beta_0 is BPZ even} in the same manner as \eqref{b_0 is BPZ even} in appendix~\ref{app:B}, using the fact that the weight of the superconformal ghost $\beta (z)$ is $3/2$. 
Then we find
\begin{equation}
\begin{split}
\omega \circ \Xi_\text{RNRN} 
= & \ {}- \alpha^2 \, \check{\Xi}_\text{RR} [h_2, h_3] \\
= & \ {}-\Xi_\text{NRNR} \,.
\end{split}
\end{equation}
 Since $\Xi_\text{NRNR}$ equals $\Xi_\text{RNRN}$, the latter equation of \eqref{Xi cyclicity for RNRN and NRNR} is also satisfied. Therefore, the three-string products in~\eqref{V3 RNRN and V3 NRNR} satisfy the cyclicity equations~\eqref{V3 cyclicity for RNRN} and \eqref{V3 cyclicity for NRNR}.

We note that $(X_s)_\text{RNRN}$ and $(X_t)_\text{NRNR}$ are related by the rotation $\omega$. For instance we can show that
\begin{equation}
\omega \circ (X_s)_\text{RNRN} 
= \, \alpha^2 \, \omega \circ X_\text{R}[h_3] 
= \, \alpha^2 \, X_\text{R}[h_4] 
= \, \alpha^2 \, X_\text{R}[h_2] 
= \, (X_t)_\text{NRNR} \,,
\end{equation}
where we used 
\begin{equation}
    X_\text{R}[h_1] = \, X_\text{R}[h_3], \quad     X_\text{R}[h_2] = \, X_\text{R}[h_4] \,,
\end{equation}
which follows from the fact that $X_\text{R}$ is BPZ even \eqref{X_R is BPZ even}. To summarize, we find
\begin{align}
\omega \circ (X_t)_\text{RNRN} = & \, (X_s)_\text{NRNR}, \quad 
\omega \circ (X_s)_\text{RNRN} = \, (X_t)_\text{NRNR} \,,
\label{X_t and X_s cyclicity for RNRN} \\
\omega \circ (X_t)_\text{NRNR} = & \, (X_s)_\text{RNRN}, \quad 
\omega \circ (X_s)_\text{NRNR} = \, (X_t)_\text{RNRN} \,.
\label{X_t and X_s cyclicity for NRNR}
\end{align}
We will use these equations in appendix~\ref{app:C3}.

\subsection{Determination of the cubic vertex including the Ramond sector \label{subsec:alpha}}
Let us consider the remaining three-string products in~\eqref{V3s for other NS R combination}. We construct the three-string product $V_3 (R_1, N_1, N_2)$ in this subsection, and the explicit construction of the rest is given in appendix~\ref{app:C}. 

We define the three-string product $V_3(R_1,N_1,N_2)$ in terms of a Grassmann-odd operator $\Xi_\text{RRNN}$ by
\begin{equation}
\langle \, R_1, Y_\text{R} V_3(R_2, N_1, N_2) \, \rangle
= \, \langle \, g_1 \circ R_1(0) \ \Xi_\text{RRNN} \, g_2 \circ R_2(0) \, g_3 \circ N_1(0) \, g_4 \circ N_2(0) \, \rangle_D \,. \label{RRNN definition}
\end{equation}
We define Grassmann-even operators $(X_t)_\text{RRNN}$ and $(X_s)_\text{RRNN}$ by
\begin{align}
& \, \langle \, R_1, Y_\text{R} V_2( V_2 (R_2, N_1), N_2) \, \rangle \notag \\ 
= & \, \langle \, (X_t)_\text{RRNN} \, g_1 \circ R_1(0) \, g_2 \circ R_2(0) \, g_3 \circ N_1(0) \, g_4 \circ N_2(0) \, \rangle_D \label{definition of X_t RRNN} \,, \\
& \, \langle \, R_1, Y_\text{R} V_2( R_2, V_2(N_1, N_2)) \, \rangle \notag \\
= & \, \langle \, (X_s)_\text{RRNN} \, g_1 \circ R_1(0) \, g_2 \circ R_2(0) \, g_3 \circ N_1(0) \, g_4 \circ N_2(0) \, \rangle_D \label{definition of X_s RRNN} \,.
\end{align}
Then the $A_\infty$ relation for $V_3 (R_1, N_1, N_2)$,
\begin{equation}
\begin{split}
0 = & \ QV_3(R_1,N_1,N_2) +V_3(QR_1,N_1,N_2)  \\
& \quad + (-1)^{R_1}V_3(R_1,QN_1,N_2) +(-1)^{R_1+N_1}V_3(R_1,N_1,QN_2) \\
& \quad - V_2(V_2(R_1, N_1), N_2) + V_2(R_1, V_2(N_1, N_2)) \,, \label{3rd A_infty for RNN}
\end{split}
\end{equation}
is translated into the following equation:
\begin{equation}
Q \cdot \Xi_\text{RRNN} = (X_t)_\text{RRNN} - (X_s)_\text{RRNN} \,.
\end{equation}
Using the definitions of the two-string products~\eqref{V2 N-N} and \eqref{V2 R-N}, we find
\begin{equation}
(X_t)_\text{RRNN} = \ \alpha^2 \, X_\text{R} [h_2] \,, \quad (X_s)_\text{RRNN} = \ \frac{\alpha}{3} X_\text{N} [h_1] + \frac{\alpha}{3} X_\text{N} [g_3] + \frac{\alpha}{3} X_\text{N} [g_4] \,.
\end{equation}
Then the equation for $\Xi_\text{RRNN}$ is 
\begin{equation}
\begin{split}
    Q \cdot \Xi_\text{RRNN} 
    = & \, (X_t)_\text{RRNN} - (X_s)_\text{RRNN} \\
    = & \, \alpha^2 \, X_\text{R} [h_2] 
    - \frac{\alpha}{3} X_\text{N} [h_1] 
    - \frac{\alpha}{3} X_\text{N} [g_3] 
    - \frac{\alpha}{3} X_\text{N} [g_4] \\
    = & \,\frac{\alpha}{3} \Bigl[ 
    (X_\text{R}[h_2] - \alpha X_\text{N} [h_1]) 
    - (X_\text{R}[h_2] - \alpha  X_\text{N} [g_3]) 
    + (X_\text{R}[h_2] - \alpha  X_\text{N} [g_4]) 
    \Bigr] \,.
\label{3rd A_infty for Xi_RRNN explicit}
\end{split}
\end{equation}
The right-hand side of \eqref{3rd A_infty for Xi_RRNN explicit} consists of the difference of $X_\text{N}$ and $ X_\text{R}$. If the coefficient of the Ramond cubic interaction $\alpha$ is given by
\begin{equation}
    \alpha = 1, \label{alpha = 1}
\end{equation}
we can interpolate between the picture-changing operators by introducing the following operators:
\begin{align}
& \Xi_\text{RN} [a, b] = \, \int_0^1 dt \int d\tilde{t} \int d\zeta d\tilde{\zeta} \ e^{-\{ Q', t\zeta \beta_0 [a] \} } \ e^{-\{ Q', (1-t) \zeta \beta_{-1/2} [b]\} } \,,\\
& \Xi_\text{NR} [a, b] = \, \int_0^1 dt \int d\tilde{t} \int d\zeta d\tilde{\zeta} \ e^{-\{ Q', t\zeta \beta_{-1/2} [a] \} } \ e^{-\{ Q', (1-t) \zeta \beta_0 [b]\} } \,,
\end{align}
where $a, b$ are some integration contours $g_i, h_i$. The operators $\Xi_\text{NR}$ and $\Xi_\text{RN}$ are related by
\begin{align}
 \Xi_\text{RN} [a, b] = {}-\Xi_\text{NR} [b, a] \,, \label{Xi_NR and Xi_RN antisym}
\end{align}
and they interpolate between $X_\text{N}$ and $X_\text{R}$: 
\begin{align}
 Q \cdot \Xi_\text{RN} [a, b] = & \, X_\text{R} [a] - X_\text{N} [b] \,, \label{Xi_RN interpolate}\\ 
 Q \cdot \Xi_\text{NR} [a, b] = & \, X_\text{N} [a] - X_\text{R} [b] \,. \label{Xi_NR interpolate}
\end{align}
As before, we avoid $\Xi_\text{RN}[h_i, h_j]$ by introducing $\check{\Xi}_\text{RN}[h_i, h_j]$ given by
\begin{equation}
\begin{split}
\check{\Xi}_\text{RN} [h_i, h_j]
= & \, \frac{1}{4} \, ( \Xi_\text{RR} [h_i, g_1] + \Xi_\text{RN} [g_1, h_j] ) 
+ \frac{1}{4} \, ( \Xi_\text{RR} [h_i, g_2] + \Xi_\text{RN} [g_2, h_j] ) \\
& + \frac{1}{4} \, ( \Xi_\text{RR} [h_i, g_3] + \Xi_\text{RN} [g_3, h_j] )
+ \frac{1}{4} \, ( \Xi_\text{RR} [h_i, g_4] + \Xi_\text{RN} [g_4, h_j] ) \,.
\label{Xi_RN check}
\end{split}
\end{equation}
Using the operators $\Xi_\text{RN}$ and $\check{\Xi}_\text{RN}$, we find 
\begin{equation}
    \Xi_\text{RRNN} = \ \frac{1}{3} \Bigl( \, \check{\Xi}_\text{RN}[h_2, h_1] + \Xi_\text{RN}[h_2, g_3] + \Xi_\text{RN}[h_2, g_4] \, \Bigr) \,.
    \label{Xi RRNN answer}
\end{equation}
We also introduce $\check{\Xi}_\text{NR}$ by
\begin{equation}
\begin{split}
\check{\Xi}_\text{NR} [h_i, h_j]
= & \, \frac{1}{4} \, ( \Xi_\text{NR} [h_i, g_1] + \Xi_\text{RR} [g_1, h_j] ) 
+ \frac{1}{4} \, ( \Xi_\text{NR} [h_i, g_2] + \Xi_\text{RR} [g_2, h_j] ) \\
& + \frac{1}{4} \, ( \Xi_\text{NR} [h_i, g_3] + \Xi_\text{RR} [g_3, h_j] )
+ \frac{1}{4} \, ( \Xi_\text{NR} [h_i, g_4] + \Xi_\text{RR} [g_4, h_j] ) \,. 
\label{Xi_NR check}
\end{split}
\end{equation}
Note that the operators $\check{\Xi}_\text{RN} [h_i, h_j]$ and $\check{\Xi}_\text{NR} [h_i, h_j]$ also interpolate between $X_\text{N}$ and $X_\text{R}$:
\begin{align}
 Q \cdot \check{\Xi}_\text{RN} [h_i, h_j] = & \ X_\text{R} [h_i] - X_\text{N} [h_j] \,, \label{check_Xi_RN interpolate}\\
  Q \cdot \check{\Xi}_\text{NR} [h_i, h_j] = & \ X_\text{N} [h_i] - X_\text{R} [h_j] \,. \label{check_Xi_NR interpolate}
\end{align}
We can show these equations using the equations~\eqref{Xi_RR interpolate}, \eqref{Xi_RN interpolate}, and \eqref{Xi_NR interpolate}. Furthermore, the operators $\check{\Xi}_\text{RN} [h_i, h_j]$ and $\check{\Xi}_\text{NR} [h_i, h_j]$ are related by
\begin{align}
\check{\Xi}_\text{RN} [h_i, h_j] = & \ {}-\check{\Xi}_\text{NR} [h_j, h_i] \,, \label{check_Xi_RN and NR antisym}
\end{align}
which follows from the equations~\eqref{Xi_RR antisym} and  \eqref{Xi_NR and Xi_RN antisym}.

In appendix~\ref{app:C}, we will explicitly construct the remaining three-string products  \eqref{V3s for other NS R combination} in the same manner. We will show that the three-string products in~\eqref{V3s for other NS R combination} satisfy the cyclicity equations~\eqref{V3 cyclicity for RRNN}, \eqref{V3 cyclicity for NRRN}, \eqref{V3 cyclicity for NNRR}, and \eqref{V3 cyclicity for RNNR}. This completes the construction of the three-string products including the Ramond sector with a cyclic $A_\infty$ structure up to quartic order.

We comment on the correlation functions which contain any one of the operators $\Xi_\text{RR}$, $\Xi_\text{RN}$, and $\Xi_\text{NR}$. For instance, the $\beta \gamma$ ghost part of the correlation function
\begin{equation}
\langle \, \Xi_\text{RNRN} \, g_1 \circ R_1(0) \, g_2 \circ N_1(0) \,
g_3 \circ R_2(0) \, g_4 \circ N_2(0) \, \rangle_D
\end{equation}
consists of insertions of two vertex operators $\delta(\gamma)$ in the $-1$ picture, two vertex operators $\Theta_{-1/2}$ in the $-1/2$ picture, and one $\Xi_{RR}$. 
In the case of the Neveu-Schwarz sector~\cite{Ohmori:2017wtx}, as we mentioned before, the correlation function including the operator $\Xi[a,b]$~\eqref{Xi_NN explicit}
\begin{equation}
    \Xi[a,b]
    = \, \int_0^1 dt \, (\beta_{-1/2}[a] - \beta_{-1/2} [b]) \, \delta (t\beta_{-1/2}[a] +(1-t)\beta_{-1/2} [b])
\end{equation}
may develop a double pole, but it was shown that it does not develop a single pole \cite{Ohmori:2017wtx}.
The operator
\begin{equation}
    \Xi_\text{RR}[a,b]
    = \, \int_0^1 dt \, (\beta_0[a] - \beta_0 [b]) \, \delta (t\beta_0[a] +(1-t)\beta_0 [b]) \,,
\end{equation}
has the same structure as the operator $\Xi[a,b]$, and the proof of appendix~A.5 of \cite{Ohmori:2017wtx} can be applied to our case.
Although the vertex operator $\Theta_{-1/2}$ in the Ramond sector changes the boundary condition of the $\beta \gamma$ path integral,\footnote{The author would like to thank Kantaro Ohmori for answering the question about this issue.}
$\,$ we conclude that correlation functions which include any one of the operators $\Xi_\text{RR}$, $\Xi_\text{RN}$, and $\Xi_\text{NR}$ may develop a double pole, but do not develop a single pole, and they are unambiguously defined as in the case of the Neveu-Schwarz sector~\cite{Ohmori:2017wtx}.

\section{Open superstring field theory including the Ramond sector with stubs \label{sec:NSRstub}}
\setcounter{equation}{0}
In this section, we construct an action for open superstring field theory with stubs including both the Neveu-Schwarz sector and the Ramond string sector. We note that the calculation is separable into the stub part and the stubless superstring part, and as a consequence of it, the action is easily constructed by combining the results of section~\ref{sec:stub} and section~\ref{sec:NSR}. This is an advantage of our method.

\subsection{Construction of cubic vertices}
Let us consider the cubic vertices. We define the two-string products with stubs by
\begin{align}
 V_2 (N_1, N_2) 
= & \ \frac{1}{3}( X_\text{N}^\star V_2^\text{stub} (N_1, N_2) + V_2^\text{stub} (X_\text{N}N_1, N_2) + V_2^\text{stub} (N_1, X_\text{N}N_2) ) \label{V2-NN with stub}  \\
= & \ \frac{1}{3} X_\text{N}^\star e^{-wL_0} ((e^{-wL_0} N_1)\ast(e^{-wL_0} N_2)) 
 + \frac{1}{3} e^{-wL_0} ((e^{-wL_0} X_\text{N}N_1)\ast(e^{-wL_0} N_2)) \notag \\
& \quad + \frac{1}{3} e^{-wL_0} ((e^{-wL_0} N_1)\ast(e^{-wL_0} X_\text{N}N_2)) \notag
\end{align}
for the Neveu-Schwarz sector and 
\begin{align}
V_2 (N_1, R_1) 
= & \ X_\text{R} V_2^\text{stub} (N_1, R_1) \label{V2-NR with stub}\\
= & \ X_\text{R} e^{-wL_0} ((e^{-wL_0} N_1)\ast(e^{-wL_0} R_1)) \,, \notag  \\
V_2 (R_1, N_1) 
= & \ X_\text{R} V_2^\text{stub} (R_1, N_1) \label{V2-RN with stub} \\
= & \ X_\text{R} e^{-wL_0} ((e^{-wL_0} R_1)\ast(e^{-wL_0} N_1)) \,, \notag\\
V_2 (R_1, R_2) 
= & \ V_2^\text{stub} (R_1, R_2) \label{V2-RR with stub} \\
= & \ e^{-wL_0} ((e^{-wL_0} R_1)\ast(e^{-wL_0} R_2))  \notag
\end{align}
for the Ramond sector. In the above definitions, we replaced the star product in the two-string products \eqref{V2 N-N}, \eqref{V2 R-R}, \eqref{V2 N-R}, and \eqref{V2 R-N} with the two-string stub product $V_2^\text{stub}$  \eqref{two-string product with stubs}. 

The cyclicity equation for the two-string product $V_2 (R_1, R_2)$ \eqref{V2-RR with stub} follows from that of the stub product $V_2^\text{stub}$ \eqref{cyclicity_of_stub_product}. Let us consider the cyclicity equation for the two-string product $V_2 (N_1, R_1)$ \eqref{V2-NR with stub}. We find
\begin{equation}
\begin{split}
\langle \, R_1, Y_\text{R} V_2 (N_1, R_2) \, \rangle 
= & \, \langle \, R_1, Y_\text{R} X_\text{R} V_2^\text{stub} (N_1, R_2) \, \rangle \\
= & \ \langle \, R_1, V_2^\text{stub} (N_1, R_2) \, \rangle \\
= & \ \langle \, V_2^\text{stub} (R_1, N_1), R_2 \, \rangle \\
= & \ \langle \, X_\text{R} V_2^\text{stub} (R_1, N_1), Y_\text{R} R_2 \, \rangle \\
= & \ \langle \, V_2 (R_1, N_1), Y_\text{R} R_2 \, \rangle \,,
\end{split}
\end{equation}
where we used the cyclicity of the two-string stub product $V_2^\text{stub}$ \eqref{cyclicity_of_stub_product}, the equation~\eqref{XYX=X} and the fact that $X_\text{R}$ is BPZ even \eqref{X_R is BPZ even}. We can show that the remaining cyclicity equations hold in the same manner. 

The $A_\infty$ relation for the two-string product $V_2 (R_1, R_2)$ \eqref{V2-RR with stub} follows from the $A_\infty$ relation for the stub product $V_2^\text{stub}$ \eqref{Leibnitz_Q_stub}. Let us consider the $A_\infty$ relation for the two-string product $V_2 (N_1, R_1)$ \eqref{V2-NR with stub}. We find
\begin{equation}
\begin{split}
Q V_2 (N_1, R_1) 
= & \, QX_\text{R} V_2^\text{stub} (N_1, R_1) \\
= & \, X_\text{R} QV_2^\text{stub} (N_1, R_1) \\
= & \, X_\text{R} (V_2^\text{stub} (QN_1, R_1) 
    + (-1)^{N_1} V_2^\text{stub} (N_1, QR_1)) \\
= & \ V_2 (QN_1, R_1) + (-1)^{N_1} V_2 (N_1, QR_1) \,,
\end{split}
\end{equation}
where we used the equations~\eqref{Leibnitz_Q_stub} and \eqref{Q commutes with X_R}. We can show that the remaining $A_\infty$ relations hold in the same manner. 

We find that these two-string products are not associative. We note that the two-string product is not associative for inputs of three Ramond string fields: 
\begin{equation}
V_2 (V_2 (R_1,R_2),R_3) \neq V_2 (R_1,V_2 (R_2,R_3)) \,.
\end{equation}
This non-associativity is due to the fact that the stub product $V_2^\text{stub}$ \eqref{two-string product with stubs} is not associative, while the two-string product $V_2(R_1, R_2)$ \eqref{V2 R-R} constructed from the star product in section~\ref{sec:NSR} is associative.

\subsection{Construction of quartic vertices}
Let us consider quartic vertices for open superstring field theory including the Ramond sector with stubs. We express quartic vertices for various Neveu-Schwarz and Ramond ordering in terms of the following three-string products:
\begin{equation}
\begin{split}
& V_3 (N_1, N_2, N_3) \,, \quad V_3 (N_1, N_2, R_1) \,, 
\quad V_3 (N_1, R_1, N_2) \,, \quad  V_3 (R_1, N_1, N_2)\,, \\
& V_3 (N_1, R_1, R_2) \,, \, \quad V_3 (R_1, N_1, R_2) \,,
\quad V_3 (R_1, R_2, N_1)\,, \quad V_3 (R_1, R_2, R_3) \,. \label{V3 with stubs}
\end{split}
\end{equation}
In this subsection, we illustrate the construction of $V_3 (N_1, R_1, N_2)$. Explicit constructions of other three-string products in~\eqref{V3 with stubs} are presented in appendix~\ref{app:C3}. 

We express the three-string product $V_3 (N_1, R_1, N_2)$ 
in terms of a Grassmann-odd operator $\Xi_\text{RNRN}^\text{stub}$:
\begin{equation}
\langle \, R_1, Y_\text{R} V_3(N_1, R_2, N_2) \, \rangle 
= \, \langle \, g_1 \circ R_1(0) \, \Xi_\text{RNRN}^\text{stub} \, g_2 \circ N_1(0) \, g_3 \circ R_2(0) g_4 \circ N_2(0) \, \rangle_D \,.
\label{RNRN with stub definition}
\end{equation}
We introduce Grassmann-even operators $(X_t)_\text{RNRN}^\text{stub}$ 
and $(X_s)_\text{RNRN}^\text{stub}$ by
\begin{align}
\langle \, R_1, Y_\text{R} V_2( V_2 (N_1, R_2), N_2) \, \rangle 
= \, & \langle \, (X_t)_\text{RNRN}^\text{stub} \, g_1 \circ R_1(0) \, g_2 \circ N_1(0) \, g_3 \circ R_2(0) \, g_4 \circ N_2(0) \, \rangle_D \label{definition of X_t RNRN with stub} \,,  \\
\langle \, R_1, Y_\text{R} V_2( N_1, V_2(R_2, N_2)) \, \rangle
= \, & \langle \, (X_s)_\text{RNRN}^\text{stub} \, g_1 \circ R_1(0) \, g_2 \circ N_1(0) \, g_3 \circ R_2(0) \, g_4 \circ N_2(0) \, \rangle_D \label{definition of X_s RNRN with stub} \,.
\end{align}
Using the definition of two-string products \eqref{V2-NR with stub} and \eqref{V2-RN with stub}, we find
\begin{align}
(X_t)_\text{RNRN}^\text{stub} 
= \ & X_\text{R}[h_4] \, e^{-wL_0[g_1]} \, e^{-2wL_0[h_2]} \, e^{-wL_0[g_2]} \, e^{-wL_0[g_3]} \, e^{-wL_0[g_4]} \,, \\
(X_s)_\text{RNRN}^\text{stub} 
= \ & X_\text{R}[h_3] \, e^{-wL_0[g_1]} \, e^{-wL_0[g_2]} \, e^{-2wL_0[h_3]} \, e^{-wL_0[g_3]} \, e^{-wL_0[g_4]}\,.
\end{align}
We find that the operator $\mathcal{L}_t^\text{stub}$ can be expressed as the product of the stubless superstring part $(X_t)_\text{RNRN}$ \eqref{definition of X_t RNRN} and the stub part $\mathcal{L}_t^\text{stub}$ \eqref{definition of L_t}:
\begin{equation}
(X_t)_\text{RNRN}^\text{stub} = \, (X_t)_\text{RNRN} \, 
\mathcal{L}_t^\text{stub} \,. \label{t channel for the stub factorization}
\end{equation}
Similarly, we find
\begin{equation}
(X_t)_\text{RNRN}^\text{stub} = \, (X_s)_\text{RNRN} \, 
\mathcal{L}_s^\text{stub} \,. \label{s channel for the stub factorization}
\end{equation}
As before, the $A_\infty$ relation for the three-string product $V_3 (N_1, R_1, N_2)$,
\begin{equation}
\begin{split}
0 = & \ QV_3(N_1,R_1,N_2) +V_3(QN_1,R_1,N_2)  \\
& \quad + (-1)^{N_1}V_3(N_1,QR_1,N_2) +(-1)^{N_1+R_1}V_3(N_1,R_1,QN_2) \\
& \quad - V_2(V_2(N_1, R_1), N_2) + V_2(N_1, V_2(R_1, N_2)) \,,
\end{split}
\end{equation}
is translated into the following equation:
\begin{equation}
Q \cdot \Xi_\text{RNRN}^\text{stub} = (X_t)_\text{RNRN}^\text{stub} -(X_s)_\text{RNRN}^\text{stub} \,.    
\label{3rd A_infty for Xi_RNRN with stubs}
\end{equation}
The point is that we can decompose the right-hand side of \eqref{3rd A_infty for Xi_RNRN with stubs} as follows:
\begin{equation}
\begin{split}
& \, (X_t)_\text{RNRN}^\text{stub} - (X_s)_\text{RNRN}^\text{stub} \\
= & \ \Bigl( \, (X_t)_\text{RNRN} \, e^{-2wL_0[h_2]} - (X_s)_\text{RNRN} \,  e^{-2wL_0[h_3]} \, \Bigr) \, e^{-wL_0[g_1]} \, e^{-wL_0[g_2]} \, e^{-wL_0[g_3]} \, e^{-wL_0[g_4]} \\
= & \ \Bigl( \, (X_t)_\text{RNRN} (e^{-2wL_0[h_2]}-1) 
 + (X_t)_\text{RNRN} - (X_s)_\text{RNRN} + (X_s)_\text{RNRN} 
(1-e^{-2wL_0[h_3]}) \, \Bigr) \\
& \quad \times e^{-wL_0[g_1]} \, e^{-wL_0[g_2]} \, e^{-wL_0[g_3]} \, e^{-wL_0[g_4]} \,. \\
\end{split}
\end{equation}
Using the stub creation operator $\mathcal{B}[f]$ \eqref{stub_creation_relation}, the stub annihilation operator ${}-\mathcal{B}[f]$ \eqref{stub_annihilation_relation}, and the operator  $\Xi_\text{RNRN}$ \eqref{Xi RNRN answer}, we find
\begin{equation}
\begin{split}
 \ (X_t)_\text{RNRN}^\text{stub} - (X_s)_\text{RNRN}^\text{stub} 
= & \ Q \cdot \Bigl[ \, \Bigl( (X_t)_\text{RNRN} \, (-\mathcal{B}[h_2]) + \Xi_\text{RNRN}
+ (X_s)_\text{RNRN} \, \mathcal{B}[h_3] \Bigr) \\
& \qquad \qquad  \times e^{-wL_0[g_1]} \, e^{-wL_0[g_2]} \, e^{-wL_0[g_3]} \, e^{-wL_0[g_4]} 
\, \Bigr] \,.\\
\end{split}
\end{equation}
Finally, we obtain the following result as a solution to the equation~\eqref{3rd A_infty for Xi_RNRN with stubs}:
\begin{equation}
\begin{split}
\Xi_\text{RNRN}^\text{stub} = & \ \Bigl( (X_t)_\text{RNRN} \, (-\mathcal{B}[h_2]) + \Xi_\text{RNRN}
+ (X_s)_\text{RNRN} \, \mathcal{B}[h_3] \Bigr) \\
& \qquad \qquad  \times e^{-wL_0[g_1]} \, e^{-wL_0[g_2]} \, e^{-wL_0[g_3]} \, e^{-wL_0[g_4]} \,. \label{Xi_RNRN^stub answer}
\end{split}
\end{equation}
A pictorial representation of this interpolation is given in Figure~\ref{fig:interpolation_superstring_with_stub}. In appendix~\ref{app:C3}, we will show that the remaining three-string products in~\eqref{V3 with stubs} can be constructed in the same manner.

\begin{figure}[htb]
    \centering
    \includegraphics[width=7cm]{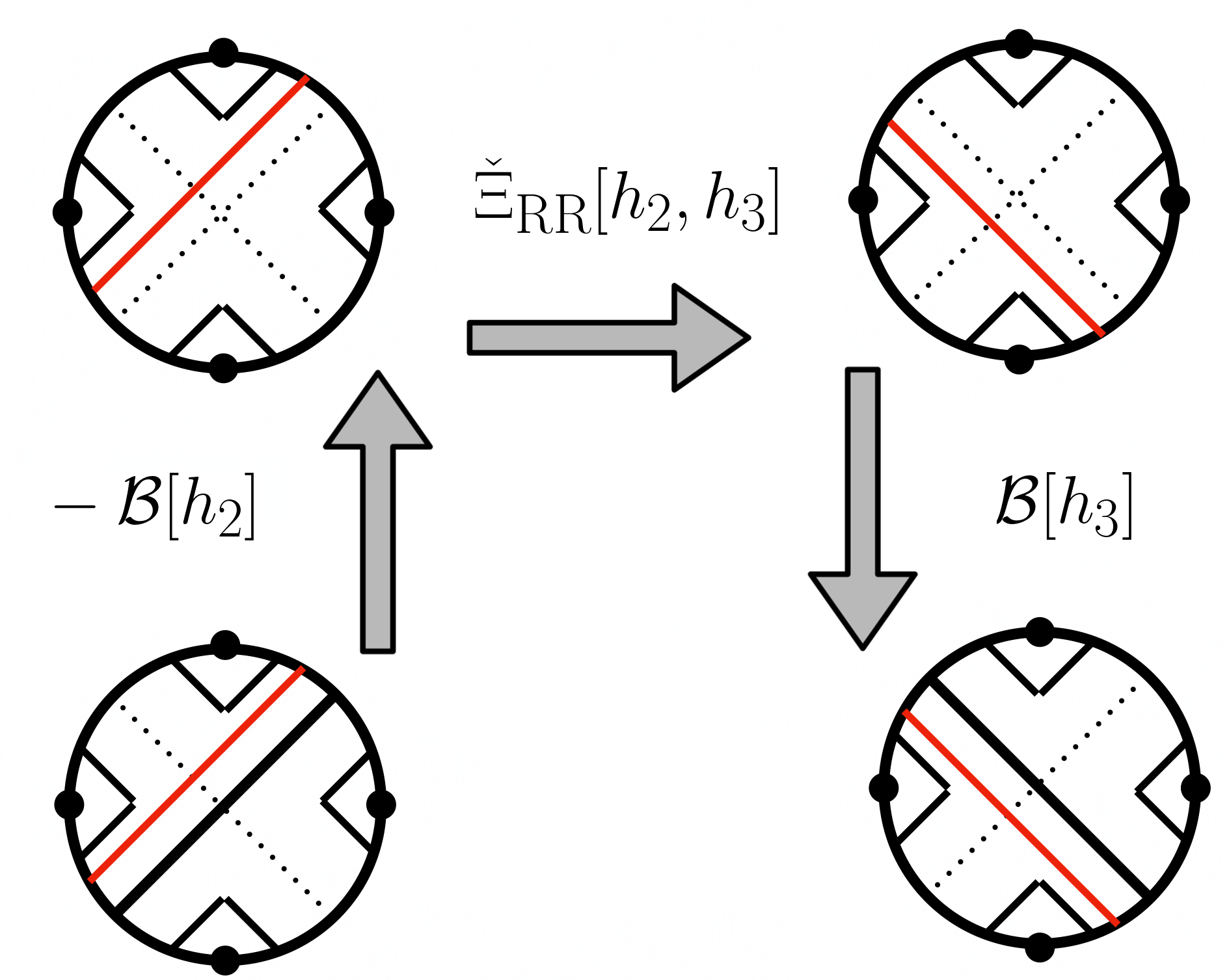}
    \caption{A pictorial representation of the interpolation from the $t$-channel to the $s$-channel. The insertions of the line integrals $X_\text{R} [h_2]$ and $X_\text{R} [h_3]$ are represented by red lines. 
    In the first step, $-\mathcal{B}[h_2]$ of the first term annihilates the stub on the $t$-channel. In the second step,  $\Xi_\text{RNRN} = \check{\Xi}_\text{RR}[h_2, h_3]$ of the second term 
    changes the configuration of the picture-changing operator. In the third step, $\mathcal{B}[h_3]$ of the third term creates the stub on the $s$-channel.}
    \label{fig:interpolation_superstring_with_stub}
\end{figure}

\subsection{Cyclicity equations}
Let us show that the three-string product $V_3(N_1,R_1,N_2)$ constructed from the operator $\Xi_\text{RNRN}^\text{stub}$ satisfies the cyclicity equation:
\begin{equation}
    \langle \, R_1, Y_\text{R} V_3(N_1, R_2, N_2) \, \rangle 
    = \, (-1)^{R_1} \, \langle \, V_3(R_1, N_1, R_2), N_2 \, \rangle \,.
\end{equation}
This equation is translated into the following form:
\begin{equation}
\omega \circ \Xi_\text{RNRN}^\text{stub} = {}-\Xi_\text{NRNR}^\text{stub} \,,
\label{Xi cyclicity for RNRN with stubs}
\end{equation}
where $\Xi_\text{NRNR}^\text{stub}$ is given in~\eqref{Xi_NRNR^stub answer}. 
This equation can be confirmed in the following way: 
\begin{equation}
\begin{split}
\omega \circ \Xi_\text{RNRN}^\text{stub} 
= & \ \omega \circ \Bigl[ \Bigl( \, 
(X_t)_\text{RNRN} \, (-\mathcal{B}[h_2]) 
+ \Xi_\text{RNRN}
+ (X_s)_\text{RNRN} \, \mathcal{B}[h_3] \, \Bigr)  \\
&  \quad \qquad \times e^{-wL_0[g_1]} \, e^{-wL_0[g_2]} \, e^{-wL_0[g_3]} \, e^{-wL_0[g_4]} \Bigr] \\
= & \ \Bigl( \, 
(X_s)_\text{NRNR} \, (-\mathcal{B}[h_3]) 
- \Xi_\text{NRNR}
+ (X_t)_\text{NRNR} \, \mathcal{B}[h_4] \, \Bigr) \\
& \quad \times e^{-wL_0[g_2]} \, e^{-wL_0[g_3]} \, e^{-wL_0[g_4]} \, e^{-wL_0[g_1]} \,,
\end{split}
\end{equation}
where we used the equations~\eqref{Xi cyclicity for RNRN and NRNR} and \eqref{X_t and X_s cyclicity for RNRN}. Using the equation~\eqref{B is BPZ even}, we find
\begin{equation}
\begin{split}
\omega \circ \Xi_\text{RNRN}^\text{stub} 
= & \ {}- \Bigl( \, (X_s)_\text{NRNR} \, \mathcal{B}[h_3] 
+\Xi_\text{NRNR}
- (X_t)_\text{NRNR} \, \mathcal{B}[h_2] \, \Bigr) \, \\
& \quad \qquad \times  e^{-wL_0[g_1]} \, e^{-wL_0[g_2]} \, e^{-wL_0[g_3]} \, e^{-wL_0[g_4]} \,.
\end{split}
\end{equation}
According to the result in appendix~\ref{app:C}, this equation is exactly the cyclicity equation~\eqref{Xi cyclicity for RNRN with stubs}.\\
For the other three-string products in~\eqref{V3 with stubs}, we can show that the cyclicity equations hold in the same manner. See appendix~\ref{app:C}. This completes the construction of the three-string products with stubs including the Ramond sector with the cyclic $A_\infty$ structure up to quartic order.

\section{Conclusion and discussions}
\setcounter{equation}{0}
First, we have constructed an action for open superstring field theory including the Ramond sector up to quartic order.
Our construction is based on the supermoduli space of super-Riemann surfaces, extending the approach presented in the construction of an action for the Neveu-Schwarz sector of open superstring field theory by Ohmori and Okawa in~\cite{Ohmori:2017wtx}. 
We constructed the cubic vertex~\eqref{the cubic vertex for the Ramond sector} for one Ramond string field and two Neveu-Schwarz string fields, and the quartic vertices for two Ramond string fields and two Neveu-Schwarz string fields.
The complete list of the quartic vertices are in appendix~\ref{app:C1}. 
The associated two-string and three-string products satisfy the $A_\infty$ relation~\eqref{3rd A infty} and the cyclicity equation~\eqref{V3 cyclicity}.

The cubic vertex is constructed using the star product, and the quartic vertexes are constructed from the following operators:
\begin{align}
    \Xi_\text{RR} [a, b] = & \, \int_0^1 dt \int d\tilde{t} \int d\zeta d\tilde{\zeta} \ e^{-\{ Q', t\zeta \beta_0 [a] \} } \ e^{-\{ Q', (1-t) \zeta \beta_0 [b]\} } \,, 
    \label{Xi_RR in summary}\\
    \Xi_\text{NR} [a, b] = & \, \int_0^1 dt \int d\tilde{t} \int d\zeta d\tilde{\zeta} \ e^{-\{ Q', t\zeta \beta_{-1/2} [a] \} } \ e^{-\{ Q', (1-t) \zeta \beta_0 [b]\} } \,, 
    \label{Xi_NR in summary}\\
    \Xi_\text{RN} [a, b] = & \, \int_0^1 dt \int d\tilde{t} \int d\zeta d\tilde{\zeta} \ e^{-\{ Q', t\zeta \beta_0 [a] \} } \ e^{-\{ Q', (1-t) \zeta \beta_{-1/2} [b]\} } 
    \label{Xi_RN in summary} \,.
\end{align}
These operators consist of an integral over one even modulus $t$ and one odd modulus $\zeta$.
These operators interpolate the different configurations of the picture-changing operators $X_\text{R}$ and $X_\text{N}$.
We can interpret the operators $\Xi_\text{RNRN}$, $\Xi_\text{NRNR}$, $\Xi_\text{RRNN}$, $\Xi_\text{RNNR}$, $\Xi_\text{NNRR}$, and $\Xi_\text{NRRN}$ as the integtral over the region of the supermoduli space of disks with two Neveu-Schwarz punctures and two Ramond punctures which is not covered by Feynman diagrams with two cubic vertices and one propagator.
The $A_\infty$ relation follows from this construction.
We have constructed an action for open superstring field theory including the Ramond sector up to quartic order, and this is one of our new results.

Second, we have also constructed an action for open superstring field theory including the Ramond sector with stubs up to quartic order. 
The construction is done by combining the stub part in section~\ref{sec:stub} and the superstring part in section~\ref{sec:NSR}.
For the stub part, we considered the cubic vertex \eqref{cubic vertex with stubs}:
\begin{equation}
\langle \, e^{-wL_0}A_1,  (e^{-wL_0} A_2) \ast (e^{-wL_0} A_3) \, \rangle \,. \label{cubic vertex with stubs in conclusion}
\end{equation}
The quartic vertex is constructed from the stub creation operator~\eqref{stub_creation_operator}:
\begin{equation}
\mathcal{B}[f] = \ \int_0^1 dt \int d\tilde{t} \, e^{\{Q', t(-2w)b_0[f] \} } \,,
\end{equation}
and the stub annihilation operator $-\mathcal{B}[f]$.
These operators consist of an integration over one even modulus $t$. 
As their names indicate, the stub creation operator interpolates between the unit operator $1$ and the stub operator $e^{-2wL_0[f]}$ \eqref{stub_creation_relation}:
\begin{equation}
 \, Q \cdot \mathcal{B}[f] 
= \, 1-e^{-2wL_0[f]} \,,
\end{equation}
and the stub annihilation operator interpolates between the unit operator and the stub operator in the reverse direction \eqref{stub_annihilation_relation}:
\begin{equation}
 \, Q \cdot (-\mathcal{B}[f]) 
= \, e^{-2wL_0[f]} -1 \,.
\end{equation}
We use the stub creation operator and the stub annihilation operator in the construction of an action for open superstring field theory with stubs.

The cubic vertices for the Ramond sector is constructed from the cubic vertex \eqref{cubic vertex with stubs in conclusion}, and the cubic vertex for the Neveu-Schwarz sector is constructed from the cubic vertex \eqref{cubic vertex with stubs in conclusion} with insertions of the picture-changing operator $X_\text{N}$.
The quartic vertex for open superstring field theory including the Ramond sector with stubs is constructed by an insetion of operator \eqref{Xi_RNRN^stub answer}:
\begin{equation}
\begin{split}
\Xi_\text{RNRN}^\text{stub} = & \ \Bigl( \, (X_t)_\text{RNRN} \, (-\mathcal{B}[h_2]) + \Xi_\text{RNRN}
+ (X_s)_\text{RNRN} \, \mathcal{B}[h_3] \, \Bigr) \\
& \qquad \qquad  \times e^{-wL_0[g_1]} \, e^{-wL_0[g_2]} \, e^{-wL_0[g_3]} \, e^{-wL_0[g_4]} \,. 
\end{split}
\end{equation}
This operator changes the configulation of the picture-changing operator by creating and annihilating stub operators. 
The interpolation by this operator is pictorially represented in Figure~\ref{fig:interpolation_superstring_with_stub}. 
The complete list of the quartic vertices for open superstring with stubs is in appendix~\ref{app:C3}. 
While the equation of motion for open superstring field theory with stubs was constructed in~\cite{Erler:2015lya}, an action with stubs is not yet constructed even at quartic order.
This is because solving the entire recursive system of multi-string products consistent with the cyclicity equations is a challenging problem. 
We have constructed an action for open superstring field theory including the Ramond sector with stubs up to quartic order, and this is one of our new results.

Extending our construction to higher orders has no obstacles in principle, but it would be difficult in practice.
The construction would be combinatorially complicated, and there would be a lot of ambiguities.
For the construction of the quintic vertex $\langle \, N_1, V_4 (R_1, N_2, N_3, R_2)\, \rangle$, for instance, we will need to express $\langle \, N_1, V_3 (R_1, V_2(N_2, N_3), R_2) \, \rangle$ and other terms in terms of operators of the form $\Xi_\text{RN}[f_1, f_2] \, X_\text{N}$.
We then need to construct an operator such that its BRST transformation gives the resulting operators of the form $\Xi_\text{RN}[f_1, f_2] \, X_\text{N}$ and consistent with the cyclicity equations.
It is not obvious a priori that whether we can construct the quintic vertex only by the operators used in the construction of the quartic vertices. 
Extending the construction of vertices to all orders is an open problem.

It may be useful to consider the relationship between the recursive construction in~\cite{Erler:2016ybs} and our approach.
In the recursive construction, a set of multi-string products is constructed to all orders using the large Hilbert space.
The key relation is
\begin{equation}
    X_0 = \, \{ \, Q \,, \xi_0 \, \} \,, \label{X=[Q,xi]}
\end{equation}
where $X_0$ is the zero mode of the local picture-changing operator $X(z)$ 
\begin{equation}
    X_0 = \oint \, \frac{dz}{2 \pi i} \, \frac{1}{z} \, X(z) \,,
\end{equation}
and $\xi_0$ is the zero mode of the bosonized superconformal ghost $\xi(z)$.
While $\xi_0$ is not in the small Hilbert space, the resulting multi-string products were constructed to be in the small Hilbert space. 
In the recursive construction, the equation~\eqref{X=[Q,xi]} was used to pull out $Q$ from both sides of the $A_\infty$ relation~\eqref{3rd A infty}, whereas we used the equations~\eqref{Xi_RR in summary}, \eqref{Xi_NR in summary}, and \eqref{Xi_RN in summary}. 
An operator $\Theta(\beta(z))$, which corresponds to $\xi(z)$, can be written in the following form:
\begin{equation}
    \Theta (\beta(z)) = \, {}-\int d\tau \frac{1}{\tau} e^{-\beta(z) \tau} \,,
\end{equation}
where $\tau$ is a Grassmann-even variable of ghost number $1$. The singular factor $1/\tau$ corresponds to the fact that $\xi(z)$ is not in the small Hilbert space.
When we consider the difference between $\Theta (\beta_{-1/2} [f_1])$ and $\Theta (\beta_{-1/2} [f_2])$, the singular factor cancels, and the resulting operator is in the small Hilbert space.
In~\cite{Ohmori:2017wtx}, it was pointed out that the following equation holds:
\begin{equation}
    \Xi [f_1, f_2] = \, \Theta(\beta_{-1/2}[f_1]) - \Theta(\beta_{-1/2}[f_2]) \,, \label{Xi_NN and Theta}
\end{equation}
where $\Theta(\beta_{-1/2} [f])$ is defined by
\begin{equation}
    \Theta (\beta_{-1/2}[f]) 
    = \, {}-\int d\tau \frac{1}{\tau} e^{-\beta_{1/2} [f] \tau} \,.
\end{equation}
In the same manner, we can show that the following equations hold:
\begin{align}
        \Xi_\text{RR}[f_1, f_2] 
        = & \ \Theta(\beta_0 [f_1]) - \Theta(\beta_0 [f_2]) \,, \label{Xi_RR and Theta} \\
        \Xi_\text{RN}[f_1, f_2] 
        = & \ \Theta(\beta_0 [f_1]) - \Theta(\beta_{-1/2} [f_2]) \,, \label{Xi_RN and Theta} \\
        \Xi_\text{NR}[f_1, f_2] 
        = & \ \Theta(\beta_{-1/2} [f_1]) - \Theta(\beta_0 [f_2]) \,, \label{Xi_NR and Theta}
\end{align}
where $\Theta(\beta_0 [f])$ is defined by
\begin{equation}
    \Theta (\beta_0 [f]) 
    = \, {}-\int d\tau \frac{1}{\tau} e^{-\beta_0 [f] \tau} \,.
\end{equation}
The multi-string products in the recursive construction~\cite{Erler:2013xta} consist of the star product, the line integral of the picture-changing operator, and the operator $\xi_0$ in the large Hilbert space.
Obviously, we can obtain an action in the large Hilbert space by the following replacement
\begin{equation}
    \Xi [f_1,f_2] \ \to \ \xi_0 [f_1] - \xi_0 [f_2] \,,
\end{equation}
where $\xi_0[f]$ is defined by
\begin{equation}
    \xi_0 [f] = \, \oint \frac{dz}{2 \pi i} 
    \, (f^{-1}(z))^{-1} \,
    \left( \frac{df^{-1}(z)}{dz} \right)^{-1} \xi (z) \,.
\end{equation}
On the other hand, if we write the multi-string products in the recursive construction in terms of the picture-changing operator and the difference $\xi_0 [f_1] - \xi_0 [f_2]$ in a canonical way, it might be a hint to the extension of our approach to all orders.
Furthermore, we can consider the same problem including the Ramond sector to obtain some insight into an all-order construction.

\subsection*{Acknowledgements}
The author would like to thank his supervisor Yuji Okawa for useful discussions and careful reading of the manuscript.
The results of this paper were presented during the workshop ``Discussion Meeting on String Field Theory and String Phenomenology'' at Harish-Chandra Research Institute.
The author would like to thank the members of the institute for their hospitality during the visit.

\appendix
\section{Integral over Grassmann-even variables \label{app:A}}
\setcounter{equation}{0}
In this appendix, we review the algebraic treatment of an integral over the Grassmann-even variables. 
After describing the relation between the integral of Grassmann-even variable and the integral of the Grassmann-odd variable,\footnote{The contents of appendix~\ref{app:A1} is based on the paper by Ohmori and Okawa~\cite{Ohmori:2017wtx}.}
$\,$ we give an explicit proof of the equations~\eqref{XYX=X} and \eqref{YXY=Y} regarding on operators $X_\text{R}$ and $Y_\text{R}$ .

\subsection{The definition of the Gaussian integral of Grassmann-even variables \label{app:A1}}
It is emphasized in~\cite{Witten:2012bh} that the $\beta \gamma$ ghosts are Grassmann-even but they do not obey any reality condition. 
An integration of the Grassmann-even variables should be understood as an algebraic operation. 
We require that the integration of a Grassmann-even variable $x$ has the following properties:
\begin{align}
    \int dx \partial_{x} f(x) = & \, 0\,, 
    \label{integration over total derivative of Grassmann-even variable equals 0} \\
    \int dx f(x+a) = & \, \int dx f(x) \,, 
    \label{translational invariance of the integration of Grassmann-even variable}
\end{align}
where $a$ is an arbitrary Grassmann-even constant. Let us consider a Gaussian integral for Grassmann-even variables $x_1$, $\dots$, $x_{2n}$:
\begin{equation}
    \int d^{2n} x \exp{\left( \, -\frac{1}{2} \sum_{i,j} x_i N_{ij} x_j \, \right)} \,.
    \label{Gaussian integral for the Grassmann-even variables}
\end{equation}
We consider the case where the integration variables consist of $\beta_i$ of ghost number $-1$ and $\gamma_i$ of ghost number $1$. We assume that the symmetric matrix $N$ takes the following form:
\begin{equation}
    \tilde{N} = 
    \left(
    \begin{array}{cc}
      0 &  M \\
      M^t &  0
    \end{array}
  \right) \,,
\end{equation}
where $M^t$ is the transpose of $M$. Then the Gaussian integral is given by
\begin{equation}
    \int d^{2n} x \exp{\left( \, -\frac{1}{2} \sum_{i,j} x_i N_{ij} x_j \, \right)}
    = \, \int d^n \beta d^n \gamma \exp{\left( \, - \sum_{i,j} \beta_i M_{ij} \gamma_j \, \right)} \,.
\end{equation}
up to a possible sign by reordering from $d^{2n}x$ to $d^n \beta d^n \gamma$, where we defined $d^n \beta d^n \gamma$ by $d \beta_1 d\gamma_1 $ $d \beta_2 d\gamma_2 $ $\cdots $ $ d \beta_n d\gamma_n$. We define the Gaussian integral as follows:
\begin{equation}
    \int d^n \beta d^n \gamma \exp{\left( \, - \sum_{i,j} \beta_i M_{ij} \gamma_j \, \right)}
    = \, \frac{1}{\det M} \,. 
    \label{normalization of the Gaussian integral for the Grasmann even variables}
\end{equation}
We note that overall sign of the determinant depends on the order of the variables. If we order the variables, for instance, as
\begin{equation}
{}-
\begin{pmatrix}
\beta_{1} &\beta_{2} 
\end{pmatrix}
\begin{pmatrix}
M_{11} &M_{12} \\
M_{21} &M_{22}
\end{pmatrix}
\begin{pmatrix}
\gamma_{1} \\
\gamma_{2} 
\end{pmatrix}
\,,
\end{equation}
the determinant is given by
\begin{equation}
\begin{vmatrix}
M_{11} &M_{12} \\
M_{21} &M_{22}
\end{vmatrix}
= \, M_{11}M_{22}-M_{12}M_{21} \,,
\end{equation}
but if we order the variables as
\begin{equation}
{}-
\begin{pmatrix}
\beta_{2} &\beta_{1} 
\end{pmatrix}
\begin{pmatrix}
M_{21} &M_{22} \\
M_{11} &M_{12} 
\end{pmatrix}
\begin{pmatrix}
\gamma_{1} \\
\gamma_{2} 
\end{pmatrix}
\,,
\end{equation}
the determinant is given by
\begin{equation}
\begin{vmatrix}
M_{21} &M_{22} \\
M_{11} &M_{12}
\end{vmatrix}
= \, {}-(M_{11}M_{22}-M_{12}M_{21}) \,.
\end{equation}
We follow the prescription in~\cite{Ohmori:2017wtx}, that is, to correlate this ordering of the quadratic form with the measure $d\beta_1 d\gamma_1 d\beta_2 d\gamma_2 \cdots d\beta_n d\gamma_n $ and to treat $d\beta_i$ and $d\gamma_i$ as Grassmann-odd objects.

We can relate the Gaussian integral for the Grassmann-even variables \eqref{Gaussian integral for the Grassmann-even variables} to the Gaussian integral for Grassmann-odd variables $x_1^\ast$, $\cdots$, $x_{2n}^\ast$:
\begin{equation}
    \int d^{2n} x^\ast \exp{\left( \, -\frac{1}{2} \sum_{i,j} x_i^\ast \tilde{N}_{ij} x_j^\ast \, \right)} \,,
\end{equation}
whrer $\tilde{N}$ is an anti-symmetric matrix. The integration of a Grassmann-odd variable $x^\ast$ has the following properties:
\begin{align}
    \int dx^\ast \partial_{x^\ast} f(x^\ast) = & \ 0\,, 
    \label{integration over total derivative of Grassmann-odd variable equals 0}\\
    \int dx^\ast f(x^\ast + a^\ast) = & \, \int dx^\ast f(x^\ast) \,,
    \label{translational invariance of the integration of Grassmann-odd variable}
\end{align}
where $a^\ast$ is an arbitrary Grassmann-odd constant. We consider the case where the integration variables consist of $\beta_i^\ast$ of ghost number $-1$ and $\gamma_i^\ast$ of ghost number $1$, and the anti-symmetric matrix takes the following form:
\begin{equation}
    \tilde{N} = 
    \left(
    \begin{array}{cc}
      0 &  M \\
      -M^t &  0
    \end{array}
  \right) \,.
\end{equation}
The Gaussian integral is given by
\begin{equation}
    \int d^{2n} x^\ast \exp{\left( \, -\frac{1}{2} \sum_{i,j} x_i^\ast \tilde{N}_{ij} x_j^\ast \, \right)}
    = \, \int d^n \beta^\ast d^n \gamma^\ast \exp{\left( \, - \sum_{i,j} \beta_i^\ast M_{ij} \gamma_j^\ast \, \right)}
\end{equation}
up to a possible sign by reordering from $d^{2n}x$ to $d^n \beta^\ast d^n \gamma^\ast$, where we defined $d^n \beta^\ast d^n \gamma^\ast$ by $d \beta^\ast_1 d\gamma^\ast_1 $ $d \beta^\ast_2 d\gamma^\ast_2 $ $\cdots $ $ d \beta^\ast_n d\gamma^\ast_n$. We normalize the Gaussian integral as follows:
\begin{equation}
    \int d^n \beta^\ast d^n \gamma^\ast \exp{\left( \, - \sum_{i,j} \beta_i^\ast M_{ij} \gamma_j^\ast \, \right)}
    = \, \det M \,.
    \label{normalization of the Gaussian integral for the Grasmann odd variables}
\end{equation}
Then we can calculate integrals of Grassmann-even variables in terms of integrals of Grassmann-odd variables. We find
\begin{equation}
    \int d^n \beta d^n \gamma \exp{\left( \, - \sum_{i,j} \beta_i M_{ij} \gamma_j \, \right)} 
    = \, \left[ \, \int d^n \beta^\ast d^n \gamma^\ast \exp{\left( \, - \sum_{i,j} \beta_i^\ast M_{ij} \gamma_j^\ast \, \right)} \, \right]^{-1} \,.
    \label{relation between integral over Grassmann-even and Grassmann-odd variables}
\end{equation}
For the detail of the calculation of correlation functions, see~\cite{Ohmori:2017wtx}. 

\subsection{Proof of the formulas regarding on \texorpdfstring{$X_\text{R}$}{Lg} and \texorpdfstring{$Y_\text{R}$}{Lg}
}
The delta functions for $\beta_0$ and $\gamma_0$ are defined by
\begin{align}
    \delta (\beta_0) = & \, \int d\tau \, e^{-\tau \beta_0} \,, \quad 
    \delta' (\beta_0) =  {}-\int d\tau \tau \, e^{-\tau \beta_0} \,, \\
    \delta (\gamma_0) = & \, \int d\sigma \, e^{\sigma \gamma_0} \,, \quad 
    \delta' (\gamma_0) =  \, \int d\sigma \sigma \, e^{\sigma \gamma_0} \,,
\end{align}
where $\tau$ and $\sigma$ are Grassmann-even variables of ghost number $1$ and $-1$, respectively. We find
\begin{equation}
    \beta_0 \, \delta(\beta_0) 
    = \, \int d\tau \, \beta_0 \, e^{-\tau \beta_0}
    = \, {}-\int d\tau \, \partial_\tau e^{-\tau \beta_0} 
    = \, 0 \,.
\end{equation}
Similarly, we can show that
\begin{equation}
    \gamma_0 \, \delta (\gamma_0) = \, 0 \,.
\end{equation}
On the other hand, we find
\begin{equation}
\begin{split}
    \gamma_0 \delta'(\gamma_0)
    = & \ \gamma_0 \int d\sigma \sigma \, e^{\sigma \gamma_0} \,
    = \, \int d\sigma \sigma \partial_\sigma \, e^{\sigma \gamma_0} \\
    = & \, \int d\sigma \, \Bigl( 
    \partial_\sigma (\sigma e^{\sigma \gamma_0}) 
    - e^{\sigma \gamma_0} \Bigr) \,
    = \, -\delta(\gamma_0) \,.
\end{split}
\end{equation}
By a similar calculation, we can show that
\begin{equation}
    \beta_0 \, \delta'(\beta_0) 
    = \, -\delta(\beta_0) \,.
\end{equation}
We can express $\delta'(\beta_0)$ and $\delta'(\beta_0)$ as
\begin{align}
    \delta'(\beta_0) = \, [ \, \gamma_0 \,, \delta(\beta_0) \, ] \,, \quad
    \delta'(\gamma_0) = \, [ \, \delta(\gamma_0) \,, \beta_0 \, ] \,.
\end{align}

We show that the following formulas hold:
\begin{equation}
    \, \delta (\beta_0) \, \delta (\gamma_0) \, \delta (\beta_0)
    = \, \delta (\beta_0) \,, \quad 
    \delta (\gamma_0) \, \delta (\beta_0) \, \delta (\gamma_0)
    = \, \delta (\gamma_0) \,.
    \label{formulas on the product of delta functions}
\end{equation}
Let us prove the first equation of \eqref{formulas on the product of delta functions}. We find
\begin{equation}
\begin{split}
    \delta (\beta_0) \, \delta (\gamma_0) \, \delta (\beta_0)
    = & \, \int d\tau_1 e^{-\beta_0 \tau_1} 
    \int d\tau_1 e^{\sigma \gamma_0}
    \int d\tau_2 e^{-\beta_0 \tau_2} \\
    = & \, \int d\tau_1 d\sigma d\tau_2 
    e^{-\beta_0 \tau_1} e^{\sigma \gamma_0} e^{\beta_0 \tau_1} 
    e^{-\beta_0 (\tau_1 + \tau_2)} \\
    = & \, \int d\tau_1 d\sigma d\tau_2 
    e^{\sigma (\gamma_0+\tau_1)} e^{-\beta_0 (\tau_1 + \tau_2)} \,,
\end{split}
\end{equation}
where we used
\begin{equation}
    e^A e^B e^{-A}= \, e^{B} e^{[ \, A \,, B \, ]} \,,
\end{equation}
which holds when the commutator $[ \, A \,, B \, ]$ equals constant. Finally, we find
\begin{equation}
\begin{split}
    \delta (\beta_0) \, \delta (\gamma_0) \, \delta (\beta_0)
    = & \, \int d\tau_1 d\sigma d\tau_2 
    e^{\sigma \tau_1} e^{-\beta_0 (\tau_1 + \tau_2)} \, 
    = \, \int d\tau_1 d\sigma d\tau_2 
    e^{(\sigma -\beta_0)\tau_1 -\beta_0 \tau_2} \, \\
    = & \, \int d\tau_1 d\sigma d\tau_2 
    e^{\sigma \tau_1} e^{-\beta_0 \tau_2} \, 
    = \, \int d\tau_2 e^{-\beta_0 \tau_2} \,
    = \ \delta(\beta_0) \,,
\end{split}
\end{equation}
where we used the translational invariance of the integration \eqref{translational invariance of the integration of Grassmann-even variable}. We can show that the second equation of \eqref{formulas on the product of delta functions} holds in the same manner.

Let us consider the equation~\eqref{YXY=Y}. We find
\begin{equation}
\begin{split}
    \, Y_\text{R} X_\text{R} Y_\text{R}
    = & \ c_0 \delta'(\gamma_0) G_0 \delta(\beta_0) c_0 \delta'(\gamma_0)
    + c_0 \delta'(\gamma_0) b_0 \delta'(\beta_0) c_0 \delta'(\gamma_0) \\
    = & \, {}-2c_0 \delta'(\gamma_0) b_0 \gamma_0 \delta(\beta_0) c_0 \delta'(\gamma_0)
    - c_0 \delta'(\gamma_0) \delta'(\beta_0) \delta'(\gamma_0) \,, \\
\end{split}
\end{equation}
where we replaced 
\begin{equation}
    G_0 = {}-\sum_{n \in \mathbb{Z}} 2b_n \gamma_{-n}
\end{equation}
by ${}-2b_0\gamma_0$ since $G_0$ is sandwiched by $c_0$. We find
\begin{equation}
\begin{split}
     \, Y_\text{R} X_\text{R} Y_\text{R}
    = & \ 2c_0 \delta'(\gamma_0) \gamma_0 \delta(\beta_0) \delta'(\gamma_0)
    - c_0 \delta'(\gamma_0) \delta'(\beta_0) \delta'(\gamma_0) \\
    = & \, -2c_0 \delta(\gamma_0) \delta(\beta_0) \delta'(\gamma_0)
    - c_0 \delta'(\gamma_0) \delta'(\beta_0) \delta'(\gamma_0) \,.
    \label{YXY intermediate}
\end{split}
\end{equation}
The first term on the right-hand side of \eqref{YXY intermediate} can be written as
\begin{equation}
\begin{split}
    -2c_0 \delta(\gamma_0) \delta(\beta_0) \delta'(\gamma_0)
    = & \, -2c_0 \delta(\gamma_0) \delta(\beta_0) [ \, \delta(\gamma_0) \,, \beta_0 \, ] \\
    = & \, -2c_0 \delta(\gamma_0) \delta(\beta_0) \delta(\gamma_0) \beta_0 \\
    = & \, -2c_0 \delta(\gamma_0) \beta_0 \,,
\end{split}
\end{equation}
and the second term on the right-hand side of \eqref{YXY intermediate} can be written as
\begin{equation}
\begin{split}
    - c_0 \delta'(\gamma_0) \delta'(\beta_0) \delta'(\gamma_0) 
    = & \, - c_0 \delta'(\gamma_0) [ \, \gamma_0 \,, \delta (\beta_0) \, ] \delta'(\gamma_0) \\
    = & \ c_0 \delta(\gamma_0) \delta (\beta_0) \delta'(\gamma_0)
    -c_0 \delta'(\gamma_0) \delta (\beta_0) \delta(\gamma_0) \\
    = & \ c_0 \delta(\gamma_0) \delta (\beta_0) [ \, \delta(\gamma_0) \,, \beta_0 \, ]
    -c_0 [ \, \delta(\gamma_0) \,, \beta_0 \, ] \delta (\beta_0) \delta(\gamma_0) \\
    = & \ c_0 \delta(\gamma_0) \delta (\beta_0) \delta(\gamma_0) \beta_0 
    + c_0 \beta_0 \delta(\gamma_0) \delta (\beta_0) \delta(\gamma_0)
     \\
    = & \ c_0 \delta(\gamma_0) \beta_0 +c_0 \beta_0 \delta(\gamma_0) \,.
\end{split}
\end{equation}
Finally, we find
\begin{equation}
    Y_\text{R} X_\text{R} Y_\text{R} 
    = \, -c_0 \delta(\gamma_0) \beta_0 + c_0 \beta_0 \delta(\gamma_0)
    = \, -c_0 \delta' (\gamma_0) 
    = \, Y_\text{R} \,.
\end{equation}
This completes the proof of \eqref{YXY=Y}. 

Let us consider \eqref{XYX=X}. We find
\begin{equation}
\begin{split}
    \, X_\text{R} Y_\text{R} X_\text{R} 
    = & \, -G_0 \delta(\beta_0) c_0 \delta'(\gamma_0) G_0 \delta(\beta_0)
    -b_0 \delta'(\beta_0) c_0 \delta'(\gamma_0) G_0 \delta(\beta_0) \\
    & \,
    -G_0 \delta(\beta_0) c_0 \delta'(\gamma_0) b_0 \delta'(\beta_0)
    -b_0 \delta'(\beta_0) c_0 \delta'(\gamma_0) b_0 \delta'(\beta_0) \\
    = & \ G_0 c_0 \delta(\beta_0) \delta'(\gamma_0) G_0 \delta(\beta_0) 
    +b_0 c_0 \delta'(\beta_0) \delta'(\gamma_0) G_0 \delta(\beta_0) \\
    & \,
    +G_0 c_0 b_0  \delta(\beta_0)  \delta'(\gamma_0) \delta'(\beta_0)
    +b_0 \delta'(\beta_0) \delta'(\gamma_0) \delta'(\beta_0) \,.
    \label{XYX intermediate1} 
    \end{split}
\end{equation}
The first term on the right-hand side of \eqref{XYX intermediate1} can be written as
\begin{equation}
\begin{split}
    & \ G_0 c_0 \delta(\beta_0) \delta'(\gamma_0) G_0 \delta(\beta_0) \\
    = & \, G_0 c_0 \delta(\beta_0) \delta'(\gamma_0) (-2b_0\gamma_0) \delta(\beta_0)
    +G_0 c_0 \delta(\beta_0) \delta'(\gamma_0) (G_0 +2b_0\gamma_0) \delta(\beta_0) \,, \label{XYX intermediate2}
\end{split}
\end{equation}
where we decomposed $G_0$ into $-2b_0 \gamma_0$ and $G_0 +2b_0\gamma_0$. Then the second term on the right-hand side of \eqref{XYX intermediate2} vanishes since $\delta'(\gamma_0) = [ \, \delta(\gamma_0) \,, \beta_0]$. Then we find
\begin{equation}
\begin{split}
    G_0 c_0 \delta(\beta_0) \delta'(\gamma_0) G_0 \delta(\beta_0)
    = & \ G_0 c_0 \delta(\beta_0) \delta'(\gamma_0) (-2b_0\gamma_0) \delta(\beta_0) \\
    = & \ 2G_0 c_0 b_0 \delta(\beta_0) \delta(\gamma_0) \delta(\beta_0) \\
    = & \ 2G_0 c_0 b_0 \delta(\beta_0) \,.
\end{split}
\end{equation}
The second term on the right-hand side of \eqref{XYX intermediate1} can be written as
\begin{equation}
\begin{split}
    & \ b_0 c_0 \delta'(\beta_0) \delta'(\gamma_0) G_0 \delta(\beta_0) \\
    = & \ b_0 c_0 \delta'(\beta_0) \delta'(\gamma_0) (-2b_0\gamma_0) \delta(\beta_0) 
    +b_0 c_0 \delta'(\beta_0) \delta'(\gamma_0) (G_0+2b_0\gamma_0) \delta(\beta_0) \\
    = & \ 2b_0 \delta'(\beta_0) \delta(\gamma_0) \delta(\beta_0)
    +(G_0+2b_0\gamma_0) b_0 c_0 \delta'(\beta_0) \delta'(\gamma_0) \delta(\beta_0) \\
    = & \ 2b_0 [\, \gamma_0 \,, \delta(\beta_0) \,] \delta(\gamma_0) \delta(\beta_0)
    + G_0 b_0 c_0  [\, \gamma_0 \,, \delta(\beta_0) \,] [\, \delta(\gamma_0) \,, \beta_0 \,] \delta(\beta_0) \\
    = & \ 2b_0 \gamma_0 \delta(\beta_0) \delta(\gamma_0) \delta(\beta_0)
    + G_0 b_0 c_0  \delta(\beta_0) \gamma_0  
    \beta_0 \delta(\gamma_0) \delta(\beta_0) \\
    = & \ 2b_0 \gamma_0 \delta(\beta_0)
    +G_0 b_0 c_0  \delta(\beta_0)  \,.
\end{split}
\end{equation}
The third term on the right-hand side of \eqref{XYX intermediate1} can be written as
\begin{equation}
\begin{split}
    G_0 c_0 b_0 \delta(\beta_0) \delta'(\gamma_0) \delta'(\beta_0)
    = & \ G_0 c_0 b_0 \delta(\beta_0) [ \, \delta(\gamma_0) \,, \beta_0 \, ] \delta'(\beta_0) \\
    = & \, -G_0 c_0 b_0 \delta(\beta_0) \delta(\gamma_0) \delta (\beta_0) \\
    = & \, -G_0 c_0 b_0 \delta(\beta_0) \,.
\end{split}
\end{equation}
The fourth term on the right-hand side of \eqref{XYX intermediate1} can be written as
\begin{equation}
\begin{split}
    b_0 \delta'(\beta_0) \delta'(\gamma_0) \delta'(\beta_0) 
    = & \, b_0 \delta'(\beta_0) [ \, \delta (\gamma_0) \,, \beta_0 \, ] \delta'(\beta_0) \\
    = & \, -b_0 \delta'(\beta_0) \delta (\gamma_0) \delta (\beta_0) 
    +b_0 \delta(\beta_0) \delta (\gamma_0) \delta'(\beta_0) \\
    = & \, -b_0 [ \, \gamma_0 \,, \delta(\beta_0) \, ] \delta (\gamma_0) \delta (\beta_0)
    +b_0 \delta(\beta_0) \delta (\gamma_0) [ \, \gamma_0 \,, \delta(\beta_0) \, ] \\
    = & \, -b_0 \gamma_0 \delta(\beta_0) \delta (\gamma_0) \delta (\beta_0) 
    -b_0 \delta(\beta_0) \delta (\gamma_0) \delta(\beta_0) \gamma_0 \\
    = & \, -b_0 \gamma_0 \delta (\beta_0) -b_0 \delta(\beta_0) \gamma_0 \,.\\
\end{split}
\end{equation}
Finally, we find
\begin{equation}
    X_\text{R} Y_\text{R}X_\text{R} 
    = \, G_0 (b_0 c_0+c_0 b_0) \, \delta(\beta_0) 
    + b_0 (\gamma_0 \delta (\beta_0) - \delta(\beta_0) \gamma_0)
    = \, G_0 \delta(\beta_0) + b_0 \delta'(\beta_0)
    = \, X_\text{R} \,.
\end{equation}
This completes the proof of \eqref{XYX=X}.

\section{Cyclicity equation for open bosonic string field theory with stubs \label{app:B}}
\setcounter{equation}{0}
In this appendix, we show that the three-string product $V_3^\text{stub}$ constructed in section~\ref{sec:stub} satisfies the cyclicity equation~\eqref{V3 cyclicity for stubs}. This cyclicity equation is translated into the following equation:
\begin{equation}
    \omega \circ \mathcal{B}_3^\text{stub} = {}-\mathcal{B}_3^\text{stub} \,.
\end{equation}
In the proof of the cyclicity equation, we use the following relations:
\begin{align}
    b_0 [h_1] = b_0 [h_3], \quad b_0 [h_2] = b_0 [h_4] \,,
    \label{b_0 is BPZ even}
\end{align}
and we prove them. We note that $h_i(\xi)$ defined in~\eqref{integration contour h} satisfies 
\begin{equation}
        h_1(\xi) = {}-h_3(\xi), \quad h_2(\xi) = {}-h_4(\xi) \,,
\end{equation}
and 
\begin{equation}
        h_1^{-1}(\xi) = \, h_3^{-1}(-\xi), \quad h_2^{-1}(\xi) = \, h_4^{-1}(-\xi) \,.
\end{equation}
Let us consider $b_0 [h_1]$. We find
\begin{equation}
    b_0 [h_1] = \, \oint \frac{dz}{2\pi i} \, 
    h_1^{-1}(z)
    \left( \frac{dh_1^{-1}(z)}{dz}\right)^{-1} \,
    b (z) \quad
    \text{with} \ z = \, h_1 (\xi) \,.
\end{equation}
Under the change of variables
\begin{equation}
    z' = \, h_3(\xi) = \, -z\,,
\end{equation}
we find
\begin{equation}
\begin{split}
      b_0 [h_1] 
    = & \, \oint \frac{-dz'}{2\pi i} \, 
    h_1^{-1}(-z')
    \left( \frac{dh_1^{-1}(-z')}{-dz'}\right)^{-1} \,
     b (-z') \\
    = & \, \oint \frac{-dz'}{2\pi i} \, 
    h_1^{-1}(z')
    \left( \frac{dh_3^{-1}(z')}{-dz'}\right)^{-1} \,
     b (-z') \quad
    \text{with} \ z' = \, h_3 (\xi) \,. 
\end{split}
\end{equation}
Since the weight of the $b$ ghost $b (z)$ is $2$, we find
\begin{equation}
\begin{split}
    b_0 [h_1]
     = & \, \oint \frac{dz'}{2\pi i} \, 
    h_1^{-1}(z')
    \left( \frac{dh_3^{-1}(z')}{dz'}\right)^{-1} \,
     b(z') \quad
    \text{with} \ z' = \, h_3 (\xi) \, \\
    = & \ b_0 [h_3] \,.
\end{split}
\end{equation}
By the same calculation, we find
\begin{equation}
b_0[h_2] = b_0[h_4] \,.
\end{equation}
Therefore, we obtain
\begin{equation}
    \mathcal{B}[h_1] = \, \mathcal{B}[h_3] \,, \quad
    \mathcal{B}[h_2] = \, \mathcal{B}[h_4] \,.
    \label{B is BPZ even}
\end{equation}
On the other hand, the action of $\omega$ on $\mathcal{B}[h_i]$ is
\begin{equation}
\begin{split}
\omega \circ \mathcal{B}[h_i] 
= & \, \int_0^1 dt \int d\tilde{t} \, e^{\{Q', t(-2w) \omega \, \circ \, b_0[h_i] \} } \\
= & \, \int_0^1 dt \int d\tilde{t} \, e^{\{Q', t(-2w) b_0[h_{i+1}] \} } \\
= & \, \mathcal{B}[h_{i+1}] \,.
\end{split}
\end{equation}
Finally, we find
\begin{equation}
\begin{split}
\omega \circ \mathcal{B}_3^\text{stub} 
= \, & \ \omega \circ \Bigl( (-\mathcal{B}[h_2] + \mathcal{B}[h_3])
e^{-wL_0[g_1]} e^{-wL_0[g_2]} e^{-wL_0[g_3]} \, e^{-wL_0[g_4]} \,\Bigr) \\
= \, & \ (-\mathcal{B}[h_3] + \mathcal{B}[h_4]) \, e^{-wL_0[g_2]} \, e^{-wL_0[g_3]} \, e^{-wL_0[g_4]} \, e^{-wL_0[g_1]} \\
= \, & \ (-\mathcal{B}[h_3] + \mathcal{B}[h_2]) \, e^{-wL_0[g_2]} \, e^{-wL_0[g_3]} \, e^{-wL_0[g_4]} \, e^{-wL_0[g_1]} \\
= \, & \ {}-\mathcal{B}_3^\text{stub} \,.
\end{split}
\end{equation}
This completes the proof of \eqref{V3 cyclicity}. We note that $\mathcal{L}_t^\text{stub}$ and $\mathcal{L}_s^\text{stub}$ are related by 
\begin{equation}
\omega \circ \mathcal{L}_t^\text{stub} = \ \mathcal{L}_s^\text{stub} \,, \quad 
\omega \circ \mathcal{L}_s^\text{stub} = \ \mathcal{L}_t^\text{stub} \,.
\end{equation}
We will use these equations in appendix~\ref{app:C3}.

\section{Completing the construction of the quartic vertices \label{app:C}}
\setcounter{equation}{0}
In this section, we complete the construction of the quartic vertices for open superstring field theory including the Ramond sector in section~\ref{sec:NSR}. We also complete the construction of the quartic vertices of open superstring field theory including the Ramond sector with stubs in section~\ref{sec:NSRstub}.
\subsection{Quartic vertices including the Ramond sector \label{app:C1}}
In this subsection, we construct the remaining three-string products in~\eqref{V3s for other NS R combination}. 
For completeness, we include the results of three-string products already constructed in section~\ref{sec:NSR}. We express three-string products in terms of Grassmann-odd operators:
\begin{align}
\langle \, R_1, Y_\text{R} V_3(N_1, R_2, N_2) \, \rangle
= & \ \langle \, g_1 \circ R_1(0) \ \Xi_\text{RNRN} \, g_2 \circ N_1(0) \, g_3 \circ R_2(0) \, g_4 \circ N_2(0) \, \rangle_D \,, \\
\langle \, N_1, V_3(R_1, N_2, R_2) \, \rangle
= & \ \langle \, g_1 \circ N_1(0) \ \Xi_\text{NRNR} \, g_2 \circ R_1(0) \, g_3 \circ N_2(0) \, g_4 \circ R_2(0) \, \rangle_D \,
\end{align}
for the quartic vertices in the RNRN group and
\begin{align}
\langle \, R_1, Y_\text{R} V_3(R_2, N_1, N_2) \, \rangle
= & \ \langle \, g_1 \circ R_1(0) \ \Xi_\text{RRNN} \, g_2 \circ R_2(0) \, g_3 \circ N_1(0) \, g_4 \circ N_2(0) \, \rangle_D \,, \\
\langle \, N_1, V_3(R_1, R_2, N_2) \, \rangle
= & \ \langle \, g_1 \circ N_1(0) \ \Xi_\text{NRRN} \, g_2 \circ R_1(0) \, g_3 \circ R_2(0) \, g_4 \circ N_2(0) \, \rangle_D \,, \label{NRRN definition} \\
\langle \, N_1, V_3(N_2, R_1, R_2) \, \rangle 
= & \ \langle \, g_1 \circ N_1(0) \ \Xi_\text{NNRR} \, g_2 \circ N_2(0) \, g_3 \circ R_1(0) \, g_4 \circ R_2(0) \, \rangle_D \,, \label{NNRR definition} \\
\langle \, R_1, Y_\text{R} V_3(N_1, N_2, R_2) \, \rangle
= & \ \langle \, g_1 \circ R_1(0) \ \Xi_\text{RNNR} \, g_2 \circ N_1(0) \, g_3 \circ N_2(0) \, g_4 \circ R_2(0) \, \rangle_D  \label{RNNR definition}
\end{align}
for the quartic vertices in the RRNN group.

We define Grassmann-even operators $X_t$ and $X_s$ for each ordering of Neveu-Schwarz and Ramond string fields:
\begin{align}
    \langle \, R_1, Y_\text{R} V_2( V_2 (N_1, R_2), N_2) \, \rangle
    = & \ \langle \, (X_t)_\text{RNRN} \, g_1 \circ R_1(0) \, g_2 \circ N_1(0) \, g_3 \circ R_2(0) \, g_4 \circ N_2(0) \, \rangle_D \,, \notag \\ 
    \langle \, R_1, Y_\text{R} V_2( N_1, V_2(R_2, N_2)) \, \rangle
    = & \ \langle \, (X_s)_\text{RNRN} \, g_1 \circ R_1(0) \, g_2 \circ N_1(0) \, g_3 \circ R_2(0) \, g_4 \circ N_2(0) \, \rangle_D \,, \notag \\
    \langle \, N_1, V_2( V_2 (R_1, N_2), R_2) \, \rangle
    = & \ \langle \, (X_t)_\text{NRNR} \, g_1 \circ N_1(0) \, g_2 \circ R_1(0) \, g_3 \circ N_2(0) \, g_4 \circ R_2(0) \, \rangle_D \,, \notag \\ 
    \langle \, N_1, V_2( R_1, V_2(N_2, R_2)) \, \rangle
    = & \ \langle \, (X_s)_\text{NRNR} \, g_1 \circ N_1(0) \, g_2 \circ R_1(0) \, g_3 \circ N_2(0) \, g_4 \circ R_2(0) \, \rangle_D \notag  
\end{align}
for the quartic vertices in the RNRN group and
\begin{align}
\langle \, R_1, Y_\text{R} V_2( V_2 (R_2, N_1), N_2) \, \rangle
= & \ \langle \, (X_t)_\text{RRNN} \, g_1 \circ R_1(0) \, g_2 \circ R_2(0) \, g_3 \circ N_1(0) \, g_4 \circ N_2(0) \, \rangle_D \,, \notag \\ 
\langle \, R_1, Y_\text{R} V_2( R_2, V_2(N_1, N_2)) \, \rangle
= & \ \langle \, (X_s)_\text{RRNN} \, g_1 \circ R_1(0) \, g_2 \circ R_2(0) \, g_3 \circ N_1(0) \, g_4 \circ N_2(0) \, \rangle_D \,, \notag \\
\langle \, N_1, V_2( V_2 (R_1, R_2), N_2) \, \rangle
= & \ \langle \, (X_t)_\text{NRRN} \, g_1 \circ N_1(0) \, g_2 \circ R_1(0) \, g_3 \circ R_2(0) \, g_4 \circ N_2(0) \, \rangle_D \notag \,, \\ 
\langle \, N_1, V_2( R_1, V_2(R_2, N_2)) \, \rangle
= & \ \langle \, (X_s)_\text{NRRN} \, g_1 \circ N_1(0) \, g_2 \circ R_1(0) \, g_3 \circ R_2(0) \, g_4 \circ N_2(0) \, \rangle_D \notag \,, \\
\langle \, N_1, V_2( V_2 (N_2, R_1), R_2) \, \rangle 
= & \ \langle \, (X_t)_\text{NNRR} \, g_1 \circ N_1(0) \, g_2 \circ N_2(0) \, g_3 \circ R_1(0) \, g_4 \circ R_2(0) \, \rangle_D \notag \,, \\
\langle \, N_1, V_2( N_2, V_2(R_1, R_2)) \, \rangle
= & \ \langle \, (X_s)_\text{NNRR} \, g_1 \circ N_1(0) \, g_2 \circ N_2(0) \, g_3 \circ R_1(0) \, g_4 \circ R_2(0) \, \rangle_D \notag \,, \\
\langle \, R_1, Y_\text{R} V_2( V_2 (N_1, N_2), R_2) \, \rangle
= & \ \langle \, (X_t)_\text{RNNR} \, g_1 \circ R_1(0) \, g_2 \circ N_1(0) \, g_3 \circ N_2(0) \, g_4 \circ R_2(0) \, \rangle_D \notag \,, \\
\langle \, R_1, Y_\text{R} V_2( N_1, V_2(N_2, R_2)) \, \rangle 
= & \ \langle \, (X_s)_\text{RNNR} \, g_1 \circ R_1(0) \, g_2 \circ N_1(0) \, g_3 \circ N_2(0) \, g_4 \circ R_2(0) \, \rangle_D \notag 
\end{align}
for the quartic vertices in the RRNN group.

In the same way as in section~\ref{sec:NSR}, we obtain the following results:
\begin{equation}
    (X_t)_\text{RNRN} = \ X_\text{R}[h_2] \,, \quad
    (X_t)_\text{RNRN} = \ X_\text{R}[h_3] \,,
\end{equation}
\begin{equation}
    (X_t)_\text{NRNR} = \ X_\text{R}[h_2] \,, \quad
    (X_t)_\text{NRNR} = \ X_\text{R}[h_3] 
\end{equation}
for the quartic vertices in the RNRN group, and
\begin{equation} 
(X_t)_\text{RRNN} = X_\text{R}[h_2] \,,\quad
(X_s)_\text{RRNN} = \ \frac{1}{3} X_\text{N} [h_1] + \frac{1}{3} X_\text{N} [g_3] + \frac{1}{3} X_\text{N} [g_4] \,, 
\end{equation}
\begin{equation}
(X_t)_\text{NRRN} = \ \frac{1}{3} X_\text{N} [g_1] + \frac{1}{3} X_\text{N} [h_2] + \frac{1}{3} X_\text{N} [g_4] \,, \quad (X_s)_\text{NRRN} = X_\text{R}[h_3] \,,
\end{equation}
\begin{equation}
(X_t)_\text{NNRR} = \ X_\text{R} [h_2] \,, \quad (X_s)_\text{NNRR} = \ \frac{1}{3} X_\text{N} [g_1] + \frac{1}{3} X_\text{N} [g_2] + \frac{1}{3} X_\text{N} [h_3] \,,
\end{equation}
\begin{equation}
(X_t)_\text{RNNR} = \ \frac{1}{3} X_\text{N} [h_4] + \frac{1}{3} X_\text{N} [g_2] + \frac{1}{3} X_\text{N} [g_3] \,, \quad (X_s)_\text{RNNR} = \ X_\text{R}[h_3] 
\end{equation}
for the quartic vertices in the RRNN group.
The $A_\infty$ relations for the three-string products $V_3(N_1, R_1, N_2)$ and $V_3(R_1, N_1, R_2)$ are translated into the following equations:
\begin{align}
    Q \cdot \Xi_\text{RNRN} = \ & (X_t)_\text{RNRN} - (X_s)_\text{RNRN} \,, \\
    Q \cdot \Xi_\text{NRNR} = \ & (X_t)_\text{NRNR} - (X_s)_\text{NRNR} \,, 
\end{align}
and the $A_\infty$ relations for $V_3(R_1, N_1, N_2)$, $V_3(R_1, R_2, N_1)$, $V_3(N_1, R_1, R_2)$, and $V_3(N_1, N_2, R_1)$ are translated into the following equations:
\begin{align}
    Q \cdot \Xi_\text{RRNN} = \ & (X_t)_\text{RRNN} - (X_s)_\text{RRNN} \,, \\
    Q \cdot \Xi_\text{NRRN} = \ & (X_t)_\text{NRRN} - (X_s)_\text{NRRN} \,, \\
    Q \cdot \Xi_\text{NNRR} = \ & (X_t)_\text{NNRR} - (X_s)_\text{NNRR} \,, \\
    Q \cdot \Xi_\text{RNNR} = \ & (X_t)_\text{RNNR} - (X_s)_\text{RNNR} \,.
\end{align}
We find the following results:
\begin{align}
    & \Xi_\text{RNRN} = \ \check{\Xi}_\text{RN}[h_2, h_3] \,, \\
    & \Xi_\text{NRNR} = \ \check{\Xi}_\text{RN}[h_2, h_3] 
\end{align}
for the quartic vertices in the RNRN group and
\begin{align}
& \Xi_\text{RRNN} = \ \frac{1}{3} \Bigl( \, \check{\Xi}_\text{RN}[h_2, h_1] + \Xi_\text{NR}[h_2, g_3] + \Xi_\text{NR}[h_2, g_4] \, \Bigr) \,, \\
& \Xi_\text{NRRN} = \ \frac{1}{3} \Bigl( \, \Xi_\text{NR}[g_1, h_3] + \check{\Xi}_\text{NR}[h_2, h_3] + \Xi_\text{NR}[g_4, h_3] \, \Bigr) \,, \label{Xi NRRN answer} \\
& \Xi_\text{NNRR} = \ \frac{1}{3} \Bigl( \, \Xi_\text{RN}[h_2, g_1] + \Xi_\text{RN}[h_2, g_2] + \check{\Xi}_\text{RN}[h_2, h_3] \, \Bigr) \,, \label{Xi NNRR answer} \\
& \Xi_\text{RNNR} = \ \frac{1}{3} \Bigl( \, \check{\Xi}_\text{NR}[h_4, h_3] + \Xi_\text{NR}[g_2, h_3] + \Xi_\text{NR}[g_3, h_3] \, \Bigr) \label{Xi RNNR answer} 
\end{align}
for the quartic vertices in the RRNN group.
Note that we used the operators $\check{\Xi}_\text{NR}[h_i, h_j]$ and $\check{\Xi}_\text{RN}[h_i, h_j]$ to avoid $\Xi_\text{NR}[h_i, h_j]$ and $\Xi_\text{RN}[h_i, h_j]$. This completes the construction of the three-string products including the Ramond sector~\eqref{V3s for other NS R combination}.

\subsection{Cyclicity equations including the Ramond sector}
In this subsection, we show that the three-string products constructed in the previous subsection satisfy the cyclicity equations~\eqref{V3 cyclicity for RRNN}, \eqref{V3 cyclicity for NRRN}, \eqref{V3 cyclicity for NNRR}, and \eqref{V3 cyclicity for RNNR}. In terms of the Grassmann-odd operators, they are translated into
\begin{equation}
\begin{split}
\omega \circ \Xi_\text{RRNN} = & {}-\Xi_\text{NRRN} \,, \quad
    \omega \circ \Xi_\text{NRRN} = {}-\Xi_\text{NNRR} \,, \\
    \omega \circ \Xi_\text{NNRR} = & {}-\Xi_\text{RNNR} \,, \quad
    \omega \circ \Xi_\text{RNNR} = {}-\Xi_\text{RRNN} \,.
    \label{cyclicity equations for other NS and R orderings}
\end{split}
\end{equation}
Let us consider the first one. 
The action of $\omega$ on the operators $\Xi_\text{RN}$ and $\Xi_\text{NR}$ are
\begin{align}
\omega \circ \Xi_\text{RN}[g_i,g_j] = & \  \Xi_\text{RN}[g_{i+1},g_{j+1}] \,, \quad
\omega \circ \Xi_\text{NR}[g_i,g_j] = \,  \Xi_\text{RN}[g_{i+1},g_{j+1}] \,, \\
\omega \circ \Xi_\text{RN}[g_i,h_j] = & \  \Xi_\text{RN}[g_{i+1},h_{j+1}] \,, \quad
\omega \circ \Xi_\text{NR}[g_i,h_j] =  \, \Xi_\text{RN}[g_{i+1},h_{j+1}] \,, \\
\omega \circ \Xi_\text{RN}[h_i,g_j] = & \  \Xi_\text{RN}[h_{i+1},g_{j+1}] \,, \quad
\omega \circ \Xi_\text{NR}[h_i,g_j] =  \, \Xi_\text{RN}[h_{i+1},g_{j+1}] \,.
\end{align}
It is straightforward to show that the following relations hold:
\begin{align}
\omega \circ \check{\Xi}_\text{RN}[h_i,h_j] = \, \check{\Xi}_\text{RN}[h_{i+1},h_{j+1}] \,, \quad
\omega \circ \check{\Xi}_\text{NR}[h_i,g_j] = \, \check{\Xi}_\text{NR}[h_{i+1},h_{j+1}] \,.
\end{align}
Then the first equation of~\eqref{cyclicity equations for other NS and R orderings} can be shown in the following way:
\begin{equation}
\begin{split}
    \omega \circ \Xi_\text{RRNN} 
    = & \ \omega \circ \left[\frac{1}{3}\Bigl( \, \check{\Xi}_\text{RN}[h_2, h_1] + \Xi_\text{RN}[h_2, g_3] + \Xi_\text{RN}[h_2, g_4] \, \Bigr) \right] \\
    = & \ \frac{1}{3}\Bigl( \, \check{\Xi}_\text{RN}[h_3, h_2] + \Xi_\text{RN}[h_3, g_4] + \Xi_\text{RN}[h_3, g_1] \, \Bigr) \\
    = & \ {}-\frac{1}{3}\Bigl( \, \check{\Xi}_\text{NR}[h_2, h_3] + \Xi_\text{NR}[g_4, h_3] + \Xi_\text{NR}[g_1, h_3] \, \Bigr) \\
    = & \ {}-\Xi_\text{NRRN} \,,
\end{split}
\end{equation}
where we used the relation between $\Xi_\text{RN}$ and $\Xi_\text{NR}$~\eqref{Xi_NR and Xi_RN antisym} and the relation between $\check{\Xi}_\text{RN}$ and $\check{\Xi}_\text{NR}$~\eqref{check_Xi_RN and NR antisym}. We can show that the remaining cyclicity equations hold in the same manner. This completes the proof of~\eqref{cyclicity equations for other NS and R orderings}.

We summarize the action of $\omega$ on $X_t$ and $X_s$ for various orderings of Neveu-Schwarz and Ramond string fields. We find
\begin{align}
& \omega \circ (X_t)_\text{RRNN} = \, (X_s)_\text{NRRN} \,, \quad
\omega \circ (X_s)_\text{RRNN} = \, (X_t)_\text{NRRN} \,, \\
& \omega \circ (X_t)_\text{NRRN} = \, (X_s)_\text{NNRR} \,, \quad
\omega \circ (X_s)_\text{NRRN} = \, (X_t)_\text{NNRR} \,, \\
& \omega \circ (X_t)_\text{NNRR} = \, (X_s)_\text{RNNR} \,, \quad
\omega \circ (X_s)_\text{NNRR} = \, (X_t)_\text{RNNR} \,, \\
& \omega \circ (X_t)_\text{RNNR} = \, (X_s)_\text{RRNN} \,, \quad
\omega \circ (X_s)_\text{RNNR} = \, (X_t)_\text{RRNN} \,. 
\end{align}
We will use these equations in appendix~\ref{app:C3}.

\subsection{Quartic vertices for open superstring field theory with stubs \label{app:C3}}
In this subsection, we construct the remaining three-string products in~\eqref{V3 with stubs}, and we show that they satisfy the cyclicity equations. For completeness, we include the results of the three-string products already constructed in section~\ref{sec:NSRstub}.

We express the three-string products in~\eqref{V3 with stubs} in terms of Grassmann-odd operators:
\begin{align}
\langle \, N_1, V_3(N_2, N_3, N_4) \, \rangle
= & \ \langle \, g_1 \circ N_1(0) \, \Xi_\text{NNNN}^\text{stub} \, g_2 \circ N_2(0) \, g_3 \circ N_3(0) g_4 \circ N_4(0) \, \rangle_D \,, \label{NNNN with stub definition} \\
\langle \, R_1, Y_\text{R} V_3(N_1, R_2, N_2) \, \rangle
= & \ \langle \, g_1 \circ R_1(0) \, \Xi_\text{RNRN}^\text{stub} \, g_2 \circ N_1(0) \, g_3 \circ R_2(0) g_4 \circ N_2(0) \, \rangle_D \,, \\
\langle \, N_1, V_3(R_1, N_2, R_2) \, \rangle
= & \ \langle \, g_1 \circ N_1(0) \, \Xi_\text{NRNR}^\text{stub} \, g_2 \circ R_1(0) \, g_3 \circ N_2(0) g_4 \circ R_2(0) \, \rangle_D \,, \label{NRNR with stub definition} \\
\langle \, R_1, Y_\text{R} V_3(R_2, N_1, N_2) \, \rangle
= & \ \langle \, g_1 \circ R_1(0) \, \Xi_\text{RRNN}^\text{stub} \, g_2 \circ R_2(0) \, g_3 \circ N_1(0) g_4 \circ N_2(0) \, \rangle_D\,, \label{RRNN with stub definition} \\
\langle \, N_1, V_3(R_1, R_2, N_2) \, \rangle
= & \ \langle \, g_1 \circ N_1(0) \, \Xi_\text{NRRN}^\text{stub} \, g_2 \circ R_1(0) \, g_3 \circ R_2(0) g_4 \circ N_2(0) \, \rangle_D\,, \label{NRRN with stub definition} \\
\langle \, N_1, V_3(N_2, R_1, R_2) \, \rangle
= & \ \langle \, g_1 \circ N_1(0) \, \Xi_\text{NNRR}^\text{stub} \, g_2 \circ N_2(0) \, g_3 \circ R_1(0) g_4 \circ R_2(0) \, \rangle_D\,, \label{NNRR with stub definition} \\
\langle \, R_1, Y_\text{R} V_3(N_1, N_2, R_2) \, \rangle
= & \ \langle \, g_1 \circ R_1(0) \, \Xi_\text{RNNR}^\text{stub} \, g_2 \circ N_1(0) \, g_3 \circ N_2(0) g_4 \circ R_2(0) \, \rangle_D\,, \label{RNNR with stub definition} \\
\langle \, R_1, Y_\text{R} V_3(R_2, R_3, R_4) \, \rangle
= & \ \langle \, g_1 \circ R_1(0) \, \Xi_\text{RRRR}^\text{stub} \, g_2 \circ R_2(0) \, g_3 \circ R_3(0) g_4 \circ R_4(0) \, \rangle_D \,. \label{RRRR with stub definition}
\end{align}
Then we define Grassmann-even operators $X_t^\text{stub}$ and $X_s^\text{stub}$ for each ordering of Neveu-Schwarz and Ramond string fields:
\begin{align}
    \langle \, N_1, V_2( V_2 (N_2, N_3), N_4) \, \rangle
    = & \ \langle \, (X_t)_\text{NNNN}^\text{stub} \, g_1 \circ N_1(0) \, g_2 \circ N_2(0) \, g_3 \circ N_3(0) \, g_4 \circ N_4(0) \, \rangle_D \,, \notag \\ 
    \langle \, N_1, V_2( N_2, V_2(N_3, N_4)) \, \rangle
    = & \ \langle \, (X_s)_\text{NNNN}^\text{stub} \, g_1 \circ N_1(0) \, g_2 \circ N_2(0) \, g_3 \circ N_3(0) \, g_4 \circ N_4(0) \, \rangle_D \,, \notag \\
    \langle \, R_1, Y_\text{R} V_2( V_2 (N_1, R_2), N_2) \, \rangle
    = & \ \langle \, (X_t)_\text{RNRN}^\text{stub} \, g_1 \circ R_1(0) \, g_2 \circ N_1(0) \, g_3 \circ R_2(0) \, g_4 \circ N_2(0) \, \rangle_D \,, \notag \\ 
    \langle \, R_1, Y_\text{R} V_2( N_1, V_2(R_2, N_2)) \, \rangle
    = & \ \langle \, (X_s)_\text{RNRN}^\text{stub} \, g_1 \circ R_1(0) \, g_2 \circ N_1(0) \, g_3 \circ R_2(0) \, g_4 \circ N_2(0) \, \rangle_D \,, \notag \\
    \langle \, N_1, V_2( V_2 (R_1, N_2), R_2) \, \rangle
    = & \ \langle \, (X_t)_\text{NRNR}^\text{stub} \, g_1 \circ N_1(0) \, g_2 \circ R_1(0) \, g_3 \circ N_2(0) \, g_4 \circ R_2(0) \, \rangle_D \,, \notag \\ 
    \langle \, N_1, V_2( R_1, V_2(N_2, R_2)) \, \rangle
    = & \ \langle \, (X_s)_\text{NRNR}^\text{stub} \, g_1 \circ N_1(0) \, g_2 \circ R_1(0) \, g_3 \circ N_2(0) \, g_4 \circ R_2(0) \, \rangle_D \,, \notag \\
    \langle \, R_1, Y_\text{R} V_2( V_2 (R_2, N_1), N_2) \, \rangle
    = & \ \langle \, (X_t)_\text{RRNN}^\text{stub} \, g_1 \circ R_1(0) \, g_2 \circ R_2(0) \, g_3 \circ N_1(0) \, g_4 \circ N_2(0) \, \rangle_D \,, \notag \\ 
    \langle \, R_1, Y_\text{R} V_2( R_2, V_2(N_1, N_2)) \, \rangle
    = & \ \langle \, (X_s)_\text{RRNN}^\text{stub} \, g_1 \circ R_1(0) \, g_2 \circ R_2(0) \, g_3 \circ N_1(0) \, g_4 \circ N_2(0) \, \rangle_D \,, \notag \\
    \langle \, N_1, V_2( V_2 (R_1, R_2), N_2) \, \rangle
    = & \ \langle \, (X_t)_\text{NRRN}^\text{stub} \, g_1 \circ N_1(0) \, g_2 \circ R_1(0) \, g_3 \circ R_2(0) \, g_4 \circ N_2(0) \, \rangle_D \notag \,, \\ 
    \langle \, N_1, V_2( R_1, V_2(R_2, N_2)) \, \rangle
    = & \ \langle \, (X_s)_\text{NRRN}^\text{stub} \, g_1 \circ N_1(0) \, g_2 \circ R_1(0) \, g_3 \circ R_2(0) \, g_4 \circ N_2(0) \, \rangle_D \notag \,, \\
    \langle \, N_1, V_2( V_2 (N_2, R_1), R_2) \, \rangle 
    = & \ \langle \, (X_t)_\text{NNRR}^\text{stub} \, g_1 \circ N_1(0) \, g_2 \circ N_2(0) \, g_3 \circ R_1(0) \, g_4 \circ R_2(0) \, \rangle_D \notag \,, \\
    \langle \, N_1, V_2( N_2, V_2(R_1, R_2)) \, \rangle
    = & \ \langle \, (X_s)_\text{NNRR}^\text{stub} \, g_1 \circ N_1(0) \, g_2 \circ N_2(0) \, g_3 \circ R_1(0) \, g_4 \circ R_2(0) \, \rangle_D \notag \,, \\
    \langle \, R_1, Y_\text{R} V_2( V_2 (N_1, N_2), R_2) \, \rangle
    = & \ \langle \, (X_t)_\text{RNNR}^\text{stub} \, g_1 \circ R_1(0) \, g_2 \circ N_1(0) \, g_3 \circ N_2(0) \, g_4 \circ R_2(0) \, \rangle_D \notag \,, \\
    \langle \, R_1, Y_\text{R} V_2( N_1, V_2(N_2, R_2)) \, \rangle 
    = & \ \langle \, (X_s)_\text{RNNR}^\text{stub} \, g_1 \circ R_1(0) \, g_2 \circ N_1(0) \, g_3 \circ N_2(0) \, g_4 \circ R_2(0) \, \rangle_D \notag \,.
\end{align}
We find that $X_t^\text{stub}$ and $X_s^\text{stub}$ can be decomposed into the product of the superstring parts $X_t$ and $X_s$ and the stub parts $\mathcal{L}_t$ and $\mathcal{L}_s$:
\begin{align}
& (X_t)_\text{NNNN}^\text{stub} = X_t \, \mathcal{L}_t^\text{stub}, 
& (X_s)_\text{NNNN}^\text{stub} & = X_s \, \mathcal{L}_s^\text{stub}, \\
& (X_t)_\text{RNRN}^\text{stub} = (X_t)_\text{RNRN} \, \mathcal{L}_t^\text{stub}, 
& (X_s)_\text{RNRN}^\text{stub} & = (X_s)_\text{RNRN} \, 
\mathcal{L}_s^\text{stub}, \\
& (X_t)_\text{NRNR}^\text{stub} = (X_t)_\text{NRNR} \, \mathcal{L}_t^\text{stub}, 
& (X_s)_\text{NRNR}^\text{stub} & = (X_s)_\text{NRNR} \, 
\mathcal{L}_s^\text{stub}, \\
& (X_t)_\text{RRNN}^\text{stub} = (X_t)_\text{RRNN} \, \mathcal{L}_t^\text{stub}, 
& (X_s)_\text{RRNN}^\text{stub} & = (X_s)_\text{RRNN} \, \mathcal{L}_s^\text{stub}, \\
& (X_t)_\text{NRRN}^\text{stub} = (X_t)_\text{NRRN} \, \mathcal{L}_t^\text{stub}, 
& (X_s)_\text{NRRN}^\text{stub} & = (X_s)_\text{NRRN} \, \mathcal{L}_s^\text{stub}, \\
& (X_t)_\text{NNRR}^\text{stub} = (X_t)_\text{NNRR} \, \mathcal{L}_t^\text{stub}, 
& (X_s)_\text{NNRR}^\text{stub} & = (X_s)_\text{NNRR} \, \mathcal{L}_s^\text{stub}, \\
& (X_t)_\text{RNNR}^\text{stub} = (X_t)_\text{RNNR} \, \mathcal{L}_t^\text{stub}, 
& (X_s)_\text{RNNR}^\text{stub} & = (X_s)_\text{RNNR} \, \mathcal{L}_s^\text{stub}, \\
& (X_t)_\text{RRRR}^\text{stub} = \mathcal{L}_t^\text{stub}, 
& (X_s)_\text{RRRR}^\text{stub} & = \mathcal{L}_s^\text{stub}.
\end{align}
Since the $t$-channel contribution and the $s$-channel contribution can be expressed as the product of the stub part and the stubless superstring part, we obtain the following results:
\begin{align}
\Xi_\text{NNNN}^\text{stub} 
= & \ \Bigl( X_t \, (-\mathcal{B}[h_2]) + \Xi
+ X_s \, \mathcal{B}[h_3] \Bigr) \, e^{-wL_0[g_1]} \, e^{-wL_0[g_2]} \, e^{-wL_0[g_3]} \, e^{-wL_0[g_4]} \,,
\end{align}
\vspace{-6mm}
\begin{align}
\Xi_\text{RNRN}^\text{stub} 
= & \ \Bigl( (X_t)_\text{RNRN} \, (-\mathcal{B}[h_2]) + \Xi_\text{RNRN}
+ (X_s)_\text{RNRN} \, \mathcal{B}[h_3] \Bigr)  \\
& \qquad \times e^{-wL_0[g_1]} \, e^{-wL_0[g_2]} \, e^{-wL_0[g_3]} \, e^{-wL_0[g_4]} \,, \notag \\
\Xi_\text{NRNR}^\text{stub} 
= & \ \Bigl( (X_t)_\text{NRNR} \, (-\mathcal{B}[h_2]) + \Xi_\text{NRNR}
+ (X_s)_\text{NRNR} \, \mathcal{B}[h_3] \Bigr) \label{Xi_NRNR^stub answer} \\
& \qquad \times e^{-wL_0[g_1]} \, e^{-wL_0[g_2]} \, e^{-wL_0[g_3]} \, e^{-wL_0[g_4]} \,, \notag \\
\Xi_\text{RRNN}^\text{stub} 
= & \ \Bigl( (X_t)_\text{RRNN} \, (-\mathcal{B}[h_2]) + \Xi_\text{RRNN}
+ (X_s)_\text{RRNN} \, \mathcal{B}[h_3] \Bigr) \\
& \qquad \times e^{-wL_0[g_1]} \, e^{-wL_0[g_2]} \, e^{-wL_0[g_3]} \, e^{-wL_0[g_4]} \,, \notag \\
\Xi_\text{NRRN}^\text{stub} 
= & \ \Bigl( (X_t)_\text{NRRN} \, (-\mathcal{B}[h_2]) + \Xi_\text{NRRN}
+ (X_s)_\text{NRRN} \, \mathcal{B}[h_3] \Bigr) \\
& \qquad \times e^{-wL_0[g_1]} \, e^{-wL_0[g_2]} \, e^{-wL_0[g_3]} \, e^{-wL_0[g_4]} \,, \notag \\
\Xi_\text{NNRR}^\text{stub} 
= & \ \Bigl( (X_t)_\text{NNRR} \, (-\mathcal{B}[h_2]) + \Xi_\text{NNRR}
+ (X_s)_\text{NNRR} \, \mathcal{B}[h_3] \Bigr) \\
& \qquad \times e^{-wL_0[g_1]} \, e^{-wL_0[g_2]} \, e^{-wL_0[g_3]} \, e^{-wL_0[g_4]} \,, \notag \\
\Xi_\text{RNNR}^\text{stub} 
= & \ \Bigl( (X_t)_\text{RNNR} \, (-\mathcal{B}[h_2]) + \Xi_\text{RNNR}
+ (X_s)_\text{RNNR} \, \mathcal{B}[h_3] \Bigr) \\
& \qquad \times e^{-wL_0[g_1]} \, e^{-wL_0[g_2]} \, e^{-wL_0[g_3]} \, e^{-wL_0[g_4]} \,, \notag \\
\Xi_\text{RRRR}^\text{stub} 
= & \ \mathcal{B}_3^\text{stub} \,.
\end{align}
The cyclicity equations are translated into  
\begin{equation}
\begin{split}
&\omega \circ \Xi_\text{NNNN}^\text{stub} = {}-\Xi_\text{NNNN}^\text{stub} \,, \\
\omega \circ \Xi_\text{RNRN}^\text{stub} = {}-\Xi_\text{NRNR}^\text{stub} \,, \quad 
& \omega \circ \Xi_\text{NRNR}^\text{stub} = {}-\Xi_\text{RNRN}^\text{stub} \,, \quad 
\omega \circ \Xi_\text{RRNN}^\text{stub} = {}-\Xi_\text{NRRN}^\text{stub} \,, \\
\omega \circ \Xi_\text{NRRN}^\text{stub} = {}-\Xi_\text{NNRR}^\text{stub} \,, \quad
& \omega \circ \Xi_\text{NNRR}^\text{stub} = {}-\Xi_\text{RNNR}^\text{stub} \,, \quad
\omega \circ \Xi_\text{RNNR}^\text{stub} = {}-\Xi_\text{RRNN}^\text{stub} \,.
\label{Xi cyclicity for NS and R with stubs}
\end{split}
\end{equation}
Let us consider the first equation of~\eqref{Xi cyclicity for NS and R with stubs}. The operator $\Xi_\text{NNNN}^\text{stub}$ has the same structure as $\Xi_\text{RNRN}^\text{stub}$, and its cyclicity equation follows from the cyclicity equation
\begin{equation}
    \omega \circ \Xi = {}-\Xi \,,
\end{equation}
which is equivalent to~\eqref{V3 cyclicity for NS}, and the fact that $X_t$ and $X_s$ satisfy
\begin{equation}
    \omega \circ X_t =X_s, \quad  \omega \circ X_t =X_s \,.
\end{equation}
We can show that other equations in~\eqref{Xi cyclicity for NS and R with stubs} holds. This completes the proof of the cyclicity equations for the three-string products including the Ramond sector with stubs~\eqref{V3 with stubs}.

\end{document}